\documentclass[12pt,preprint]{aastex}
\usepackage{graphicx}
\usepackage{lscape}

\long\def\symbolfootnote[#1]#2{\begingroup%
\def\thefootnote{\fnsymbol{footnote}}\footnote[#1]{#2}\endgroup} 

\defcitealias{greene10}{Greene, Peng \& Ludwig
(2010, \citealt{greene10})}

\defcitealias{vestergaard06}{Vestergaard \& Peterson
2006, \citealt{vestergaard06}}

\defcitealias{collin06}{Collin et al. 2006}
\defcitealias{baskin05}{Baskin \& Laor, 2005}
\defcitealias{McGill08}{McGill et al. 2008}

\newcommand{\Halpha}{\ifmmode {\rm H}\alpha \else H$\alpha$\fi}
\newcommand{\Hbeta}{\ifmmode {\rm H}\beta \else H$\beta$\fi}
\newcommand{\feii}{Fe\,{\sc ii}}
\newcommand{\civ}{C\,{\sc iv}}
\newcommand{\Civ}{C\,{\sc iv}\,$\lambda 1549$}
\newcommand{\nii}{N\,{\sc ii}}
\newcommand{\niidb}{[N\,{\sc ii}]\,$\lambda \lambda 6548,6583$}
\newcommand{\oiii}{O\,{\sc iii}}
\newcommand{\ob}{[O\,{\sc iii}]\,$\lambda \lambda 4959,5007$}
\newcommand{\lam}{$\lambda$}
\newcommand{\civm}{{\rm C}\,{\scriptstyle{\rm IV}}}

\begin{document}

\title{Black Hole Mass Estimates Based on \civ\ are Consistent with
Those Based on the Balmer Lines\altaffilmark{\dagger}}

\author{
  R.J.~Assef\altaffilmark{1,2,3},
  K.D.~Denney\altaffilmark{1,4},
  C.S.~Kochanek\altaffilmark{1,5},
  B.M.~Peterson\altaffilmark{1,5},
  S.~Koz\l owski\altaffilmark{1},
  N.~Ageorges\altaffilmark{6},
  R.S.~Barrows\altaffilmark{7},
  P.~Buschkamp\altaffilmark{6},
  M.~Dietrich\altaffilmark{1},
  E.~Falco\altaffilmark{8},
  C.~Feiz\altaffilmark{9},
  H.~Gemperlein\altaffilmark{6},
  A.~Germeroth\altaffilmark{9},
  C.J.~Grier\altaffilmark{1},
  R.~Hofmann\altaffilmark{6},
  M.~Juette\altaffilmark{10},
  R.~Khan\altaffilmark{1},
  M.~Kilic\altaffilmark{8},
  V.~Knierim\altaffilmark{10},
  W.~Laun\altaffilmark{11},
  R.~Lederer\altaffilmark{6},
  M.~Lehmitz\altaffilmark{11},
  R.~Lenzen\altaffilmark{11},
  U.~Mall\altaffilmark{11},
  K.K.~Madsen\altaffilmark{12},
  H.~Mandel\altaffilmark{9},
  P.~Martini\altaffilmark{1,5},
  S.~Mathur\altaffilmark{1,5},
  K.~Mogren\altaffilmark{1},
  P.~Mueller\altaffilmark{9},
  V.~Naranjo\altaffilmark{11},
  A.~Pasquali\altaffilmark{11},
  K.~Polsterer\altaffilmark{10},
  R.W.~Pogge\altaffilmark{1,5},
  A.~Quirrenbach\altaffilmark{9},
  W.~Seifert\altaffilmark{9},
  D.~Stern\altaffilmark{2},
  B.~Shappee\altaffilmark{1},
  C.~Storz\altaffilmark{11},
  J.~Van Saders\altaffilmark{1},
  P.~Weiser\altaffilmark{13} and
  D.~Zhang\altaffilmark{1}
}

\altaffiltext{1} {Department of Astronomy, The Ohio State
  University, 140 W.\ 18th Ave., Columbus, OH 43210, USA
  [email:{\tt{rjassef@astronomy.ohio-state.edu}}]}

\altaffiltext{2} {Jet Propulsion Laboratory, California Institute of
  Technology, MS 169-530, 4800 Oak Grove Drive, Pasadena, 91109, USA}

\altaffiltext{3} {NASA Postdoctoral Program Fellow}

\altaffiltext{4} {DARK Fellow, Dark Cosmology Centre, Niels Bohr
  Institute, University of Copenhagen, Juliane Maries Vej 30, 2100
  Copenhagen, Denmark}

\altaffiltext{5} {The Center for Cosmology and Astroparticle Physics,
The Ohio State University, 191 West Woodruff Avenue, Columbus, OH
43210, USA}

\altaffiltext{6} {Max-Planck-Institut fuer Extraterrestrische
Physik, Giessenbachstr., D-85748 Garching, Germany}

\altaffiltext{7} {Arkansas Center for Space and Planetary Sciences,
  University of Arkansas, Fayetteville, AR 72701}

\altaffiltext{8} {Harvard-Smithsonian Center for Astrophysics, 60
Garden Street, Cambridge, MA 02138, USA}

\altaffiltext{9} {Landessternwarte, ZAH, Koenigstuhl 12, D-69117
Heidelberg, Germany}

\altaffiltext{10} {Astron. Institut der Ruhr Univ. Bochum,
Universitaetsstr. 150, D-44780 Bochum, Germany}

\altaffiltext{11} {Max-Planck-Institut fuer Astronomie, Koenigstuhl
  17, D-69117 Heidelberg, Germany}

\altaffiltext{12} {California Institute of Technology, 1200
  E. California Blvd, Pasadena, CA 91125, USA}

\altaffiltext{13} {Fachhochschule fuer Technik und Gestaltung,
  Windeckstr. 110, D-68163 Mannheim, Germany}

\altaffiltext{$\dagger$} {This works relies partly on observations
  the Large Binocular Telescope. The LBT is an international
  collaboration among institutions in the United States, Italy and
  Germany. LBT Corporation partners are: The Ohio State University;
  The University of Arizona on behalf of the Arizona university
  system; Istituto Nazionale di Astrofisica, Italy; LBT
  Beteiligungsgesellschaft, Germany, representing the Max-Planck
  Society, the Astrophysical Institute Potsdam, and Heidelberg
  University; and The Research Corporation, on behalf of The
  University of Notre Dame, University of Minnesota and University of
  Virginia.}

\begin{abstract}
  Using a sample of high-redshift lensed quasars from the CASTLES
  project with observed-frame ultraviolet or optical and near-infrared
  spectra, we have searched for possible biases between supermassive
  black hole (BH) mass estimates based on the \civ, H$\alpha$ and
  H$\beta$ broad emission lines. Our sample is based upon that of
  Greene, Peng \& Ludwig, expanded with new near-IR spectroscopic
  observations, consistently analyzed high $S/N$ optical spectra, and
  consistent continuum luminosity estimates at 5100\AA. We find that
  BH mass estimates based on the FWHM of \civ\ show a systematic
  offset with respect to those obtained from the line dispersion,
  $\sigma_l$, of the same emission line, but not with those obtained
  from the FWHM of H$\alpha$ and H$\beta$. The magnitude of the offset
  depends on the treatment of the He\,{\sc ii} and Fe\,{\sc ii}
  emission blended with \civ, but there is little scatter for any
  fixed measurement prescription. While we otherwise find no
  systematic offsets between \civ\ and Balmer line mass estimates, we
  do find that the residuals between them are strongly correlated with
  the ratio of the UV and optical continuum luminosities. This means
  that much of the dispersion in previous comparisons of \civ\ and
  H$\beta$ BH mass estimates are due to the continuum luminosities
  rather than any properties of the lines. Removing this dependency
  reduces the scatter between the UV- and optical-based BH mass
  estimates by a factor of approximately 2, from roughly 0.35 to
  0.18~dex. The dispersion is smallest when comparing the
  \civ\ $\sigma_l$ mass estimate, after removing the offset from the
  FWHM estimates, and either Balmer line mass estimate. The
  correlation with the continuum slope is likely due to a combination
  of reddening, host contamination and object-dependent SED
  shapes. When we add additional heterogeneous measurements from the
  literature, the results are unchanged. Moreover, in a trial
  observation of a remaining outlier, the origin of the deviation is
  clearly due to unrecognized absorption in a low $S/N$ spectrum. This
  not only highlights the importance of the quality of the
  observations, but also raises the question if whether cases like
  this one are common in the literature, further biasing comparisons
  between \civ\ and other broad emission lines.
\end{abstract}


\section{Introduction}\label{sec:intro}

It is thought that every massive galaxy has a supermassive black hole
(BH) at its center, and some physical properties of the BH appear to
be tightly correlated with those of the galaxy. In particular, the
mass of the central BH correlates well with the luminosity of the
spheroidal component of the host \citep[see,
  e.g.,][]{marconi03,graham07} and with its velocity dispersion
\citep[see,
  e.g.,][]{ferrarese00,gebhardt00,tremaine02,gultekin09,graham11}.
Both of these properties of galaxies have physical scales a few orders
of magnitude larger than the sphere of influence of the BH, so
mechanisms linking their properties are not immediately
apparent. Theoretical models try to account for the correlation
through co-evolution of the galaxy and its BH, in which accretion
induced by galaxy mergers regulates the BH's growth, and feedback from
the accretion regulates the growth of the galaxy by quenching star
formation and removing cold gas
\citep[e.g.,][]{granato04,hopkins05,hopkins06,hopkins08,somerville08,shankar09}. However,
the existence of these correlations does not necessarily imply
co-evolutionary mechanisms, as some authors argue that they can be a
simple consequence of mergers and the central limit theorem
\citep{peng07,peng10,jahnke10}.

Direct measurements of BH masses in inactive galaxies are only
possible for a small number of nearby objects because it is necessary,
or at least desirable \citep[see, e.g.,][]{merrit01,gultekin09}, to
resolve the BH's sphere of influence in order to determine the BH mass
from the kinematics of the stars and gas closest to it. Galaxies with
active nuclei (AGNs) offer a completely different means of estimating
BH masses at any distance. In particular, Type 1 AGNs show bright
broad emission lines in their spectra produced by gas in the broad
line region (BLR), which is close to the central black hole but
outside the hot accretion disk. The large line-widths are thought to
arise from the Doppler broadening due to the orbital velocity of the
gas around the BH. Thus, measuring the mass of the central BH from the
width of the broad lines is possible if the distance of the BLR from
the BH is known.

This distance can be directly measured with the reverberation mapping
(RM) technique \citep{blandford82,peterson93}. This technique works by
measuring the light travel time between the continuum and the
broad-line emitting regions, which is derived from the time lag
between changes in their respective luminosities. Unfortunately the
timescale over which appreciable variability is observed in AGNs
increases with BH mass \citep[e.g.,][]{vandenberk04, wilhite08,
  kelly09,macleod10}, making it difficult (i.e., more time intensive)
to apply RM to the luminous QSOs that possess the most massive
BHs. For example, \citet{macleod10} find that for a typical quasar
with $M_{\rm BH} = 10^8~M_{\odot}$ (typical magnitude of $M_i\approx
-23~\rm mag$), the rest-frame timescale, $\Delta t$, required to reach
an r.m.s. variability amplitude of 0.1~mag is approximately $45~\rm
days$, while for a quasar with $M_{\rm BH} = 10^9~M_{\odot}$
($M_i\approx -25.5~\rm mag$), $\Delta t$ is approximately $125~\rm
days$. This is further complicated by the time dilation due to the
higher redshift of these rare objects. It has been shown, however,
that the distance from the BH to the BLR correlates well with the
continuum luminosity of the AGN \citep[see,
  e.g.,][]{kaspi00,kaspi05,bentz06,bentz09,zu10}. Given this
correlation, BH masses can be estimated for distant broad-line quasars
for which the RM technique is not reasonably applicable. Masses
estimated in this way are usually referred to as single epoch (SE) BH
mass estimates.

Because it is generally easier to obtain optical rather than UV or IR
spectra, SE BH masses are typically estimated from the H$\beta$ and
H$\alpha$ broad emission lines and the continuum luminosity at
5100\AA\ at low redshifts ($z\lesssim 0.7$). The overlap with RM
targets has allowed for very accurate calibration of these SE mass
estimators
\citepalias{collin06,vestergaard06,McGill08}.\defcitealias{vestergaard06}{\citet{vestergaard06}}At
high redshift, however, these emission lines are shifted into the IR,
and most mass estimates are then based upon the UV Mg\,{\sc
ii}\,\lam2798 and \Civ\ broad emission lines and the continuum
luminosities at 3000\AA\ for Mg\,{\sc ii} and 1450\AA\ or 1350\AA\
for \civ. Unlike the Balmer emission lines, these UV lines lack large
local calibration samples because of the difficulty of obtaining
UV-based RM measurements. \citet{onken08} have argued that the
Mg\,{\sc ii} line can provide accurate mass estimates, but that there
is a small, but significant, dependence on the Eddington ratio of the
AGN. \civ, on the other hand, is not thought to have this bias, and
\citet{vestergaard06} have calibrated a \civ-based mass estimator
based on local RM AGNs using space-based UV spectra. However, there
are still concerns about whether the \civ\ velocity widths are
attributed solely to gravity or if there are bulk flows due to winds
of ejected material, and the impact of these effects on the accuracy
of \civ-based BH mass estimates is still debated. For example, \civ\
is typically slightly displaced in wavelength (usually blueshifted)
with respect to the rest of the quasar emission lines \citep[see,
e.g.,][]{gaskell82,tytler92,richards02}, and frequently shows broad
absorption features \citep[e.g.,][]{weymann81} and strong line
asymmetries correlated with quasar properties
\citep[e.g.,][]{wilkes84,richards02,leighly04}.

The simplest approach to test the reliability of \civ\ mass estimates
is to systematically compare them to Balmer line estimates for the
same sources \citep[see, e.g.,][]{dietrich09}. High redshift lensed
quasars are some of the best targets for such tests. Generally, the
problem is that the high redshift makes it easy to observe the
\civ\ line, but the better calibrated H$\alpha$ and H$\beta$ lines lie
in the near-IR, where it is difficult to observe them. Magnification
increases the apparent brightness of the lensed quasars, and also,
because their observed brightness is not uniquely determined by their
intrinsic luminosity and distance, it helps to mitigate any Eddington
biases in the sample or, in other words, it makes objects in the
sample unlikely to be preferentially brighter than average for their
BH mass.

In a recent work, \citetalias{greene10} presented near-IR spectral
observations for a sub-sample of lensed quasars from the CfA-Arizona
Space Telescope LEns Survey (CASTLES) of gravitational lenses
\citep{falco01} whose \civ\ or Mg\,{\sc ii} BH masses had been
estimated in a previous work by \citet{peng06}. \citet{greene10}
measured, whenever possible, the full width at half maximum (FWHM) of
the H$\beta$ and H$\alpha$ emission lines of these objects and found
no systematic biases between BH masses estimated from these lines and
those estimated from \civ. Their sample, however, did not cover a
large enough range in BH mass to decide whether there was a mass
dependent slope to the relation between the masses. This comparison
also suffered from the fact that \citet{peng06} lacked access to the
original UV/optical spectra for many targets and frequently had to
rely on the printed spectra in published papers to measure
line widths.

In this work we start from the sample of \citet{greene10} and attempt
to improve on both of these issues. First, we add Balmer line based BH
mass estimates for the lens SDSS1138+0314 and make revised estimates
based on new, higher $S/N$, spectra of HS0810+2554 and SBS0909+532. We
obtained near-IR observations for SDSS1138+0314 and HS0810+2554 using
the newly commissioned Large Binocular Telescope (LBT) NIR
Spectrograph Utility Camera and Integral Field Unit
\citep[LUCIFER;][]{seifert03,ageorges10}, while for SBS0909+532 we
use the UV through IR observations of \citet{mediavilla10}. Second, we
made consistent \civ\ BH mass estimates from high $S/N$ spectra using
the original observations analyzed by \citet{peng06}, other published
or unpublished spectra, or new spectra for all targets in the
sample. Finally, we obtained continuum luminosities at 5100\AA\ for
all objects in the sample in a consistent manner. This allows us to
include the lenses SDSS0246--0825, HS0810+2554 and Q2237+030, which
were excluded by \citet{greene10}. With these additions we expand the
sample of \citet{greene10} with both \civ\ and Balmer lines mass
estimates from 7 to 12 quasars and the mass range covered by
approximately 0.5~dex. In \S\ref{sec:sample} we describe the sample of
gravitationally lensed quasars we use in this study as well as our
observations. In \S\ref{sec:methods} we describe the methods we use to
measure emission line velocity widths and their uncertainties, the
continuum luminosities of the quasars and the SE BH masses. In
\S\ref{sec:civ_comp} we compare the different mass estimates we have
derived and determine the possible biases we measure between them
while in \S\ref{sec:comp_others} we expand our results using a
heterogeneous sample of measurements from other studies. In
\S\ref{sec:conclusions} we summarize the conclusions. In an appendix
we discuss individual objects in detail. We use a standard
$\Lambda$CDM cosmology with $\Omega_M=0.3$, $\Omega_{\Lambda}=0.7$ and
$H_0=73~\rm km\ \rm s^{-1}\ \rm Mpc^{-1}$ throughout the paper.

\section{The Sample of Lensed QSOs}\label{sec:sample}

We selected 12 lensed quasars from the CASTLES survey with high
quality UV/optical, typically ground-based, spectra of \civ\ and
either published near-IR spectra of the Balmer lines or IR magnitudes
bright enough to obtain such spectra. The targets are listed in Table
\ref{tab:magnifications}. All 12 objects have been observed by CASTLES
with {\it{HST}} in the {\it{V}} (F555W), {\it{I}} (F814W) and {\it{H}}
(F160W) bands, except for B1422+231, which was not observed in
{\it{I}}.

We start from the sample of \citet{greene10}, who observed most of
these lensed quasars in the near-IR with the Triplespec spectrograph
at the Apache Point Observatory. The wavelength range of these spectra
is 0.95--2.46~$\mu$m with $R=3500$, and either the H$\beta$ or
H$\alpha$ (or both) emission line is observable in one of the
atmospheric windows. Although \citet{greene10} considered objects with
a large span of redshift and reddening, we limit our sample to objects
with sufficiently high redshift and small enough reddening for \civ\
emission to be observable in ground based UV/optical
spectra\footnote{\citet{peng06} mistakenly quote a \civ\ BH mass
estimate for the lens J1004+1229 that, in fact, corresponds to the
lens SDSS1004+4112 (C.Y.~Peng, private communication). This error was
propagated into the analysis of \citet{greene10}. The quasar in
J1004+1229 is highly reddened and it is not possible to see its \civ\
emission in a UV/optical spectrum.}. \citet{greene10} presented FWHM
velocity width measurements for all the objects in their sample but
did not present BH mass estimates for three of them. For these three
lensed QSOs, SDSS0246--0825, HS0810+2554 and Q2237+030, we have
measured the continuum luminosity and estimated BH masses so we can
include them in our sample.

We obtained near-IR spectra in the {\it{H}} and {\it{K}} band for
SDSS1138+0314 (Fig. \ref{fg:spec}) and in the {\it{J-}}band for
HS0810+2554 (Fig. \ref{fg:spec_hs0810}) with the LBT LUCIFER
spectrograph. The first was obtained as part of the LUCIFER science
demonstration time and is discussed here, while the second was a
target of a separate project to be presented by \citet{mogren10}. We
also analyzed the near-IR {\it{J-}} and {\it{H-}}band observations of
SBS0909+532 presented by \citet{mediavilla10}, shown in Figure
\ref{fg:spec_sbs0909}.

\subsection{LUCIFER Observations of SDSS1138+0314}\label{sssec:sdss1138_luci}

We obtained a near-infrared spectrum of SDSS1138+0314 using the new
LUCIFER instrument at the LBT during its science demonstration
time. LUCIFER is a near-infrared spectrograph and imager with an
overall wavelength range of 0.85 -- 2.5~$\mu$m. We observed
SDSS1138+0314 in the longslit mode with the {\tt{OrderSep}} filter, a
0\farcs 5 slit, the \verb#200_H+K# grating and the {\tt{N1.8}} camera
for a total integration time of 840s over 7 dithered exposures during
the night of UTC 2010-01-04. This configuration gives an effective
wavelength range of 1.49 -- 2.4~$\mu$m, which includes both the
{\it{H}} and {\it{K}} bands, with a resolving power of 1880 at
{\it{H}} and 2570 at {\it{K}}. The slit was oriented to include images
A and C of the lensed quasar, as well as part of the lens galaxy. No
emission from the lens galaxy is detected in our data. The B9V star
HIP 33350 was observed with the same configuration, except for a
change in slit width from 0\farcs5 to 1\arcsec, and was used to
correct the spectrum of SDSS1138+0314 for telluric absorption
features. The difference in resolution caused by the different slit
widths degrades our telluric corrections, but has little consequence
for measuring the width of broad emission lines. We estimated the
seeing was $\sim0\farcs8$ during the observations.

We reduced the data using standard IRAF packages in combination with
the \verb+IDL+ task \verb+xtellcor_general+ of \citet{vacca03} for the
telluric absorption corrections. We performed a 2-D wavelength
calibration on each of the 7 exposures using the sky emission lines
and built a sky frame by median combining them. The sky frame was then
used to remove the sky from each exposure before extracting the
spectrum. We also did an alternate sky subtraction of the spectra
using a version of the COSMOS software modified to work on LUCIFER
data. This software, designed for reduction of spectral observations
with IMACS \citep{dressler06} and LDSS-3 \citep[upgraded from
LDSS-2,][]{smith94} on the Magellan telescopes, follows the procedures
of \citet{kelson03}. It produces an accurate model of the sky emission
by creating a sub-pixel resolution map of the sky line profiles using
the full extent of the lines in the spectrum coupled with a model of
the optical distortions. Both extractions of the spectra yield
equivalent results, and both are shown in Figure \ref{fg:spec}. While
in principle we could use the telluric standard to perform an absolute
flux calibration, it is hard to model the slit losses, especially
considering the difference in the slit widths. Instead, the flux
calibration was performed by convolving the spectrum corrected for
telluric absorption with the NICMOS F160W filter curve and matching it
to the estimated de-magnified absolute magnitude of the quasar from
the CASTLES {\it{HST}} imaging of this lens (see
\S\ref{ssec:lum_method} for details on the lens magnification). The
blue edge of the LUCIFER SDSS1138+0314 spectrum is somewhat redder
than the blue edge of the F160W band, so we extended the observed
spectrum using the AGN SED template of \citet{assef10a} assuming no
reddening. Note that the {\it{HST}} NICMOS observations were obtained
on UTC 2003-11-06, approximately 6 years before the LUCIFER
observations, so we attempt to correct for the intrinsic variability
of the quasar. However, this is typically not an important correction
(see \S\ref{ssec:lum_method}). We use the $R-$band light curves
obtained with the SMARTS 1.3~m telescope for a gravitational lens
monitoring project \citep[see][]{morgan10}. These data show that the
quasar intrinsically brightened by $56\pm 17\%$ between UTC 2004-02-03
and UTC 2010-01-09. We assume that no significant variability occurred
between the {\it{HST}} NICMOS and the first SMARTS observations and
between the LUCIFER and the last SMARTS observations.

From an optical spectrum of SDSS1138+0314, \citet{eigenbrod06}
estimated a redshift of $z=2.438$ for the quasar, while SDSS provides
$z=2.4427\pm 0.0014$. Using the narrow component of H$\alpha$ and the
[OIII]\,$\lambda\lambda$\,4959, 5007 emission lines, we obtained
$z=2.4417$, consistent with SDSS. We did not use the [NII] lines or
the narrow component of H$\beta$ as they could not be centroided
accurately because of blending with the broad H$\alpha$ and H$\beta$
profiles, respectively.

\subsection{UV/Optical Spectra}\label{ssec:uv_opt_obs}

For most of the \citet{greene10} sample, as well as for SDSS1138+0314,
we found suitable high $S/N$ optical spectroscopic observations in the
literature that the owners kindly made available for this study (see
Table \ref{tab:magnifications} for the references, where applicable,
and Appendix \ref{sssec:notes} for details on each object). When
needed, we performed an absolute flux calibration using photometry
from several different sources, as this was not always required for
the science goals of the original project. All the UV/optical spectra
compiled from the literature are shown in Figure
\ref{fg:all_uv_opt_spec}.

We could not locate suitable optical spectra for HS0810+2554 and
FQB1633+3134. Both objects were observed by the SDSS spectroscopic
survey, but these spectra did not have high enough $S/N$ to provide
accurate line-width measurements with good continuum subtraction. We
obtained new optical spectra of these objects using the MDM
observatory 2.4m Hiltner telescope with the Boller \& Chivens CCD
Spectrograph\footnote{\url{http://www.astronomy.ohio-state.edu/MDM/CCDS/}}
(CCDS). HS0810+2554 was observed on UTC 2010-02-24 with a grating
center of 5300\AA\ and was flux calibrated using the standard star
Feige 34. FBQ1633+3134 was observed on UTC 2010-03-21 and UTC
2010-03-22 with a grating center of 4700\AA\ and was flux calibrated
using the standard star Feige 98. Absolute fluxes were obtained for
both objects by performing a cross-calibration between SDSS
{\it{g-}}band photometry of other objects in the field and
{\it{g-}}band photometric observations with the RETROCAM instrument
\citep{morgan05} obtained on UTC 2010-03-06 and UTC 2010-03-22 for
HS0810+2554 and FBQ1633+3134, respectively. The reduced spectra are
shown in Figure \ref{fg:all_uv_opt_spec}.

\section{Models and Measurements}\label{sec:methods}

In this section, we briefly discuss the methods we use to measure the
line widths and estimate the black hole masses from the optical and
near-IR spectra.

\subsection{Line-width Measurements}\label{ssec:lw_method}

There is no standard prescription for measuring the line-width
characterizations of the broad \civ\ emission line in QSOs. While for
other emission lines this may not be a significant source of
uncertainties, there is a shelf-like emission feature redward of
\civ\ that blends with the line profile and is created by a
combination of broad He\,{\sc II}\,\lam1640, OIII]\,\lam1663, and a
  feature of unknown origin at 1600\AA\ usually referred to as the
  \lam1600 feature \citep{laor94,marziani96,fine10}. While the
  \lam1600 feature is commonly thought to correspond to Fe\,{\sc ii},
  \citet{fine10} argue that this cannot account for all the observed
  flux, yet it is also unlikely that \civ\ can reach large enough
  velocities to produce the feature. Different prescriptions for
  modeling the blended emission can have significant effects on line
  width estimates \citep{denney09,fine10}, so it is important to
  explore how these affect our results. \citet{fine10} explored three
  different and widely used approaches and their effects on the
  \civ\ width measurements. The three prescriptions are: (1) to assume
  that the \lam1600 feature corresponds to \civ\ emission and
  therefore remove only the He\,{\sc II}\,\lam1640 and OIII]\,\lam1663
    contributions; (2) to assume that the \lam1600 feature belongs to
    a different species from \civ\ and so removing its contribution
    along with that of the other two components on the shelf; and (3)
    to fit the \lam1600 feature as part of the continuum
    \citep[see][for details on each prescription]{fine10}. While
    \citet{fine10} selects prescription (2) as their preferred method,
    in large part because it produces symmetric \civ\ profiles, it is
    hard to apply this approach to low $S/N$ data \citep[see][for
      details]{fine10}. Moreover, it is not guaranteed to produce more
    accurate BH masses than the other two prescriptions. The simple
    prescription of (3) produces line-width characterizations that are
    systematically smaller than prescription (2) but with very low
    dispersion between individual measurements, while (1) produces
    estimates with a larger scatter relative to (2) but without a
    systematic offset. The differences between the prescriptions is
    smallest for FWHM and largest for the line dispersion, $\sigma_l$.

Based, in part, on these issues, we considered two different
prescriptions for removing the continuum and blended emission from the
\civ\ emission line profile. Both prescriptions are amenable to large
scale automated use. The first prescription, which we will refer to as
prescription A, is very similar to that used by \citet{vestergaard06},
where the shelf feature redward of \civ\ is considered part of the
\civ\ line profile, but only the region within $\pm10,000~\rm km~\rm
s^{-1}$ of the peak is considered. The continuum is fit by linearly
interpolating between the two continuum windows in the wavelength
ranges 1425--1470 and 1680-1705\AA. When these continuum windows were
affected by absorption, we slightly shifted them as detailed in Table
\ref{tab:uv_opt_lineboundaries}. Our continuum fitting is in principle
different from that of \citet{vestergaard06}, who considered 5
different continuum windows and then fit a power-law to them, but the
differences of the measured line-widths are not significant and our
approach requires a much smaller wavelength range for the spectra. In
this prescription, He\,{\sc ii} and O\,{\sc iii}] emission is not
explicitly removed, but this has negligible effects due to the limit
on the velocity range, making it analogous to prescription (1) of
\citet{fine10}. The second prescription, B, is analogous to
prescription (3) of \citet{fine10}, as we fit the \lam1600 feature as
part of the continuum. It only differs in that the red continuum
region is chosen to match the minimum between \civ\ and the \lam1600
feature. In general, prescription A will lead to broader estimates of
the \civ\ line width than prescription B.

The observed wavelength continuum windows for each object and
prescription are listed in Table \ref{tab:uv_opt_lineboundaries}.  The
\civ\ emission line flux was then measured above the fit continuum and
between the emission line wavelength regions listed in Table
\ref{tab:uv_opt_lineboundaries}. In addition, for objects that showed
mild absorption features, bad pixels, and/or significant night sky
line residuals, we used a low-order polynomial (i.e., first, second or
third order depending on the size and location of the feature) to
interpolate across the feature before measuring the line
widths. Details for the individual targets are given in Appendix
\ref{sssec:notes}. We did not attempt to remove any narrow-line
emission from \Civ, since this line is typically very weak and cannot
be reliably isolated \citep[][although see
  \citealt{sulentic07}]{Wills93}, and the separate lines of the
\civ\ doublet are unresolved in AGN spectra \citep[see][and references
  therein for further discussion]{vestergaard06}. We characterized the
line width by both its FWHM and line dispersion ($\sigma_l$, the
second moment of the line profile). The widths were measured directly
from the actual or interpolated spectrum (except where noted below and
in Appendix \ref{sssec:notes}) following the procedures described by
\citet{peterson04}.

We also fit the original or interpolated line profiles with a
sixth-order Gauss-Hermite (GH) polynomial, because making functional
fits to emission-line profiles is a common way of mitigating the
effects of low $S/N$ on line-width measurements \citep[see, e.g.,][for
  similar approaches]{Woo07, McGill08}. The Gauss-Hermite polynomials
we fit utilize the normalization of \citet{vanderMarel93} and the
functional forms of \citet{Cappellari02}. We then use a
Levenberg-Marquardt least-squares fitting procedure to determine the
best-fitting coefficients. We measured the widths of these line
profile models using the same software as was used to measure widths
directly from the data \citep[see][]{peterson04}. Ultimately we only
used the results from the line profile models for PG1115+080 (see
Appendix \ref{sssec:notes}). Instead, these fits were primarily used
to determine uncertainties in our width measurements as described in
\S\ref{ssec:lw_error}. The continuum and the Gauss-Hermite fits to the
\civ\ line profiles are shown in Figure \ref{fg:all_uv_opt_fits} for
both prescriptions. In the cases of SDSS1138+0314 and SBS0909+532,
reasonable fits could not be achieved because of the extremely high
$S/N$ and peculiar shape of these line profiles (a very narrow peak
with broad base; see Appendix \ref{sssec:notes}). No fits are shown
for these objects.

Both the FWHM and line dispersion, $\sigma_l$, measurements of the
\Civ\ emission line are listed in Table \ref{tab:velocities} for all
objects in our sample for both prescriptions. We have corrected the
widths for spectral resolution effects following \citet{peterson04},
when possible, using the resolutions given in
Table~\ref{tab:uv_opt_lineboundaries}. Except for PG1115+080, we
utilize the line widths measured directly from the data (interpolated
across gaps where noted) for the subsequent black hole mass
calculations. For objects with multiple spectra of the individually
lensed images we averaged their line widths. Our \Civ\ widths are
smaller than those given by \citet{greene10} for the objects in which
we both used the SDSS spectra (Q0142--100, SDSS0246--0825, PG1115+080,
and H1413+117). The likely origin of the discrepancy is that
\citet{greene10} fit a narrow line component as part of the \civ\
profile, which would naturally yield larger FWHM values. We note,
however, that \citet{greene10} do not use their SDSS line-width
measurements to estimate BH masses in their analysis, but always use
those determined by \citet{peng06}. The lens HE1104--1805 is the only
object in the sample for which we use the same optical spectrum as
\citet{peng06}, that of \citet{wisotzki95}, and we find a FWHM that is
smaller by $260~\rm km~\rm s^{-1}$, compared to our measured
uncertainty of $50~\rm km~\rm s^{-1}$. Although \citet{peng06} do not
quote errors in their line width measurements, the disagreement
($\sim5~\rm \AA$\ in the observed-frame) is likely within their
uncertainties.

Line widths of the H$\beta$ and H$\alpha$ broad-emission lines are
given in Table \ref{tab:velocities}, while the continuum and broad
line spectral wavelength regions used are given in Table
\ref{tab:nir_lineboundaries}. We measured them from the near-IR
spectra following a similar procedure to the \civ\ line-widths except
that (1) the best Gauss-Hermite polynomial fit was used for all
line-width measurements, with the exception of H$\alpha$ for
SBS0909+532, because the $S/N$ of the near-IR data was typically too
poor to justify measurement directly from the data, (2) blended
emission-line components were removed from each spectrum before the
line width was measured, as described in Appendix \ref{sssec:notes},
and (3) a power-law, instead of a linear, continuum was fit to the
\Hbeta\ spectrum of HS0810+2554 because it was fit simultaneously with
additional blended emission-line components over a larger wavelength
range.

For the objects where we lack the near-IR spectroscopic observations,
we rely on the published \Halpha\ and \Hbeta\ line widths of
\citet{greene10}. These measurements were done using somewhat
different methods than ours. While we consider most of the
\citet{greene10} FWHM estimates to be reliable, there are some that we
believe are suspect because (1) they were measured from very low $S/N$
spectra, (2) the lines were not fully contained in the wavelength
range of the spectrum, and/or (3) we do not agree with the narrow-line
component models subtracted before the line width was measured. In the
relevant Figures and Tables, we differentiate between the Balmer-line
velocity widths we think are reliable (group I, solid symbols) and
those we believe are affected by any of these issues (group II, open
symbols). Individual objects can be in both groups because these
issues may affect only one of the Balmer lines. We also include in
group I the H$\alpha$ and H$\beta$ line-width measurements from our
new IR spectra. The decision to split our sample is a conservative
choice, and our conclusions are not significantly modified when the
group II line widths are included.

\subsection{Line-Width Measurement Uncertainties}\label{ssec:lw_error}

Line-width measurements can be affected by sources of error that are
difficult to model, as they depend not only on the overall $S/N$
ratio, but also on the line profile and the presence of sky emission
and absorption lines, with the latter being of particular importance
in the near-IR. We use a Monte Carlo approach to determine the
uncertainties in our line-width measurements. Using the flux
uncertainty per pixel in each spectrum and the best fit Gauss-Hermite
line profile (with the exception of the optical SDSS1138+0314 and
SBS0909+532 spectra, see Appendix \ref{sssec:notes}), we produced 1000
resampled spectra by adding random Gaussian deviates based on the
error spectrum to the flux in each pixel of the GH model spectrum and
then re-measured the line width using the methods described in the
previous section. For the UV/optical spectra from the literature
without an error spectrum, we estimated one by propagating the
measured $S/N$ of a small continuum window near the \Civ\ emission
line to the overall spectrum. In this case, $\delta F_{\lambda}$, the
flux error in a pixel of wavelength $\lambda$ with flux $F_{\lambda}$,
is given by
\begin{equation}
\delta F_{\lambda}\ =\ \sqrt{\frac{\lambda_c}{\lambda}\ F_{\lambda_c}
F_{\lambda}}\ \left(\frac{S}{N}\right)^{-1}_c ,
\label{eqn:widtherror}
\end{equation}
\noindent where $\lambda_c$ and $F_{\lambda_c}$ are the average
wavelength and flux per unit wavelength of the continuum window
chosen, and $(S/N)_c$ is the signal-to-noise ratio per pixel in the
chosen continuum window. This equation is constructed by assuming that
the only source of error is Poisson fluctuations, and that the number
of detected photons is proportional to $F_{\lambda} (h c /
\lambda)^{-1}$, where the proportionality constant is empirically
determined in the continuum window from $(S/N)_c$, $\lambda_c$ and
$F_{\lambda_c}$. This approach neglects the sky background and the
presence of strong absorption or emission sky lines, which is
reasonable for the UV/optical spectra. It also neglects changes in the
instrument sensitivity as a function of wavelength and assumes a
constant pixel wavelength-width, both of which are reasonable because
the continuum $S/N$ is measured in close proximity to the emission
line of interest. While the parametric fits are not exact
representations of each line, this still provides a reasonable
estimate of the fractional uncertainties.

\subsection{Luminosity Measurements}\label{ssec:lum_method}

We estimated the continuum luminosities at 5100\AA\ by fitting the AGN
SED template of \citet{assef10a} to the unmagnified quasar magnitudes
obtained from the CASTLES project {\it{HST}} NICMOS imaging. To
correct the observed quasar fluxes for the lens magnification, we
modeled each system using the astrometry and lens galaxy photometry
from the CASTLES {\it{HST}} WFPC2 and NICMOS observations following
the procedures of \citet{lehar00}. The image is decomposed into a set
of point sources for the quasars, de Vaucouleurs models for the lens
galaxy and, if necessary, a lensed host component, convolved with
model or empirical PSFs. The resulting component positions and image
fluxes were modeled using \verb+lensmodel+ \citep{keeton01}. The lens
was modeled as a singular isothermal ellipsoid in an external shear
with the ellipsoid's orientation and ellipticity constrained by those
of the light of the lens galaxy and a weak prior on the external
shear. The models were not tightly constrained to match the observed
fluxes due to systematic errors in image flux ratios such as source
variability and microlensing. Aside from substructure, the dominant
uncertainty in the magnifications is the radial mass distribution of
the lens \citep[see][]{kochanek04}, and this is less than a factor of
two even if we allow the full range of models between a flat rotation
curve and a constant $M/L$ model. Since we have extensive evidence
that lenses have mass distributions corresponding to flat rotation
curves on these scales \citep[e.g.,][]{rusin03,jiang07,koopmans09},
the model uncertainties are considerably less than this factor, and
the uncertainties are dominated by the systematic uncertainties in the
image fluxes. Table \ref{tab:magnifications} lists the magnifications
used for each object in the sample. The only object for which a
different model was used is Q0957+561, where we used the
magnifications determined by \citet{fadely10}.

We did not apply reddening corrections other than removing Galactic
foreground extinction (see below), as the requirement that \civ\ is
observable in the UV/optical severely limits the presence of dust
absorption, especially at rest-frame 5100\AA. For all four-image
lenses, we estimated the true source flux for all images, rejected the
highest and lowest estimates and averaged the remaining two to limit
the effects of microlensing. For two-image lenses we simply averaged
the two estimates. Table \ref{tab:magnifications} shows the estimated
unmagnified $H$-band magnitude of each quasar. Note that in general we
did not apply a correction for variability. Although there is a 5 to
10 year time difference between the CASTLES and the \citet{greene10}
Triplespec observations, the typical uncertainty introduced falls well
below the systematic uncertainties in the SE BH mass estimates. An
estimate of the typical variability of a quasar can be obtained from
measurements of their structure function. Using the power-law fit of
\citet{vandenberk04} to the $i$-band structure function of SDSS
quasars, we find that the typical quasar would experience a change in
magnitude of approximately 0.2 mag for a rest-frame time-lag of
1500~days (approximately 10 years in the observer's frame for our
lowest redshift quasar). A change of 0.2 magnitudes results in a
change to the BH mass estimate of 0.04~dex, well below their typical
error bar of 0.3~dex, and we would expect the $H-$band variability to
be still smaller, as the average variability amplitude decreases with
increasing wavelength \citep[see, e.g.,][]{vandenberk04,macleod10}.

For SDSS1138+0314, HS0810+2554 and SBS0909+532, we performed an
absolute flux calibration of the near-IR spectra and measured the
5100\AA\ continuum luminosity directly. The calibration for the first
object is discussed in detail in \S\ref{sssec:sdss1138_luci}. For
HS0810+2554 we fit a power-law to the continuum of our MDM CCDS
spectrum (see \S\ref{ssec:uv_opt_obs}) and extrapolated it to
rest-frame 5100\AA. For SBS0909+532 we calibrated the spectrum using
the {\it{HST}} NICMOS {\it{H-}}band photometry, as the object did not
show significant flux variations between the two relevant epochs
(J.~Mu\~noz, private communication).

To obtain the rest-frame continuum UV luminosities at 1350\AA\ and
1450\AA, we flux calibrated the spectra whenever it was necessary and
measured the flux by fitting a straight line to the region between
rest-frame 1349\AA\ and 1355\AA\ for the estimate at 1350\AA\ and to
the region between 1440\AA\ and 1460\AA\ for the estimate at
1450\AA. We corrected these luminosities for foreground Galactic
extinctions obtained through the NASA/IPAC Extragalactic
Database\footnote{\url{http://nedwww.ipac.caltech.edu/}} from the dust
maps of \citet{schlegel98}. Errors in the continuum luminosity will be
dominated by the uncertainties in the magnification models, which are
hard to quantify. We assume a conservative error of 20\% in each
continuum luminosity estimate.

\citet{greene10} obtained continuum luminosities at 5100\AA\ for their
sample of objects by following a similar approach. They fit a
power-law to the unmagnified {\it{HST}} photometry from the CASTLES
survey, using the lensing models of \citet{peng06}. In comparison to
\citet{greene10} we observe that our luminosity estimates are, on
average, $0.20\pm0.05$~dex smaller. The offset is likely caused by a
combination of the differences in the lensing models, in the
prescription used to deal with the flux ratio anomalies, and in the
use of the AGN SED template of \citet{assef10a} instead of the
power-law fits of \citet{peng06}. We note that this offset translates
to 0.1~dex in BH mass, well below the uncertainties we estimate for
our SE mass measurements in the next section. We also note that our
conclusions are unaltered if we replace our 5100\AA\ continuum
luminosity estimates with those of \citet{greene10} for all objects
where this is possible.

\subsection{Black Hole Mass Estimates}\label{ssec:mbh_method}

The width of a given broad emission line in a Type 1 AGN is primarily
caused by the gravitational attraction of the supermassive black hole
on the gas in the broad line region (BLR). Hence, the mass of the
black hole, $M_{\rm BH}$, can be estimated from virial assumptions by
\begin{equation}\label{eq:grav_mbh}
M_{\rm BH}\ =\ f\ \frac{R_{\rm BLR} (\Delta v)^2}{G},
\end{equation}
\noindent were $\Delta v$ is the velocity dispersion of the BLR gas,
estimated from the width of the broad emission line, $G$ is the
gravitational constant and $R_{\rm BLR}$ is the distance from the
black hole to the BLR. The factor $f$ is a scale factor of order unity
that depends on the structure, kinematics and inclination of the BLR
\citep[see, e.g.,][and references therein]{collin06}. The term $R_{\rm
BLR} (\Delta v)^2/G$ is usually referred to as the virial product (VP)
and encapsulates all the observable quantities for a single
object. The radius of the BLR can only be measured through
reverberation mapping \citep[see, e.g.,][]{peterson04}, but has been
shown to correlate well with the continuum luminosity \citep[see,
e.g.,][]{kaspi05,bentz06,bentz09,zu10}.

For the broad hydrogen emission lines we estimate the BLR radius using
the $R_{\rm BLR} - \lambda L_{\lambda}(5100\rm \AA)$ relation of
\citet{bentz09}, which was calibrated using a large sample of RM
AGNs. The $f$ factor of equation (\ref{eq:grav_mbh}) depends on the
characterization of the line width, generally either the FWHM or the
line dispersion, $\sigma_l$, as well as on the emission line being
used. For estimating $M_{\rm BH}$ from the width of the H$\beta$ broad
line, we use the $f$ factor calibrations of \citet{collin06} for the
FWHM and for $\sigma_l$. While for $\sigma_l$ a unique $f$ factor of
3.85 for all AGNs suffices, \citet{collin06} argued that $f$ is
strongly dependent on the line profile shape for FWHM-based estimates,
where the shape was quantified as the ratio between the FWHM and
$\sigma_l$. We choose, however, to use the best-fit fixed $f$ factor
of 1.17 for FWHM instead of the line-shape dependent calibrations
because \citet{denney09} have shown that $\sigma_l$ is affected by
blending with other emission lines, making the correlation found by
\citet{collin06} hard to interpret. For H$\alpha$ there is no
equivalent calibration of the $f$-factor, so we cannot directly
estimate the black hole masses. Instead, we use the relation
determined by \citet{greene05} between the FWHM of H$\alpha$ and
H$\beta$,
\begin{equation}\label{eq:fwhm_relation}
\rm FWHM_{\rm H \beta}\ =\ (1.07\pm 0.07)\times 10^3\ \left(\frac{\rm
FWHM_{\rm H \alpha}}{10^3\ \rm km\ \rm s^{-1}}\right)^{(1.03\pm 0.03)}
\rm km\ \rm s^{-1} ,
\end{equation}
\noindent to estimate the H$\beta$ FWHM and then estimate $M_{\rm
BH}(\rm H\alpha)$ using the same $f-$factor and $R_{\rm BLR} - \rm L$
relation as for $M_{\rm BH}(\rm H\beta)$. Unfortunately, there is no
equivalent transformation for $\sigma_l$, so we cannot use this
measurement to estimate the mass of the black hole from H$\alpha$.

Combining equations (\ref{eq:grav_mbh}) and (\ref{eq:fwhm_relation})
with the $R_{\rm BLR} - \lambda L_{\lambda}(5100\rm \AA)$ relation of
\citet{bentz09} we get
\begin{eqnarray}
M_{\rm BH}(\rm H\beta)\ &=&\ 6.71\times 10^6\ f\ \left(\frac{\Delta
v_{\rm H\beta}}{10^3\ \rm km\ \rm s^{-1}}\right)^2\
\left(\frac{\lambda L_{\lambda}(5100\rm \AA)}{10^{44}~\rm erg\ \rm
s^{-1}} \right)^{0.52}\ M_{\odot}\label{eq:m_hbeta}\\ M_{\rm BH}(\rm
H\alpha)\ &=&\ 7.68\times 10^6\ f\ \left(\frac{\rm FWHM_{\rm
H\alpha}}{10^3\ \rm km\ \rm s^{-1}}\right)^{2.06}\ \left(\frac{\lambda
L_{\lambda}(5100\rm \AA)}{10^{44}~\rm erg\ \rm s^{-1}}
\right)^{0.52}\ M_{\odot}\label{eq:m_halpha},
\end{eqnarray}
\noindent where in equation (\ref{eq:m_hbeta}) $\Delta v_{\rm H\beta}$
can be either the line dispersion or the FWHM. Because equation
(\ref{eq:m_halpha}) is fully dependent upon the scaling relations for
H$\beta$, the $f$ factor in it is the same as for $\rm FWHM_{\rm
H\beta}$ in equation (\ref{eq:m_hbeta}). Table \ref{tab:all_objs}
shows our BH mass estimates based on H$\alpha$ and H$\beta$ for all
objects in the sample.

For the UV/optical spectra we use the empirical $M_{\rm BH}$
calibrations of \citet{vestergaard06} for the \civ\ broad emission
line, given by
\begin{equation}\label{eq:mass_civ}
M_{\rm BH}(\civm)\ =\ 10^{\kappa}\ \left(\frac{\Delta
v_{\civm}}{10^3~\rm km~\rm s^{-1}}\right)^2\ \left(\frac{\lambda
L_{\lambda}(1350\rm \AA)}{10^{44}~\rm erg~\rm s^{-1}}\right)^{0.53}\
M_{\odot} ,
\end{equation}
\noindent where $\Delta v$ is either FWHM or $\sigma_l$, and $\kappa =
6.66\pm 0.01$ or $6.73\pm 0.01$, respectively, for these line-width
characterizations. The constant $\kappa$ implicitly contains the $f$
factor, which is assumed to be a constant for all objects. Whenever
possible, we use the observed 1350\AA\ flux to determine the continuum
luminosity. Unfortunately 1350\AA\ is not within the observed
wavelength range of all the UV/optical spectra we use. In these cases
we estimate the continuum luminosity at 1350\AA\ using the observed
flux at 1450\AA, as \citet{vestergaard06} have shown $L_{\lambda}$ at
these wavelengths to be equivalent. We list our \civ\ BH mass
estimates in Table \ref{tab:all_objs} for both prescriptions used to
measure the widths of \civ. As expected, masses determined from the
FWHM are highly consistent for both prescriptions, with a mean
difference of $0.04~\rm dex$ and a scatter of $0.02~\rm dex$, with the
average prescription B based mass estimates being smaller. The
agreement is much worse for $\sigma_l$, with a mean difference of
0.23~dex, in the sense that B is smaller, and a scatter of 0.18~dex.

We estimate the uncertainties in our BH mass estimates by propagating
the errors in the velocity widths and in the continuum
luminosities. For masses based on the width of the broad Hydrogen
lines, we also propagate the uncertainties in the $f$-factor and in
$R_{\rm BLR}$. \citet{collin06} determined that the uncertainty in $f$
when using $\sigma_l$ is 30\%, while that in FWHM is 43\%. For $R_{\rm
BLR}$ we assume the intrinsic scatter of 0.11~dex estimated by
\citet{peterson10} for the radius-luminosity relation. Adding the
uncertainties in $f$ and $R_{\rm BLR}$ is not possible for the \civ\
estimates of the BH masses. Instead, we add the measurement errors and
the intrinsic scatter between \civ\ and RM BH mass estimates in
quadrature. Using the sample of \citet{vestergaard06}, we estimate
intrinsic scatters of 0.32~dex and 0.28~dex for FWHM and $\sigma_l$
respectively. \citet{vestergaard06} found that the total scatter,
including measurement errors, was 0.32~dex for both line-width
characterizations of \civ, showing that the intrinsic scatter
dominates over measurement errors, especially for FWHM estimates.

\section{Biases in \civ\ Black Hole Mass Estimates}\label{sec:civ_comp}

In this section we use the sample described in \S\ref{sec:sample} to
study biases in the \civ\ black hole mass estimates. We first compare
how the mass depends on the characterization of the \civ\ line-width,
and then we proceed to compare these rest-frame UV estimates to those
based on the H$\alpha$ and H$\beta$ emission lines. In the next
section we will compare our results with those of other studies on the
relations between \civ\ and H$\beta$ BH estimated masses.

\subsection{Comparison of FWHM and $\sigma_l$ Derived Masses}\label{ssec:civ_sigma_fwhm}

Given that we have measured both FWHM and $\sigma_l$ for \civ\ in all
our objects, the simplest test we can perform is to determine if there
are any biases between them as BH mass estimators. Both measurements
have advantages, and some contention exists in the literature as to
which constitutes a more reliable mass estimator \citep[see][and
  references therein]{peterson04}. 

\begin{figure}
  \begin{center}
    \plotone{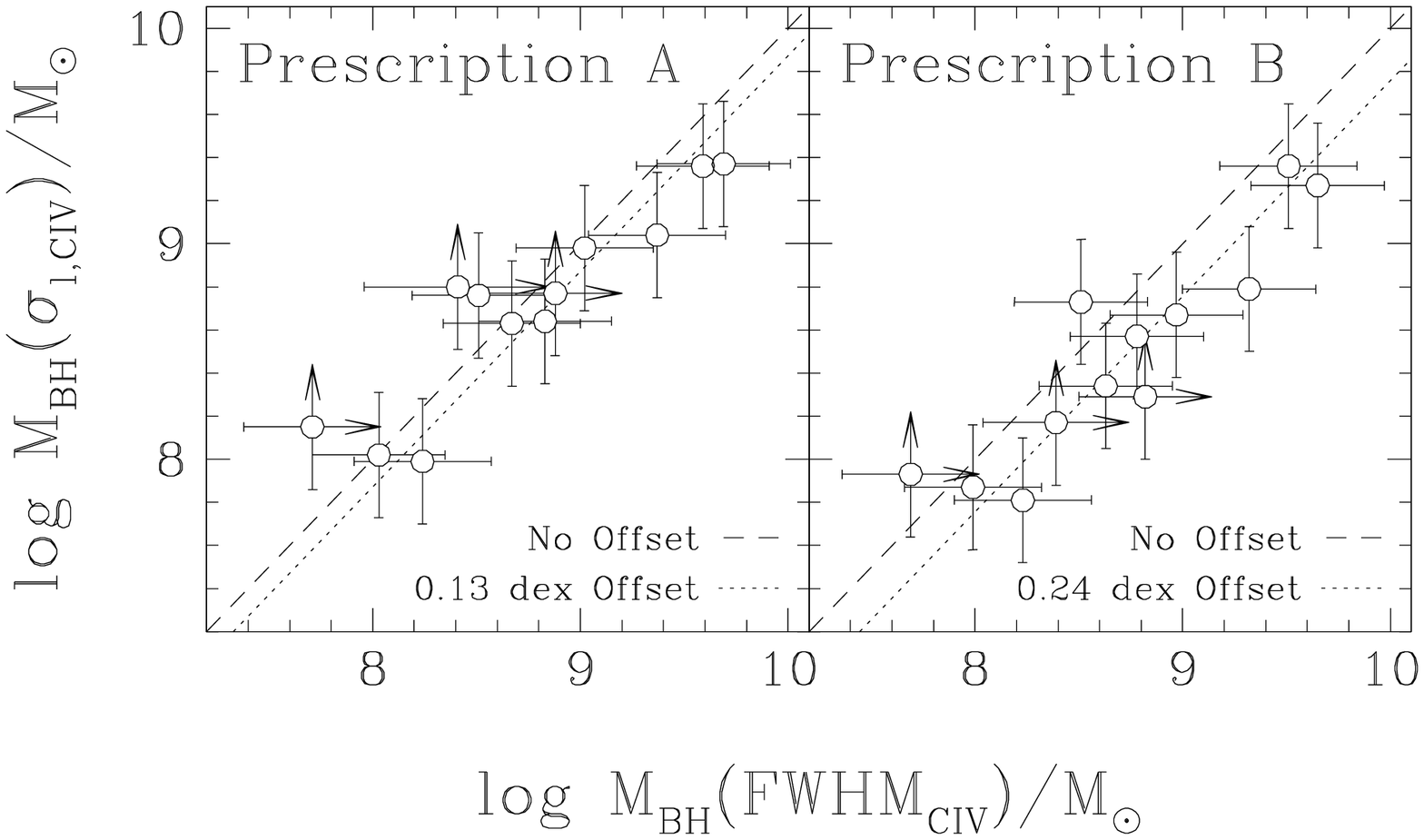}
    \caption{Comparison of \civ\ BH masses derived from the FWHM and
    $\sigma_l$ velocity widths for all objects in our sample using the
    relations of \citet{vestergaard06}. The left (right) panel
    compares the BH mass estimates based on the prescription A (B)
    line-width measurements of \civ. Masses are equal along the dashed
    line and the dotted line correspond to the best fit offset. The
    objects with arrows correspond to those for which we believe our
    \civ\ line-width measurements to be lower bounds.}
    \label{fg:FWHM_sigma_comp}
  \end{center}
\end{figure}

Figure \ref{fg:FWHM_sigma_comp} compares the \civ-based BH masses
determined for both line-width estimates and for the two continuum and
line blending prescriptions A and B, respectively. A clear bias is
observed for both prescriptions, where most objects have a lower
estimated BH mass if we use $\sigma_l$ instead of the FWHM. The bias
for prescription A (B) width measurements seems to be well represented
by a constant offset of $K = 0.13\pm 0.06~\rm dex$ ($0.24\pm 0.07~\rm
dex$) or, equivalently, a factor of 1.3 (1.7). We fit for $K$ while
simultaneously fitting for the intrinsic scatter between the two mass
estimators by adding a scatter $S$ in quadrature to the error of each
logarithmic mass difference. Note that the logarithmic mass difference
does not depend on the continuum luminosity or the intrinsic scatter
with respect to the RM estimates. In practice we maximize the
likelihood
\begin{equation}\label{eq:likelihood}
{\cal L}\ =\ \left(\langle\sigma^2\rangle+S^2\right)^{-1/2}\
\left[\prod_{i=1}^N \left(\sigma_i^2+S^2\right)^{-1/2}\right]\
e^{-\chi^2(S)/2} ,
\end{equation}
\noindent where $\sigma_i^2$ is the variance due to measurement errors
in the logarithmic mass difference of object $i$ and
$\langle\sigma^2\rangle$ is its average over all objects. We exclude
objects for which we consider the \civ-based BH mass estimates to be
lower bounds due to absorption. The leading factor in equation
(\ref{eq:likelihood}) is a logarithmic prior on the overall
dispersion. The best fit scatter is similar for both prescriptions,
with case A line-widths producing $S = 0.16~\rm dex$ while case B ones
have $S = 0.19~\rm dex$. Since the logarithmic mass difference only
depends on the line-widths and not on the continuum luminosities, the
constant BH mass offsets $K$ can also be expressed as an offset
between the line-width characterizations. As such, these values imply
an offset of $0.10\pm0.03~\rm dex$ ($0.16\pm0.04~\rm dex$) between the
FWHM and $\sigma_l$ line-width characterizations of \civ\ for
prescription A (B).

It is not surprising that prescription A provides a smaller offset
between BH masses obtained from the FWHM and $\sigma_l$ of \civ, as
this prescription is modeled after that used by \citet{vestergaard06},
who used their measurements to determine equation
(\ref{eq:mass_civ}). However, given the similarity, the presence of a
non-zero offset for prescription A is somewhat puzzling. If we examine
the sample of \citet{vestergaard06}, the scatter is larger, 0.2~dex,
and there is no offset ($-0.02\pm 0.03~\rm dex$), although the lack of
an offset is by definition small since both mass estimators were
calibrated against the same RM data set.

The large overlap in the mass and continuum luminosity ranges of our
sample and that of \citet{vestergaard06} suggest that dependence on a
secondary parameter is unlikely. Furthermore, we do not see any
correlation of this bias with BH mass, continuum luminosity or
Eddington ratio. There is also no correlation with redshift,
suggesting that it is unlikely to be an evolutionary trend. The only
other major difference between the samples is lensing by foreground
galaxies. This, however, is very unlikely to cause such an effect, as
quasars are quite compact and strong lensing affects the whole
object. Microlensing by the stars in the foreground galaxy could in
principle distort the shape of the \civ\ broad emission lines due to
the spatial dependence of their velocity structure, but this is very
unlikely for two reasons. First, the width of \civ\ is typically well
below 10,000~km/s, constraining the location of the gas to a distance
greater than $\gtrsim 10^3$ Schwarzschild radii ($R_S$) from the black
hole, while microlensing is only observed to have significant effects
on scales below $100~R_S$ \citep{morgan10}. Second, the gas moving at
the highest velocities is expected to be closest to the black hole, so
microlensing would tend to magnify the wings of the line more than the
core, and hence producing the inverse of the effect we see by making
$\sigma_l$ too large rather than too small compared to the FWHM. While
microlensing can also produce regions of demagnification in the source
plane, these are of very large spatial extent, and so it is unlikely
to see significant magnification variations across the BLR.

It is likely then that other minor differences in the method we use to
measure $\sigma_l$ as compared to \citet{vestergaard06} give rise to
the remaining bias. \citet{denney09} showed that estimates of
$\sigma_l$ depend on the exact prescription used for the line-width
measurement and the segregation of blended emission for H$\beta$. Our
investigation shows that this may be the case for \civ\ as well
\citep[see also][]{fine10}. However, the remarkably low scatter in
Figure \ref{fg:FWHM_sigma_comp} suggests that if $\sigma_l$ is
measured in a self-consistent manner it can be as accurate as the FWHM
for estimating BH masses, but the calibration will depend on the exact
prescription. In the next section we will explore the reliability of
the \civ\ FWHM and $\sigma_l$ BH mass estimates by comparing them to
those based on H$\alpha$ and H$\beta$.

\subsection{\civ\ compared to H$\alpha$ and H$\beta$}\label{ssec:civ_balmer_comp}

\begin{figure}
  \begin{center}
    \plotone{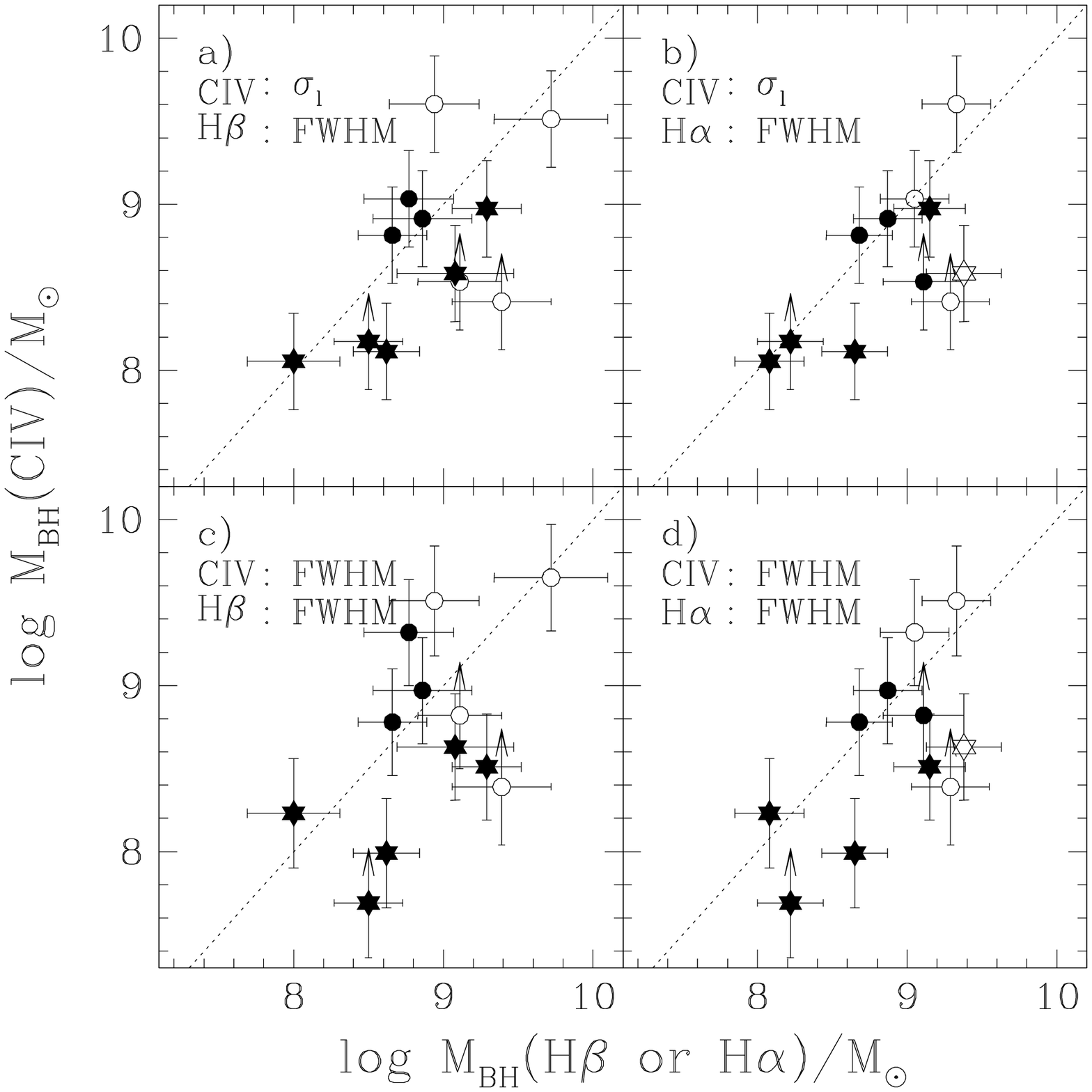}
    \caption{Comparison between BH masses estimated from the
    prescription B $\sigma_l$ and FWHM of \civ\ and from the FWHM of
    H$\alpha$ and H$\beta$. For the estimates based on the line
    dispersion of \civ\ we have added the systematic offset of
    0.24~dex described in \S\ref{ssec:civ_sigma_fwhm}. Solid symbols
    correspond to the objects with group I H$\alpha$ or H$\beta$
    line-width estimates, while open symbols correspond to those with
    group II estimates. Six-pointed stars mark the objects not
    considered in the analysis of \citet{greene10}. The dotted line
    shows where the BH masses are equal.}
    \label{fg:CIV_H_masses}
  \end{center}
\end{figure}

\begin{figure}
  \begin{center}
    \plotone{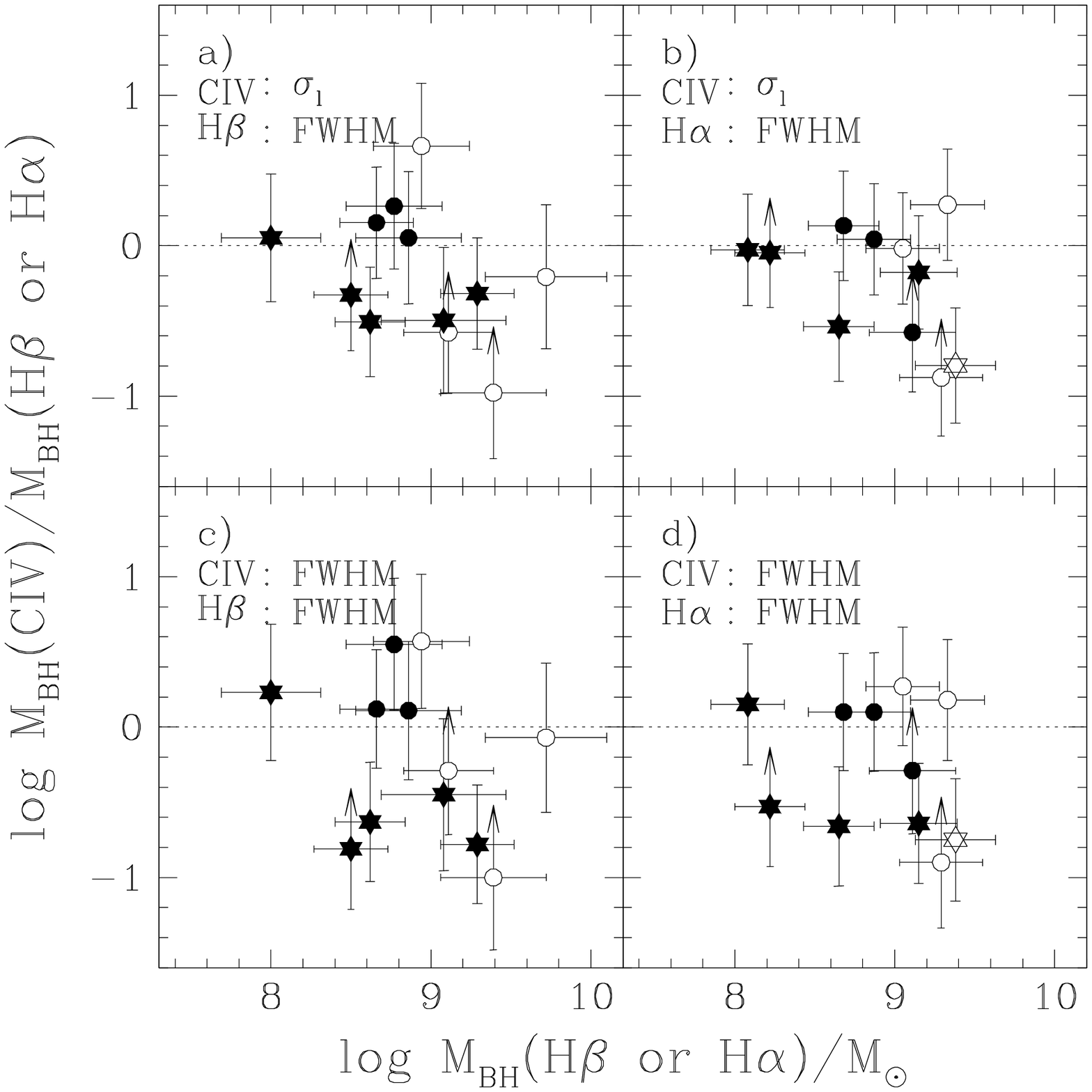}
    \caption{Ratio between the \civ\ and H$\beta$/H$\alpha$ mass
    estimates as a function of the corresponding hydrogen line mass
    estimate. Symbols and lines have the same definitions as in Figure
    \ref{fg:CIV_H_masses}.}
    \label{fg:diff_CIV_H_masses}
  \end{center}
\end{figure}

Figures \ref{fg:CIV_H_masses} and \ref{fg:diff_CIV_H_masses} compare
the mass estimates based on the H$\alpha$ and H$\beta$ lines to those
based on the width of \civ. We only show here (and for the rest of the
figures) UV BH masses based on the prescription B width measurements
of \civ. The FWHM based BH masses are almost equal for prescriptions A
and B (see \S\ref{ssec:mbh_method}), but they show a systematic offset
for the $\sigma_l$ estimates (see \S\ref{ssec:civ_sigma_fwhm}). We
adopt the prescription B masses for the rest of this section, but our
conclusions are unaltered if we instead use prescription A
measurements. We have made the assumption that the \civ\ FWHM mass
estimates are unbiased, and so those obtained from the prescription B
$\sigma_l$ measurement of \civ\ have been shifted by the systematic
offset of 0.24~dex derived in the previous section. 

We measure no significant offset between the \civ-based and H$\alpha$-
or H$\beta$-based masses when using only objects with \civ\
line-widths that are not lower bounds and have reliable Balmer line
widths (group I). We find best fit offsets of $-0.12\pm 0.15$,
$-0.11\pm 0.16$, $-0.15\pm 0.16$ and $-0.19\pm0.18~\rm dex$ for panels
a), b), c) and d), respectively, of Figure \ref{fg:diff_CIV_H_masses},
with residual scatter of 0.30, 0.23, 0.46 and 0.38~dex. Including the
objects with group II H$\alpha$ and H$\beta$ line-width estimates does
not change this conclusion, with best fit offsets of $-0.05\pm 0.14$,
$-0.13\pm 0.13$, $-0.07\pm 0.15$ and $-0.15\pm 0.14~\rm dex$,
respectively, with measured scatters of 0.36, 0.33, 0.46 and
0.41~dex. The lack of offsets confirms our assumption that \civ\ FWHM
BH masses are unbiased and that only those based on $\sigma_l$ need to
be corrected. The constant offset fits yield $\chi^2$ per degree of
freedom ($\chi^2_{\nu}$) values of 0.6, 0.5, 1.4 and 1.1 for panels
a), b), c) and d) of Figure \ref{fg:CIV_H_masses} when using only the
solid symbols. The scatter in each panel of Figure
\ref{fg:CIV_H_masses} is largely consistent with the estimated
uncertainties, although the errors in the \civ\ $\sigma_l$ masses may
be slightly overestimated. We find no evidence based on the $\chi^2$
statistic that a slope different from unity is required to describe
the relation between the logarithms of the BH masses (Figure
\ref{fg:CIV_H_masses}), independent of whether we include the group II
Balmer line-width measurements.

\begin{figure}
  \begin{center}
    \plotone{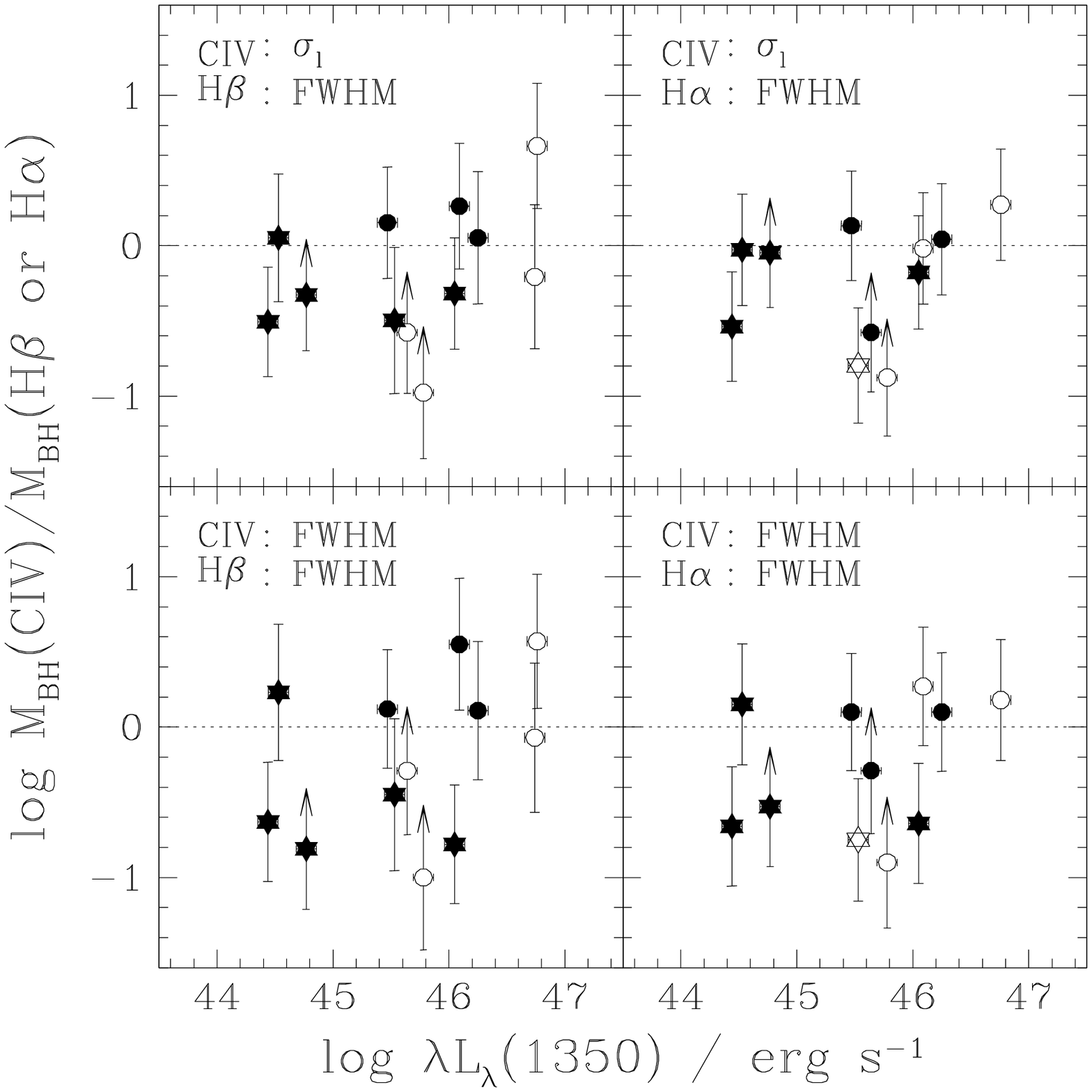}
    \caption{Same as Figure \ref{fg:diff_CIV_H_masses}, but as a
    function of the UV continuum luminosity.}
    \label{fg:diff_uv_lum}
  \end{center}
\end{figure}

\begin{figure}
  \begin{center}
    \plotone{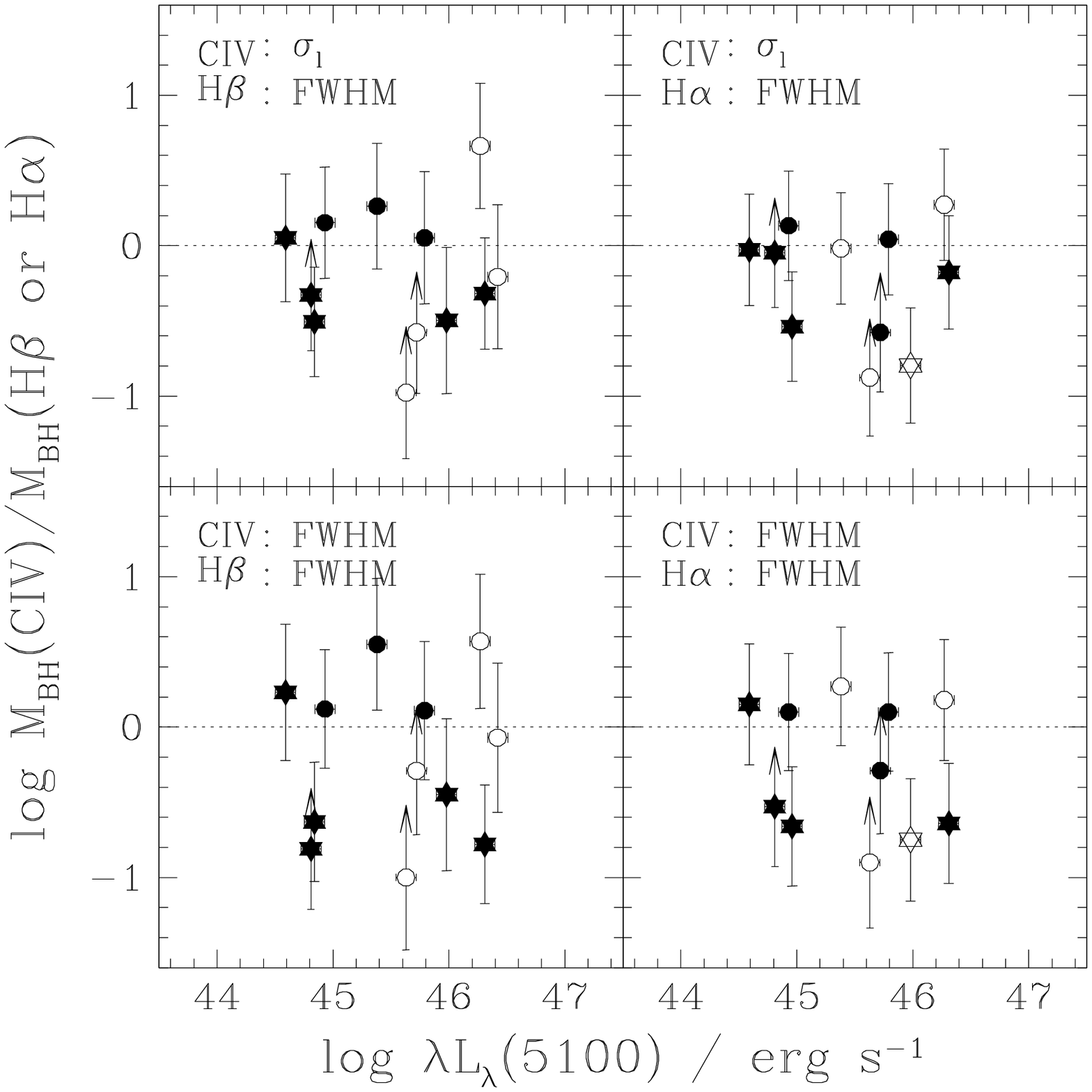}
    \caption{Same as Figure \ref{fg:diff_CIV_H_masses}, but as a
    function of the 5100\AA\ continuum luminosity.}
    \label{fg:diff_opt_lum}
  \end{center}
\end{figure}

\begin{figure}
  \begin{center}
    \plotone{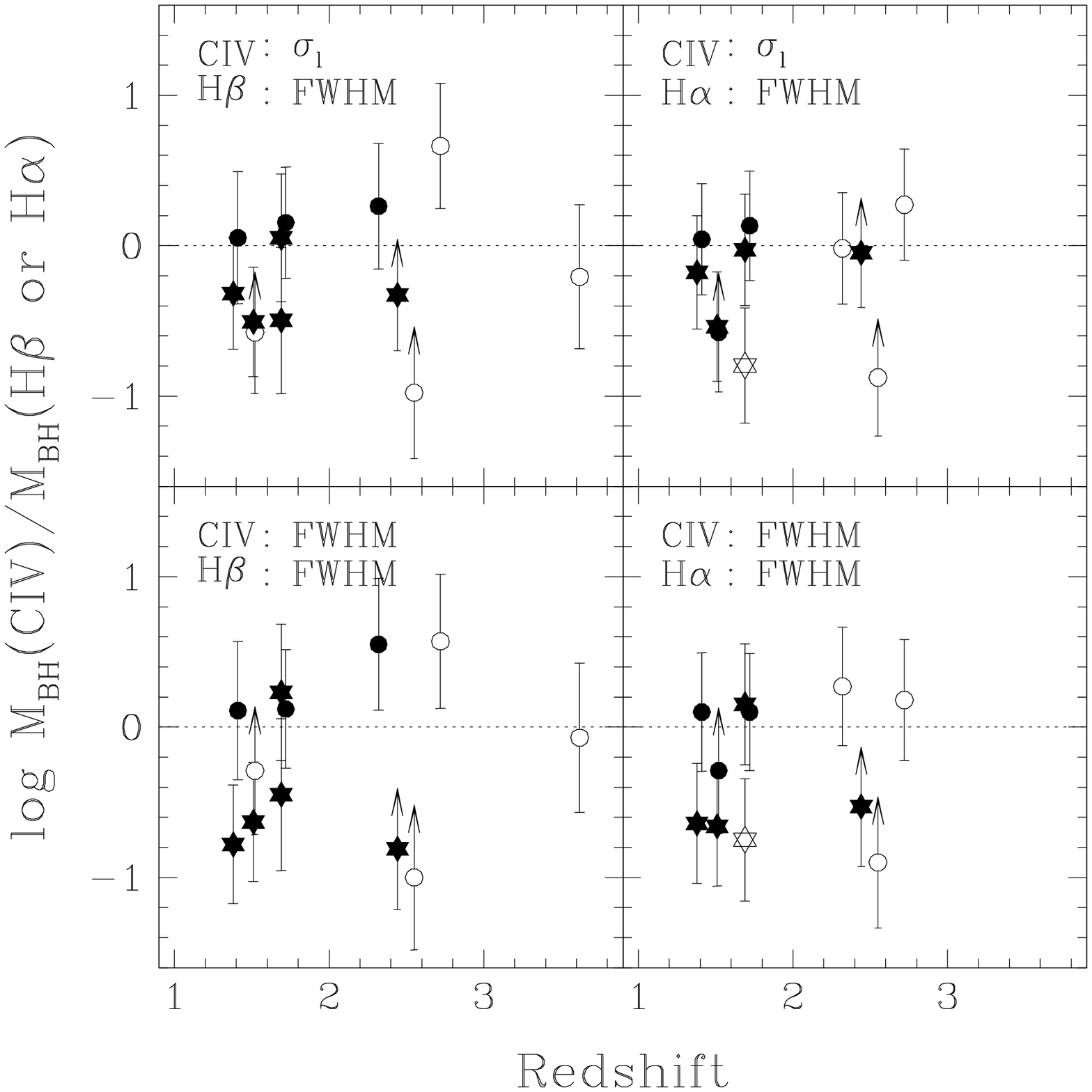}
    \caption{Same as Figure \ref{fg:diff_CIV_H_masses}, but as a
    function of redshift.}
    \label{fg:diff_redshift}
  \end{center}
\end{figure}

\begin{figure}
  \begin{center}
    \plotone{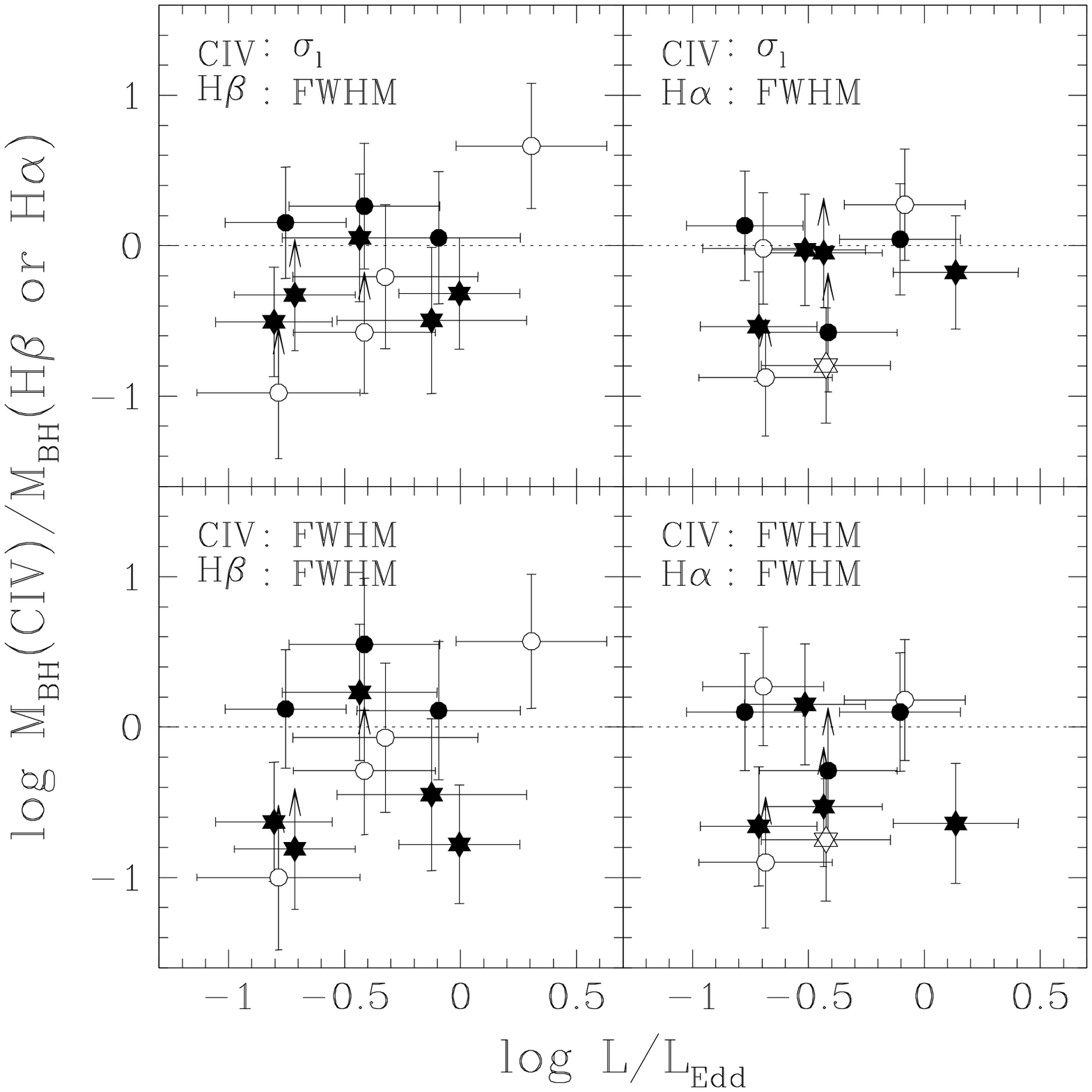}
    \caption{Same as Figure \ref{fg:diff_CIV_H_masses}, but as a
    function of the estimated Eddington ratio. We used a factor of
    11.91 to convert between $\lambda L_{\lambda}(5100\AA)$ and
    $L_{\rm Bol}$, as determined from the AGN SED of
    \citet{assef10a}.}
    \label{fg:diff_edd_rat}
  \end{center}
\end{figure}

\begin{figure}
  \begin{center}
    \plotone{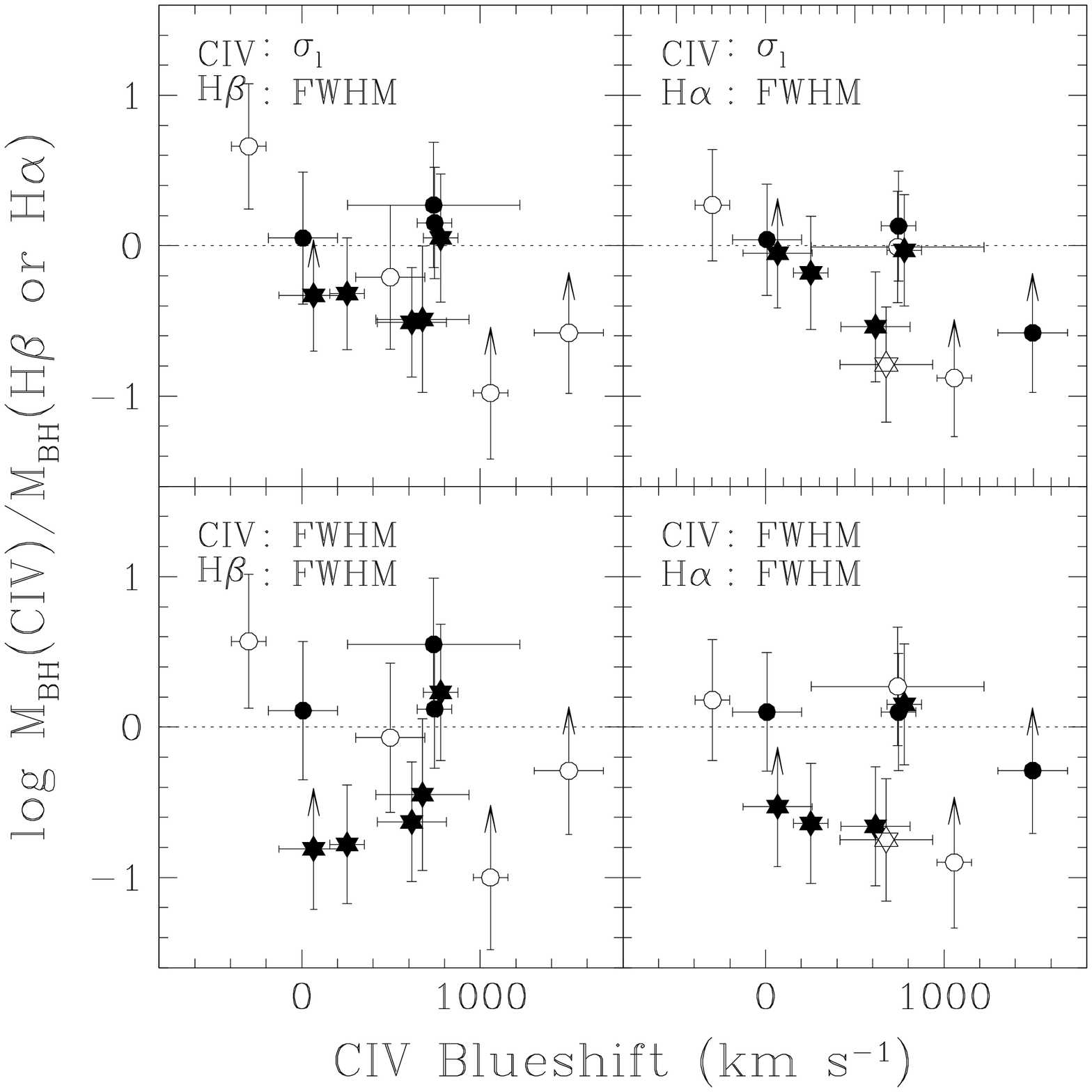}
    \caption{Same as Figure \ref{fg:diff_CIV_H_masses}, but as a
      function of the blueshift of the \civ\ line.}
    \label{fg:diff_blueshift}
  \end{center}
\end{figure}

\begin{figure}
  \begin{center}
    \plotone{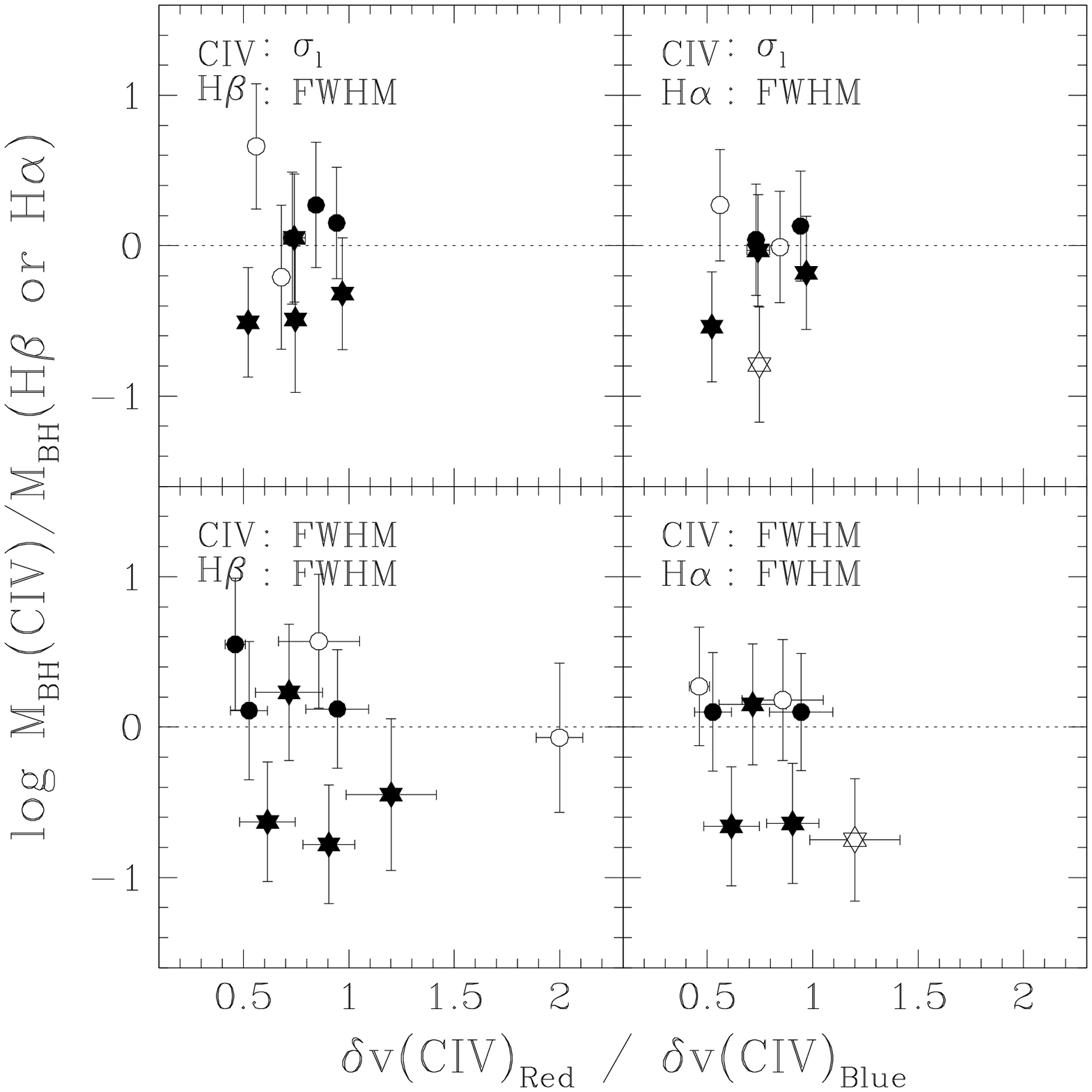}
    \caption{Same as Figure \ref{fg:diff_CIV_H_masses}, but as a
      function of the asymmetry of the \civ\ line, parametrized as the
      ratio of the widths red and blue of the line centroid. We do not
      show lower bounds on the \civ\ line-width as those object do not
      have a well defined blue side width due to the presence of
      absorption.}
    \label{fg:diff_assymetry}
  \end{center}
\end{figure}

\begin{figure}
  \begin{center}
    \plotone{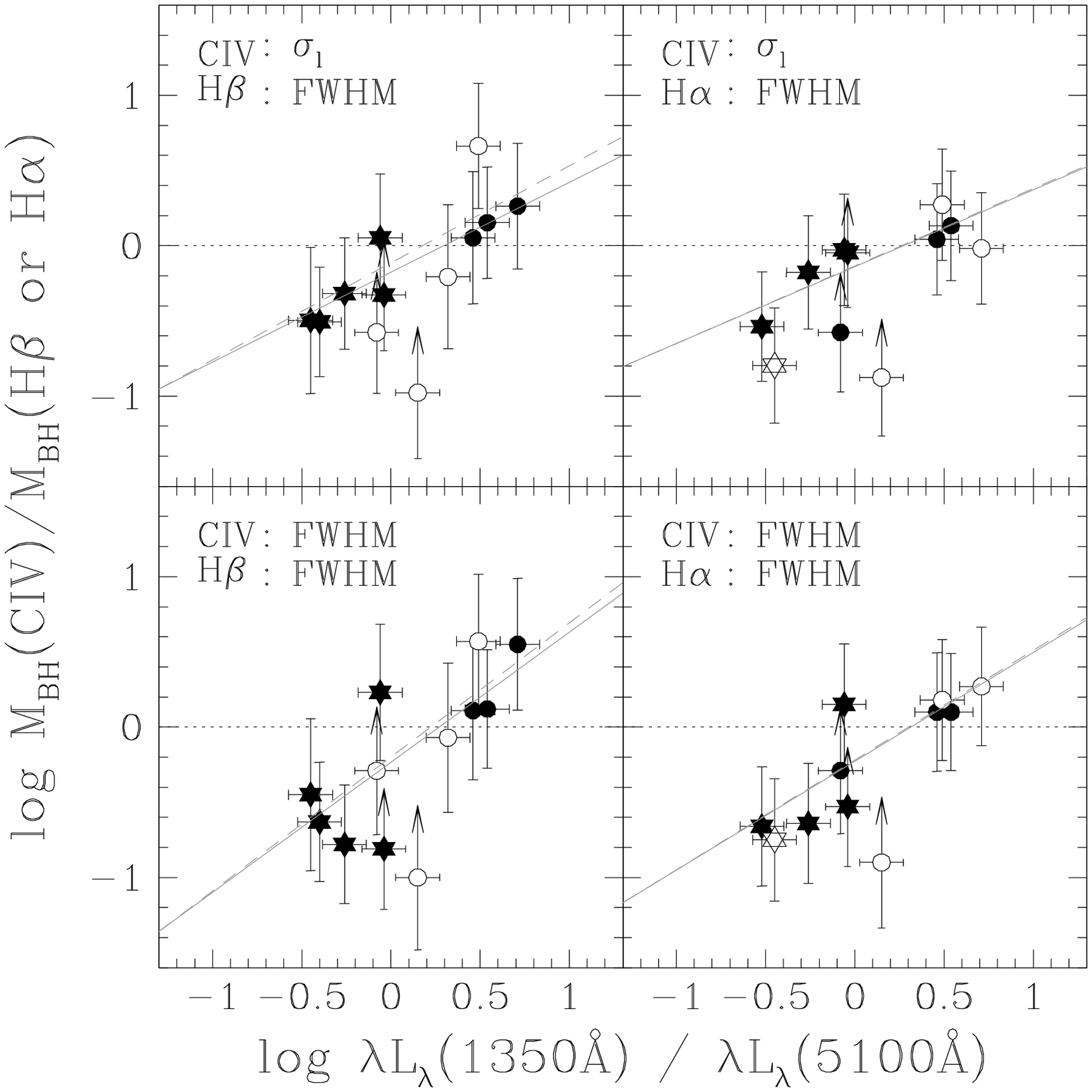}
    \caption{Same as Figure \ref{fg:diff_CIV_H_masses}, but as a
    function of the ratio of the UV to optical continuum
    luminosities. The solid line shows the best fit linear relation to
    all objects with group I Balmer-line width estimates (solid
    symbols) and the dashed line shows the linear relation obtained
    when also including object with group II estimates (open
    symbols). Note that we do not include objects with lower bound
    \civ\ line widths on the fits.}
    \label{fg:diff_color}
  \end{center}
\end{figure}

\begin{figure}
  \begin{center}
    \plotone{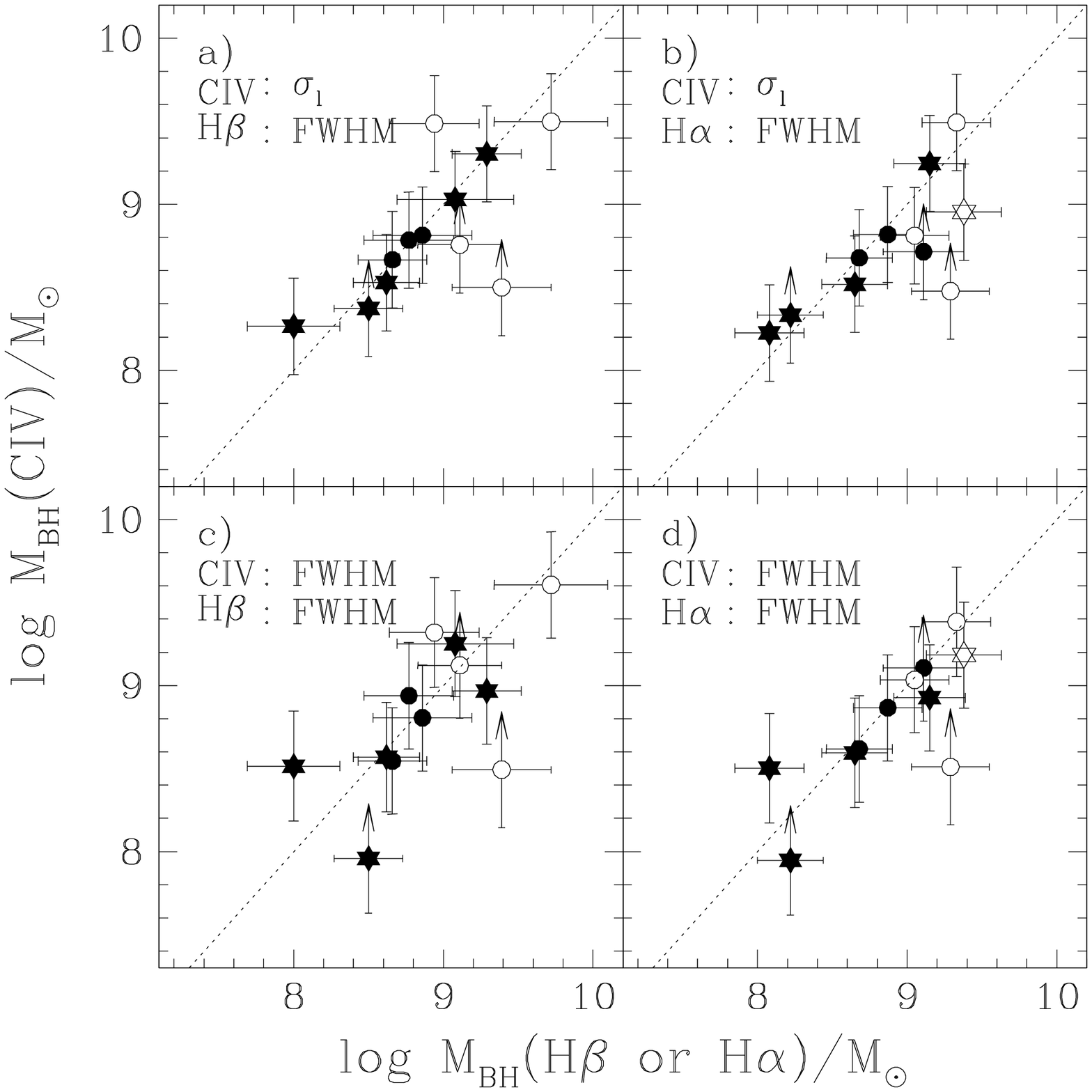}
    \caption{Same as Figure \ref{fg:CIV_H_masses}, but after
    correcting the \civ\ BH masses for the dependence on the ratio of
    the UV to optical continuum luminosities observed in Figure
    \ref{fg:diff_color}.}
    \label{fg:corrected_CIV_H_masses}
  \end{center}
\end{figure}

We next investigate if the residuals between the \civ\ and Balmer line
masses are correlated with any other observables. Figures
\ref{fg:diff_uv_lum} --- \ref{fg:diff_color} show the residuals as
a function of the 1350\AA\ and 5100\AA\ continuum luminosities,
redshift, Eddington ratio, blueshift of the \civ\ line, asymmetry of
\civ\ (parametrized by the ratio of the widths red and blue of the
centroid), and the ratio of the UV and optical continuum
luminosities. Table \ref{tab:residuals} summarizes the significance of
the correlations based on their Spearman rank-order coefficients. Only
the correlation with the ratio of the rest-frame optical and UV
continuum luminosities is significant (Figure
\ref{fg:diff_color}). Figure \ref{fg:corrected_CIV_H_masses} compares
the \civ\ and Balmer line derived BH masses after rescaling the
\civ\ masses using the best fit correlation determined from the
corresponding panel in Figure \ref{fg:diff_color}. We applied
corrections of the form
\begin{equation}\label{eq:lin_fit}
\log M_{\rm BH}^{\rm Corr}(\civm)\ =\ \log M_{\rm
BH}^{\rm VP06}(\civm)\ -\ b\ -\ a \log \frac{\lambda
L_{\lambda}(1350~\rm \AA)}{\lambda L_{\lambda}(5100~\rm \AA)} .
\end{equation}
\noindent where the coefficients $a$ and $b$ are listed in Table
\ref{tab:color_residuals}. For completeness, this Table also shows the
coefficients obtained when using the prescription A line-widths of
\civ, which are of similar magnitude and significance. Note that the
uncertainties given for these coefficients have been determined after
rescaling the errors such that the best fit has $\chi^2_{\nu} \equiv
1$. 

The agreement between the rest-frame UV and optical BH mass estimates
after applying this correction is remarkable, and the scatter of
objects with group I and non-lower bound line-widths has decreased
from 0.30 to 0.11, 0.23 to 0.10, 0.46 to 0.25 and 0.38 to 0.22~dex for
panels a--d of Figures \ref{fg:CIV_H_masses} and
\ref{fg:corrected_CIV_H_masses}, respectively. We find that the lowest
scatter is between the BH masses estimated from the $\sigma_l$ of
\civ\ $\sigma_l$ and the FWHM of either Balmer line. This supports our
conclusion in the previous section that $\sigma_l$ \civ\ BH masses
have small random errors, even if their systematic errors may be much
larger than those of the FWHM estimates due to blending of emission
lines. Such a small scatter places strong constraints on the strength
of a possible correlation between the mass residuals and any tertiary
parameter. We find again that a slope different from unity is not
required to describe the relation between the logarithm of the
\civ\ and Balmer line BH masses.

Since BH mass estimates generally scale as $\Delta v^2 L^{1/2}$
(eqns. [\ref{eq:m_hbeta}] and [\ref{eq:mass_civ}]), a naive
interpretation of the reduced scatter is that we have simply shifted
from showing $L_{1350\mathring{\rm A}}^{1/2} \rm
vs.\ L_{5100\mathring{\rm A}}^{1/2}$ to $L_{5100\mathring{\rm
    A}}^{1/2} \rm vs.\ L_{5100\mathring{\rm A}}^{1/2}$. The best fit
correction is statistically different from simply replacing
$L_{1350\mathring{\rm A}}$ by $L_{5100\mathring{\rm A}}$ by
$1$-$2\sigma$, so it is not simply swapping the
luminosities. {\textbf{More importantly, even if the slope was exactly
    $\alpha=1/2$, it reveals the crucial point that a significant
    fraction of any problems in reconciling \civ\ and Balmer line
    estimates of BH masses is due to the estimates of the continuum
    luminosities rather than any properties of either line}}.

There are 3 potential causes for a correlation of the mass ratio with
the ratio of the continuum luminosities: i) obscuration, ii) host
contamination and iii) non-universal AGN SEDs. Unfortunately, our
analysis does not allow us to determine which BH mass estimate is more
accurate. Extinction will reduce the rest-frame UV continuum
luminosity while having little effect on the rest-frame optical
luminosity. Conversely, host contamination will raise the optical
luminosity while leaving the UV unchanged, as galaxies are typically
brighter in the optical than in the UV. With respect to case iii), the
radius of the BLR is really determined by the flux of the ionizing
continuum ($\lambda < 912\rm \AA$). The $R_{\rm BLR} - L$ relations
used to construct equations (\ref{eq:m_hbeta}), (\ref{eq:m_halpha})
and (\ref{eq:mass_civ}) implicitly assume a universal SED for all
quasars, as they imply that the ionizing continuum can be uniquely
predicted from the continuum luminosity at longer wavelengths. This
approximation is likely to be better for the rest-frame UV continuum
than for the optical. All three cases discussed would produce a slope
of $a\simeq 0.5$ in equation (\ref{eq:lin_fit}), simply representing
the luminosity power indices in equations (\ref{eq:m_hbeta}),
(\ref{eq:m_halpha}) and (\ref{eq:mass_civ}). This is generally
shallower than the observed slope but within 2$\sigma$ of the best-fit
relations. A larger sample is needed to fully determine if the slope
of this correlation is statistically different from $\alpha\simeq
0.5$. We note that in order to create a slope larger than 0.5, it
would be necessary for the velocity widths of the quasar broad lines
to be dependent on the ratio of the continuum luminosities. There is
some evidence that the inclination angle of the accretion disk with
respect to the line of sight may correlate with both the SED of the
continuum \citep[][and references therein]{gallagher05} and the FWHM
of the broad H$\beta$ line \citep{wills86,wills95,jarvis06}, although
no such correlation is observed for the FWHM of \civ\ \citep[][but see
  \citealt{decarli08}]{vestergaard00}. Accretion disk inclination
corrections, however, would act in the opposite sense to the observed
correlation and hence cannot be responsible for a slope in excess of
0.5 --- disks with higher inclination angles (closer to edge-on) would
appear to have higher FWHM of H$\beta$ and bluer continua for a fixed
``true'' BH mass \citep[i.e. not estimated from spectral features;
  see][and references therein]{gallagher05}.

Our sample is likely representative of observations of the general
quasar population in terms of reddening and host contamination. It
could, in principle, have a larger typical reddening due to additional
obscuration by dust associated with the lens, but this is unlikely to
be important for our sample. Reddening by the lens galaxy will
typically vary between quasar images. \citet{falco99} studied most of
the objects in our sample and found that only two of them showed
significant differential reddening: SBS0909+523 ($\Delta
E(B-V)=0.2~\rm mag$ for image B with respect to A, see also Appendix
A) and Q2237+0305 ($\Delta E(B-V)=0.18\ \rm and\ 0.17~\rm mag$ for
images C and D with respect to A). Small but non-zero differential
reddening was also detected for three other lenses (HE1104--1805,
H1413+117 and B1422+231). The lensed quasars SDSS0246--0825,
HS0810+2554, FBQ1633+3134 and SDSS1138+0314 were not part of the
sample studied by \citet{falco99}. We studied the latter object in
\S\ref{sssec:sdss1138_luci} and concluded images B and C did not show
evidence for differential reddening between them, but there is no
information in this regard for the other three quasars. Lensing can
also alter host contamination in the quasar observations as compared
to an unlensed case. The exact amount of host contamination depends on
the size of the PSF and aperture used, the morphology of the lens and
the surface brightness profile of the quasar's host galaxy \citep[see,
  e.g.,][]{kochanek01,ross09}, however the zeroth order effect is to
not alter the amount of host contamination compared to an unlensed
quasar.

\begin{figure}
  \begin{center}
    \plotone{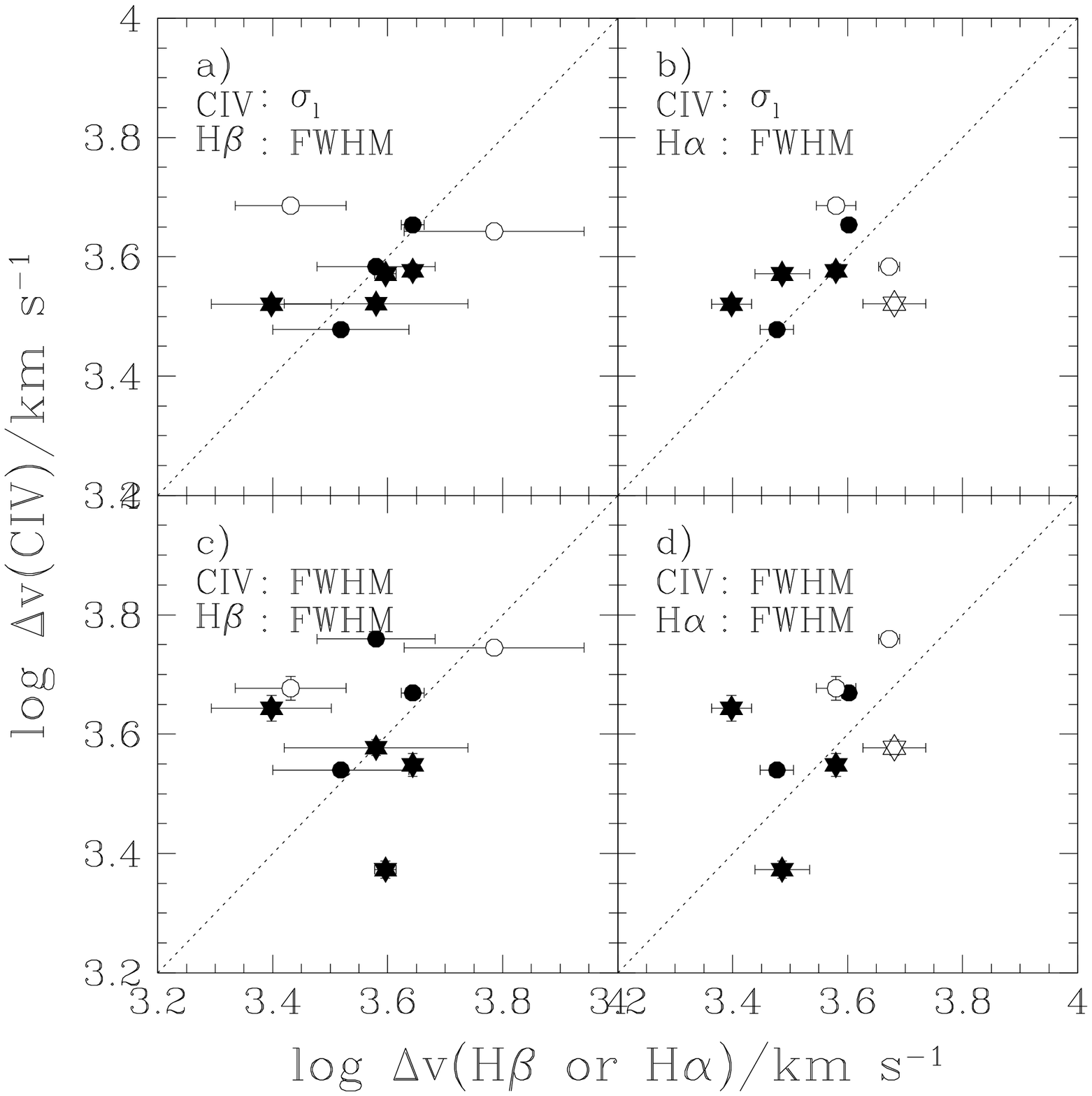}
    \caption{Comparison between the \civ\ and Balmer lines measured
      line-widths. Points and lines have the same meaning as in Figure
      \ref{fg:CIV_H_masses}. We do not show lower bounds for clarity.}
    \label{fg:raw_line_widths}
  \end{center}
\end{figure}

\begin{figure}
  \begin{center}
    \plotone{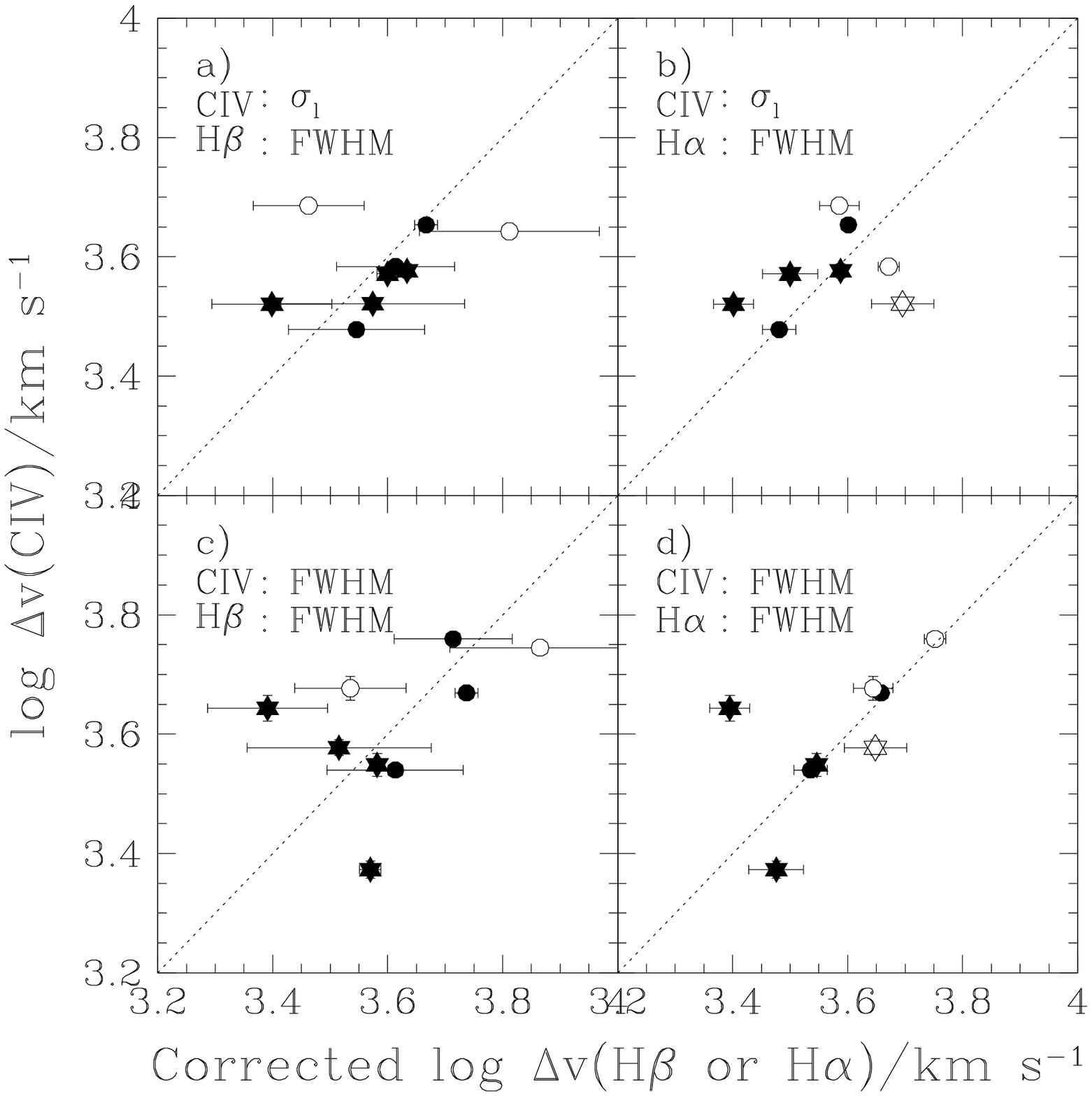}
    \caption{Comparison between the \civ\ and Balmer lines measured
      line-widths after applying correction from equation
      (\ref{eq:corr_hbeta_width}). Points and lines have the same
      meaning as in Figure \ref{fg:CIV_H_masses}. We do not show lower
      bounds for clarity.}
    \label{fg:corr_line_widths}
  \end{center}
\end{figure}

While we have shown that the dominant source of scatter in the
comparison between the BH mass estimates based on \civ\ and the Balmer
lines is due to the continuum luminosities, we still wish to assess
the relation between the widths of the different emission lines
used. Figure \ref{fg:raw_line_widths} shows the comparison between the
\civ\ and Balmer line widths. Note that we do not show measurements
for which we only have lower bounds on the \civ\ width due to
absorption.  The best agreement is between $\sigma_l$ of \civ\ and
FWHM of H$\beta$, which is expected given that these measurements also
give the lowest scatter in the BH mass estimates, however a generally
good agreement is also observed in all panels. We remind the reader,
however, that the corrections we found between the BH mass estimates
residuals and the ratio of the continuum luminosities did not have a
slope of 0.5. This implies that the ratio of the line-widths may have
a dependence on the luminosity ratio, with a power given by the excess
of the slope from 0.5. This could be a source of additional scatter in
Figure \ref{fg:raw_line_widths}, and so, instead of comparing the
line-widths directly, we also compare them after applying a correction
based on the continuum luminosity estimates, of the form
\begin{equation}\label{eq:corr_hbeta_width}
\log \Delta v({\rm H\beta\ \rm or\ \rm H\alpha})_{corr}\ =\ \log \Delta v({\rm
  H\beta\ \rm or\ \rm H\alpha})\ +\ \frac{(a-0.53)}{2} \log \frac{\lambda
  L_{\lambda}(1350~\rm \AA)}{\lambda L_{\lambda}(5100~\rm
  \AA)}\ +\ 5\times10^{-3}\log \frac{\lambda L_{\lambda}(5100\rm
  \AA)}{10^{44}~\rm erg~\rm s^{-1}} ,
\end{equation}
\noindent as shown in Figure \ref{fg:corr_line_widths}. Note that
since we don't know the origin of the corrections, applying it to
H$\beta$ rather than \civ\ is a completely arbitrary decision made for
display purposes. The agreement is now better and a correlation
between the measurements is clear, suggesting that the widths of both
lines are equally good tracers of BH mass. We have quantified the
correlation between the different line-widths in Table
\ref{tab:lw_residuals} using the Spearman rank-order coefficient, as
we did for the BH mass residuals. After applying the correction from
equation (\ref{eq:corr_hbeta_width}), we find positive correlations
which are typically statistically significant between the \civ\ and
Balmer line-widths. A weak anti-correlation, however, is measured
between the FWHMs of \civ\ and H$\beta$, but it is not statically
significant.

\section{Comparison with Other Studies}\label{sec:comp_others}

In \S\ref{ssec:civ_balmer_comp} we used a sample of lensed quasars to
compare BH masses based on observations of the \civ\ emission line and
of the Balmer lines H$\alpha$ and H$\beta$. We found that the
agreement between the rest-frame UV and rest-frame optical based BH
masses is reasonably good. We also found that this agreement is even
better once we apply an empirically determined correction based on the
ratio of the 1350\AA\ and 5100\AA\ continuum luminosities. There have
been a number of previous studies that have explored the relative
accuracy of BH masses based on \civ\ and H$\beta$, and they have
reached both similar and opposite conclusions.

The studies of \citet{vestergaard06}, which we have discussed
previously, and \citet{dietrich04} found that \civ\ derived BH masses
are consistent with those obtained from the width of H$\beta$, and
hence constitute a valid replacement as a mass
estimator. \citet{shemmer04}, \citet{netzer07}, \citet{sulentic07} and
\citet{dietrich09}, however, reached opposite
conclusions. \citet{shemmer04} concluded that BH masses derived from
\civ\ were poorly matched to those obtained from H$\beta$ and could be
systematically different. They showed that for a sample of narrow-line
Seyfert 1 galaxies, \civ\ based BH masses are larger by an average
factor of $\sim3$ with respect to those obtained from
H$\beta$. \citet{dietrich09} also found a large disagreement between
the two estimates of the BH mass, but they found that using
\civ\ tends to underestimate the BH masses by a factor of $\sim1.7$,
although the significance of this result is limited by the small
number of objects (9) in their sample. While \citet{sulentic07} also
found significant disagreement, they argue that the magnitude of the
offset depends on the spectroscopic characteristics of the
quasar. \citet{netzer07}, on the other hand, found no significant
offset between the mass estimates, but argued there was also no
discernible correlation between them. It is likely that many of the
differences between the results of these studies are due to the use of
different $R_{\rm BLR}-L$ calibrations, different $f$-factors,
different prescriptions for measuring line widths, limited mass ranges
and data quality.

Here we take their measurements, where possible, and make estimates of
the BH masses using equations (\ref{eq:m_hbeta}), (\ref{eq:m_halpha})
and (\ref{eq:mass_civ}). We caution the reader, however, that we are
not redoing the line-width and continuum luminosity measurements in a
consistent manner and that this may be a significant source of
additional scatter. We used all 21, 15 and 9 sources from
\citet{vestergaard06}, \citet{netzer07} and \citet{dietrich09} with
H$\beta$ and \civ\ line FWHM and continuum luminosity measurements. We
could not use 10, 29 and 1 sources from these studies that lack either
or both line widths, or any of the sources from \citet{dietrich04},
which lack measurements of the 5100\AA\ continuum luminosity. We also
could not use the sources of \citet{sulentic07}, as they only report
narrow-component subtracted \civ\ widths, which are not compatible
with the rest of the measurements we discuss. We note that the 29
sources we could not use from the study of \citet{netzer07} also
belong to the sample of \citet{shemmer04}, for which the \civ\ line
widths and UV continuum luminosities are not reported.

\begin{figure}
  \begin{center}
    \plotone{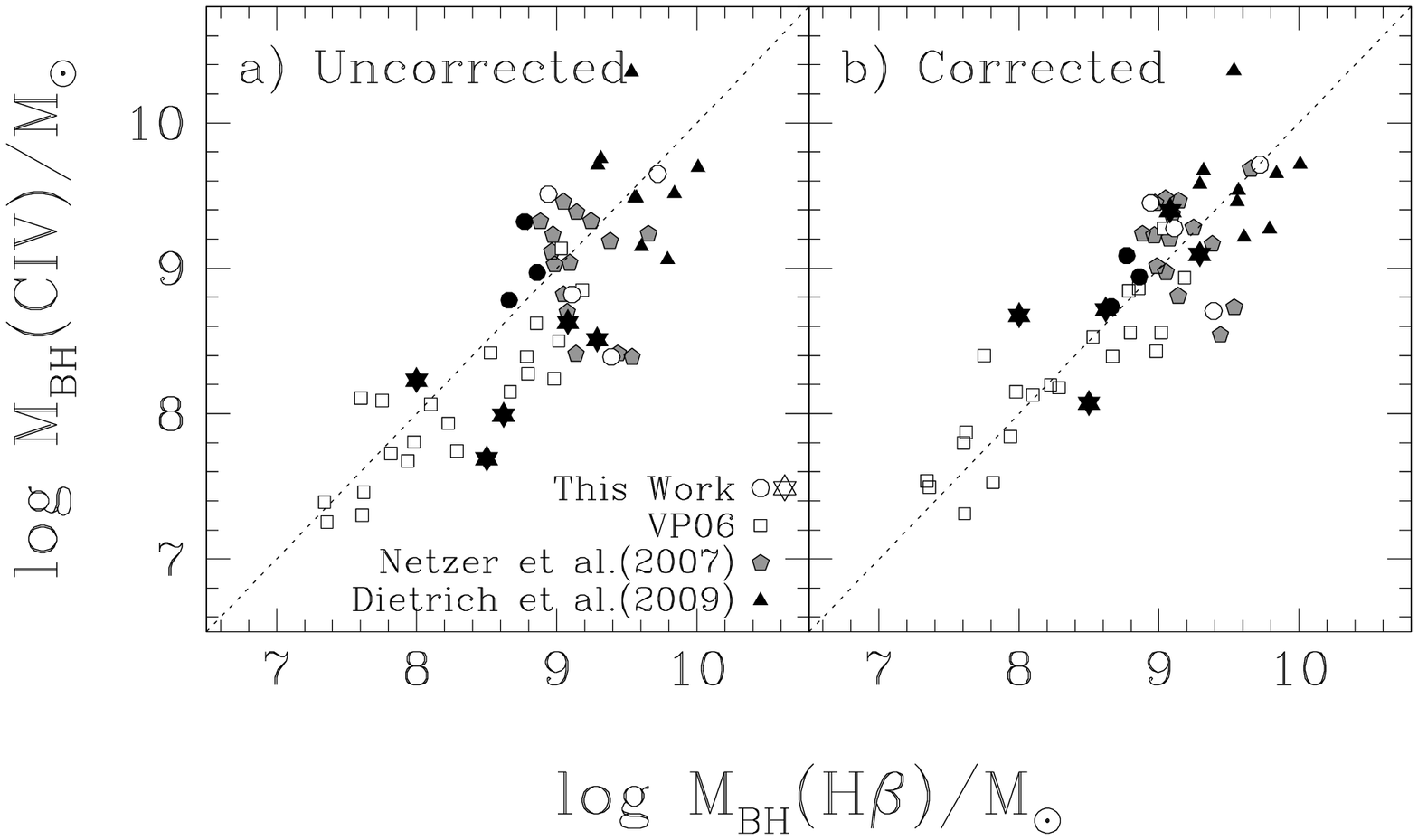}
    \caption{Panel {\it{a}}) shows the BH masses estimated using
    eqns. (\ref{eq:m_hbeta}), (\ref{eq:m_halpha}) and
    (\ref{eq:mass_civ}) for the objects in the samples of
    \citet{vestergaard06} ({\it{open squares}}), \citet{netzer07}
    ({\it{solid gray pentagons}}) and \citet{dietrich09} ({\it{solid
    black triangles}}) for which this was possible (see
    \S\ref{sec:comp_others} for details). Objects in our sample are
    shown by the solid and open six-pointed stars and circles, keeping
    the point style conventions used in previous plots. The dotted
    line shows where the masses are equal. Error-bars are not shown in
    order to make the plot more legible. Panel {\it{b}}) shows the
    results after applying the continuum slope correction from Table
    \ref{tab:color_residuals}.}
    \label{fg:all_surveys_masses}
  \end{center}
\end{figure}

\begin{figure}
  \begin{center}
    \plotone{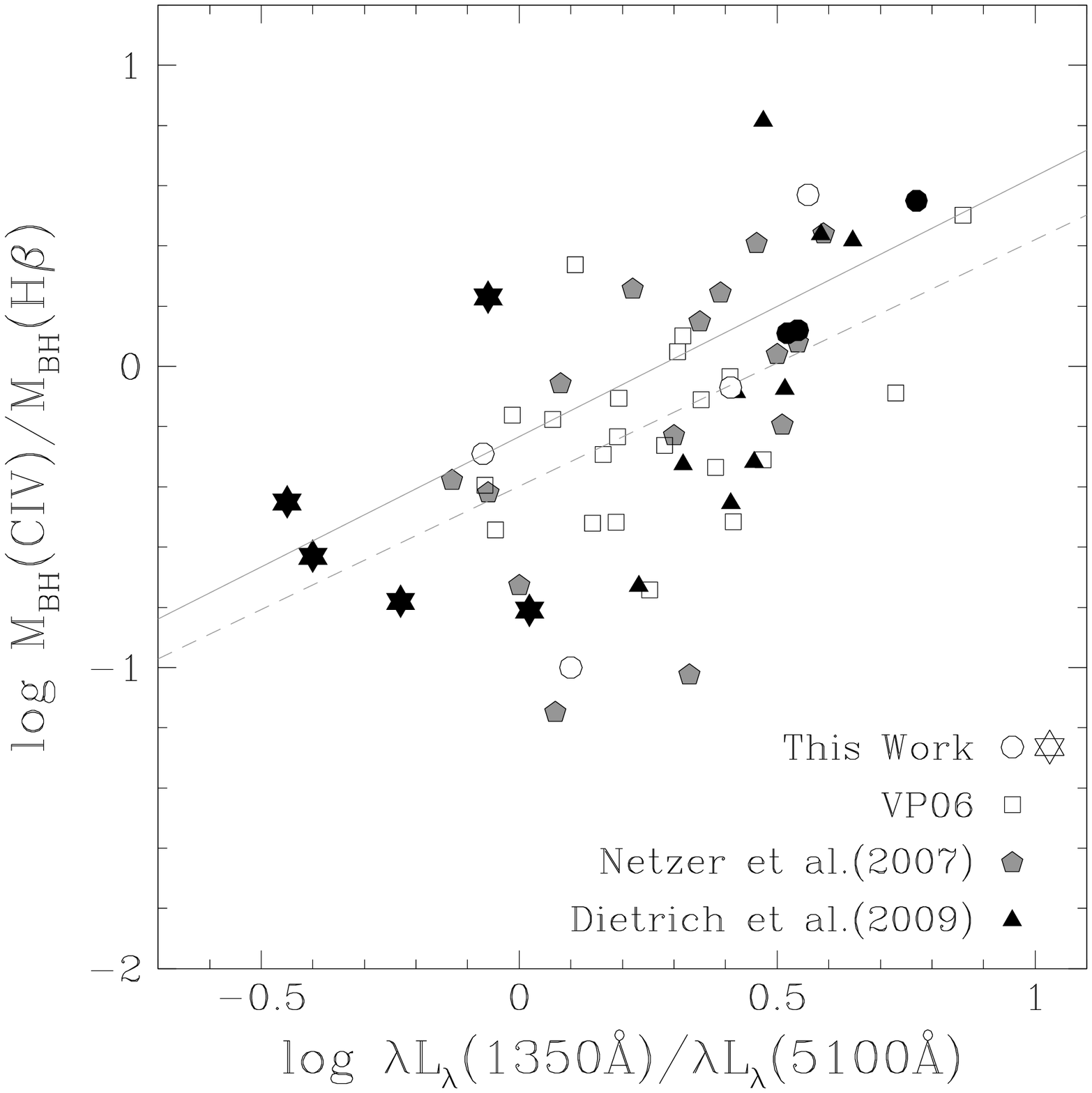}
    \caption{Residuals between BH masses estimated from the FWHM of
    the \civ\ and H$\beta$ broad emission lines as a function of the
    logarithm of ratio of the continuum luminosities at 1350\AA\ and
    5100\AA\ for the samples of \citet{vestergaard06} ({\it{open
    squares}}), \citet{netzer07} ({\it{solid gray pentagons}}) and
    \citet{dietrich09} ({\it{solid black triangles}}) as well as our
    sample ({\it{solid and open six-pointed stars and
    circles}}). Point-styles have the same definitions as in Figure
    \ref{fg:all_surveys_masses}. The solid line shows the best-fit
    linear relation for our data while the dashed line shows the
    best-fit to the combined sample. Error-bars are not shown in order
    to make the plot more legible.}
    \label{fg:all_surveys_color_residuals}
  \end{center}
\end{figure}

\begin{figure}
  \begin{center}
    \plotone{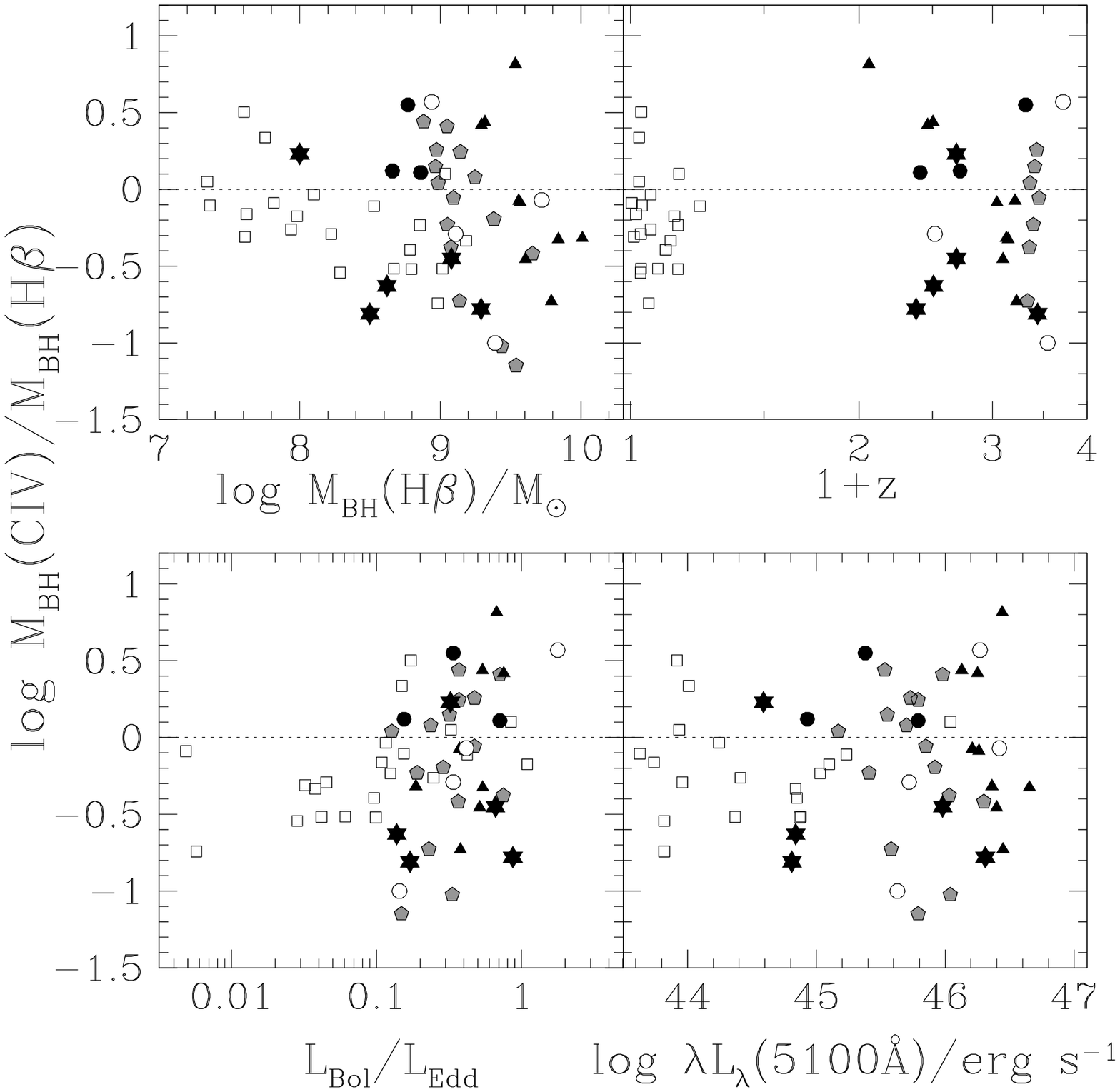}
    \caption{Residuals between BH masses estimated from the FWHM of
      the \civ\ and H$\beta$ broad emission lines as a function of the
      logarithm of H$\beta$-based BH mass ({\it{top left}}), redshift
      ({\it{top right}}), Eddington ratio ({\it{bottom left}}) and
      continuum luminosity at 5100\AA ({\it{bottom right}}). Objects
      belong to the samples of \citet{vestergaard06} ({\it{open
          squares}}), \citet{netzer07} ({\it{solid gray pentagons}})
      and \citet{dietrich09} ({\it{solid black triangles}}) as well as
      our sample ({\it{solid and open six-pointed stars and
          circles}}). Point-styles have the same definitions and in
      Figure \ref{fg:all_surveys_masses}. The dotted line shows where
      the masses are equal. Error-bars are not shown in order to make
      the plot more legible.}
    \label{fg:all_surveys_other_residuals}
  \end{center}
\end{figure}

The left panel of Figure \ref{fg:all_surveys_masses} compares the
\civ\ and H$\beta$ BH masses derived for all these objects along with
those in our sample. A clear correlation is observed for the complete
ensemble of objects, albeit with a considerable scatter of
0.41~dex. The scatter is comparable with the 0.46~dex we find for our
sample of \civ\ and H$\beta$ FWHM-based BH masses (see
\S\ref{ssec:civ_balmer_comp}). A Spearman rank-order coefficient
analysis returns $r_s=0.79$ with a probability that both mass
estimates are uncorrelated of $P_{\rm ran} = 2\times10^{-12}$. A
linear fit to the left panel of Figure \ref{fg:all_surveys_masses} of
the form 
\begin{equation}\label{eq:mass_lin_fit}
\log \frac{M_{\rm BH}(\civm)}{10^8~M_{\odot}}\ =\ m~\log \frac{M_{\rm
BH}(\rm H\beta)}{10^8~M_{\odot}}\ +\ n
\end{equation}
\noindent returns a best-fit slope of $m=0.89\pm0.08$ and intercept of
$n = -0.09\pm0.08$ (the measurement uncertainties were scaled to make
$\chi^2_{\nu}=1$ before determining the uncertainties in the fit
parameters). If we plot the residuals between the two BH mass
estimates we find, just as in \S\ref{ssec:civ_balmer_comp}, that a
significant correlation is observed with the ratio of the UV and
continuum luminosities (Figure \ref{fg:all_surveys_color_residuals}),
but not with BH mass, redshift, Eddington ratio or the continuum
luminosity at 5100\AA\ (all shown in Figure
\ref{fg:all_surveys_other_residuals}), or with the continuum
luminosity at 1350\AA\ (not shown). The best-fit linear relation to
the correlation between BH mass residuals and the ratio of the
continuum luminosities, shown in Table \ref{tab:color_residuals}, has
a slope of $a=0.82\pm0.18$ and an intercept of $b=-0.40\pm0.07$. While
the slope is consistent with the value obtained for our sample alone
($a=0.86\pm0.25$, $b=-0.23\pm0.12$), the intercept differs by
approximately 0.2~dex (approximately $1.5\sigma$). The offset is
likely produced by the different prescriptions used to measure the
width of the broad emission lines. Figure \ref{fg:all_surveys_masses}
also compares the \civ\ and H$\beta$ derived masses after correcting
for this correlation by applying equation (\ref{eq:lin_fit}) (see
\ref{ssec:civ_balmer_comp} for details). While the strength of the
correlation has not increased substantially ($r_s=0.80$, $P_{\rm ran}
= 6.4\times10^{-13}$), the scatter has decreased from 0.41~dex to
0.34~dex. This change is significantly more modest than that found for
our sample of lensed quasars, but this is likely due to the
inhomogeneous prescriptions used to measure the width of the emission
lines. A linear fit of the form of equation (\ref{eq:mass_lin_fit}) to
the relation between the BH mass estimates after applying the
correction returns very similar parameters as before, with a best-fit
slope of $m = 0.88\pm0.07$ and intercept of $n = 0.06\pm0.07$. Note
that the measurement errors have again been scaled to make
$\chi^2_{\nu}=1$ before estimating the uncertainties in the fit
parameters. Given the inhomogeneity of the measurements used, the
relatively small number statistics of the sample, their typically
large error bars and the likely intrinsic dispersion, however, we
cannot currently determine whether the deviation from a slope of unity
is significant or not.

\begin{figure}
  \begin{center}
    \plotone{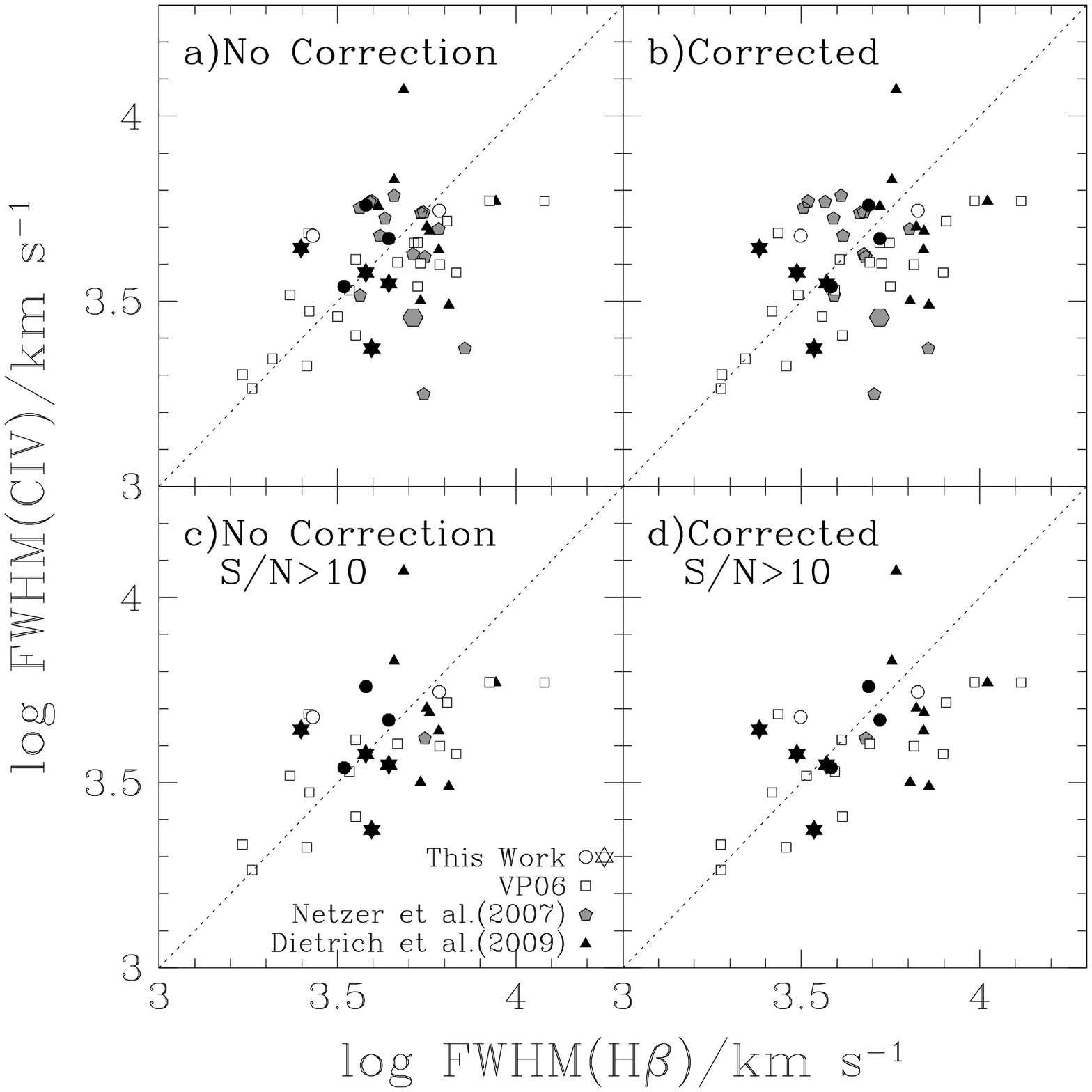}
    \caption{Comparison of the \civ\ and H$\beta$ line-widths for the
      sample we have compiled from the literature. Point types and
      line styles are the same as for
      Fig. \ref{fg:all_surveys_masses}. The large gray hexagon shows
      SDSS1151+0340. Top panels show the complete literature sample
      while bottom panels only show objects with spectra that have
      continuum $S/N>10$ in the vicinity of \civ.}
    \label{fg:all_widths}
  \end{center}
\end{figure}

As argued before, the inhomogeneity of the measurements can be a very
significant source of scatter in the comparisons discussed above. We
have shown in the previous section that with homogeneously analyzed,
high S/N spectra, the difference between the line widths is not the
dominant source of scatter in the comparison between \civ\ and Balmer
line-based BH mass estimates. We would still like to assess whether
the \civ\ and H$\beta$ line widths are correlated in this combined
sample. Figure \ref{fg:all_widths} shows this comparison for all the
objects used in this section with and without applying the correction
based on equation (\ref{eq:corr_hbeta_width}). While the scatter in
Figure \ref{fg:all_widths} is large, there is still a statistically
significant (99\%) correlation between the measured line-widths (see
Table \ref{tab:lw_residuals}). Most of the scatter is due to the
sample of \citet{netzer07}. Upon inspection of the SDSS spectra used
for that study, we find that almost all the outliers correspond to low
$S/N$ spectra. Given that the \civ\ line is typically very complex,
this can be a major source of uncertainty.

\begin{figure}
  \begin{center}
    \plotone{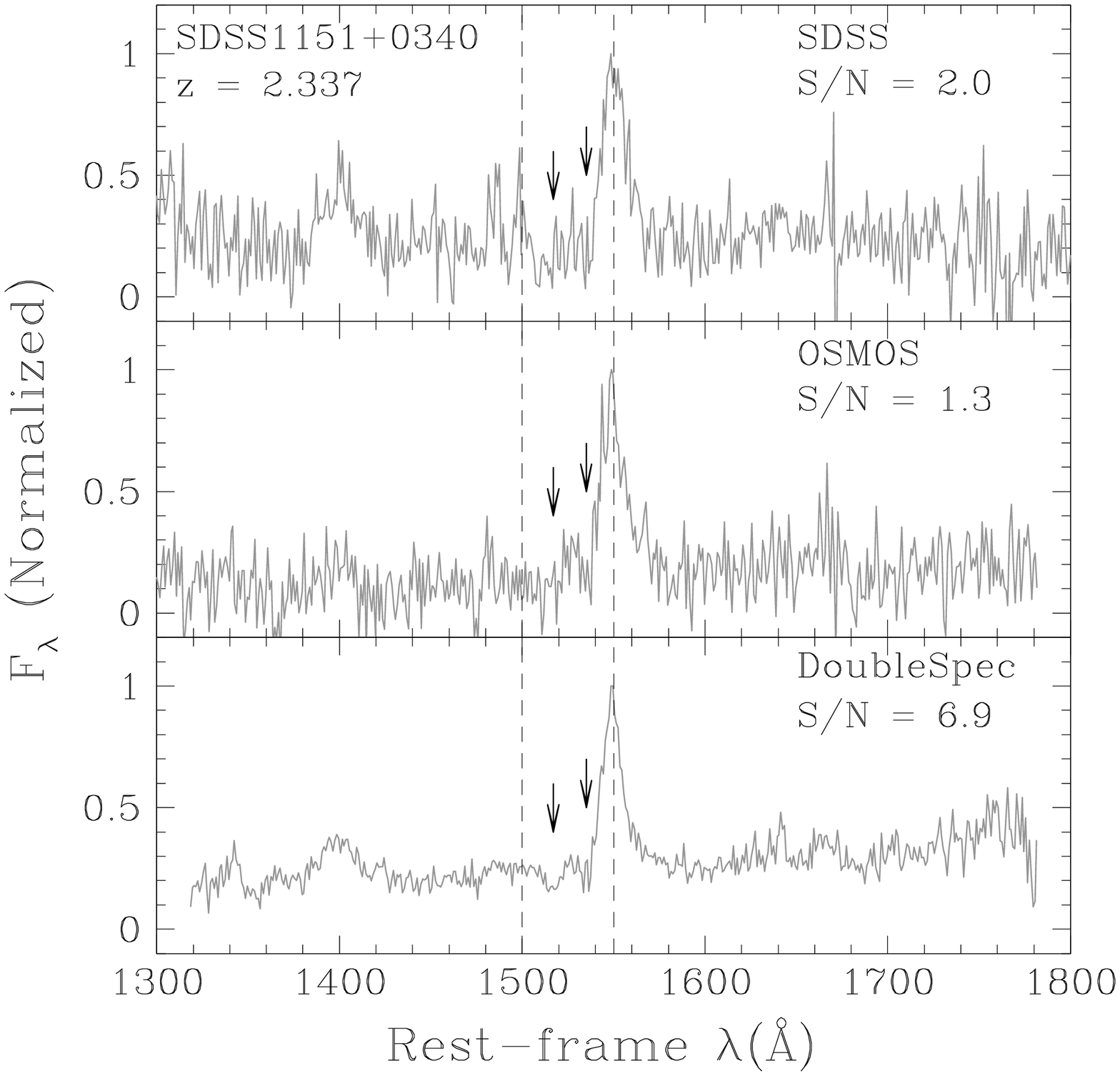}
    \caption{Spectra of the QSO SDSS1151+0340 obtained by SDSS
      ({\it{top}}), with MDM/OSMOS ({\it{middle}}) and with
      Palomar/Double Spectrograph({\it{bottom}}). The arrows mark the
      probable absorption troughs near the \civ\ line. The spectra
      have been resampled to a common resolution, and the continuum
      $S/N$ shown in the upper left corner of each panel has been
      calculated in the same way as for all other objects in our
      sample.}
    \label{fg:sdss1151_specs}
  \end{center}
\end{figure}

As an experiment, we obtained higher $S/N$ spectra for one of the
outliers, SDSS1151+0340. It has the third most discrepant line-width
ratio in the sense that the \civ\ line is too narrow compared to the
Balmer lines. We obtained two independent spectra, one with OSMOS
\citep{martini11} at the MDM 2.4m telescope and one with the Double
Spectrograph \citep{oke1982} at the Palomar 200-inch telescope. Due to
poor weather conditions and aperture size, only the Double
Spectrograph observations yielded a higher $S/N$ spectrum than that of
SDSS. All three spectra of SDSS1151+0340 are shown in Figure
\ref{fg:sdss1151_specs}. The spectrum obtained with Double
Spectrograph reveals that there is significant absorption near the
\civ\ line, with two clear absorption troughs. These can be seen in
the lower $S/N$ spectra, but are difficult to distinguish from the
noise. Due to the very substantial absorption, it is not possible to
reliably measure the width of the \civ\ line, and the width
measurement of \citet{netzer07} should only be considered as a lower
bound. While the SDSS spectrum of this source has the lowest continuum
$S/N$ in their sample ($S/N=1.5$), it is comparable to many of their
other sources. The average continuum $S/N$ of the SDSS spectra is only
6.7, with all objects having a lower $S/N$ than any optical spectra in
our lensed quasar sample. In particular, the second largest outlier in
their sample also has the second lowest $S/N$ of 4.1.

While our example comes from \citet{netzer07}, low $S/N$ spectra are
also present in all the additional samples we consider. If we
eliminate objects with continuum $S/N<10$ in the vicinity of \civ, the
statistical correlation between H$\beta$ and \civ\ line widths
increases dramatically. The bottom panels of Figure
\ref{fg:all_widths} show the comparison of the line-widths in the
absence of these objects, and a clear correlation is observed between
the \civ\ and H$\beta$ FWHM measurements, regardless of whether we
apply the continuum luminosity based correction of equation
(\ref{eq:corr_hbeta_width}). These correlations are more statistically
significant by about two orders of magnitude than when including the
low $S/N$ spectra, with a probability of not being real of $\sim
5\times 10^{-4}$ (see Table \ref{tab:lw_residuals} for details). This
suggests that the width of \civ\ is as good a tracer of BH mass as the
width of H$\beta$, with the caveat that high $S/N$ spectra of the
rest-frame UV region are fundamental to accurately model the structure
of the \civ\ emission line.

\section{Conclusions}\label{sec:conclusions}

Using a sample of high-redshift gravitationally lensed quasars
observed spectroscopically in the UV/optical and NIR, we have studied
the agreement between single-epoch BH mass estimators based on the
\civ, H$\beta$ and H$\alpha$ broad emission lines. Our sample consists
of 12 lensed quasars observed with {\it{HST}} by the CASTLES
project. In particular, we have used the sample of NIR spectroscopic
observations by \citet{greene10} as a starting point and improved on
it by (i) adding new NIR observations for 3 objects (SDSS1138+0314,
SBS0909+253 and HS0810+2554), (ii) adding high $S/N$, uniformly
analyzed, optical spectroscopic observations for all targets, and
(iii) adding the missing rest-frame $\lambda L_{\lambda}(5100\rm \AA)$
luminosity estimates for SDSS0246-0852, HS0810+2554 and Q2237+030.

We described in detail all the methods we used to measure velocity
widths and their uncertainties, rest-frame continuum luminosities and
to estimate the BH mass of each quasar using the H$\beta$, H$\alpha$
and \civ\ emission lines. We first compared the \civ\ BH mass
estimates based on the FWHM and $\sigma_l$ line-width
characterizations and the calibration of \citet{vestergaard06} and
found that, for our sample, the $\sigma_l$ based BH masses are
systematically underestimated with respect to the FWHM-based ones by
$0.13\pm0.06~\rm dex $ if using prescription A and $0.24\pm0.07~\rm
dex$ if using prescription B. A similar offset is not observed in the
\citet{vestergaard06} data set. The difference probably arises from
our treatment of the blending of the broad \civ\ emission line with
the nearby broad HeII\,$\lambda$1640 and FeII emission redward of
\civ, which is partly confirmed by the lower difference found for the
prescription A measurements. This adds to the arguments in
\citet{denney09} that $\sigma_l$ is not universally reliable for SE
mass estimates in the presence of blending, as the results obtained
are highly dependent on the exact prescription used for the line
characterization. When comparing with BH masses derived from the
H$\alpha$ and H$\beta$ broad emission lines, we find that \civ\ FWHM
based BH masses are not biased, reinforcing the conclusion that the
bias is in the $\sigma_l$ estimates. We note, however, that the
scatter between \civ\ FWHM and $\sigma_l$ derived masses is relatively
small, suggesting that if a consistent prescription for measuring
$\sigma_l$ is applied, $\sigma_l$ would be at least as accurate as
FWHM. This is important because $\sigma_l$ measurements are
significantly more reliable for complex line profile shapes and in the
presence of narrow-line component residuals
\citep{peterson04,denney09}.

We then compared the \civ\ and Balmer line BH mass estimates. After
offsetting the \civ\ $\sigma_l$ masses to agree with the FWHM
estimates, we find there is no significant offset between \civ\ and
either Balmer line BH mass estimates. Averaged over the 4 possible
\civ/Balmer line mass comparisons (see, for example, Figure
\ref{fg:CIV_H_masses}), the offset is $-0.15\pm 0.17$~dex and the
scatter is 0.35~dex. Note that the error in the mean offset
corresponds to the average of the errors of the four estimates, which
is representative given that the estimates are not truly
independent. The scatter of 0.35~dex is very close to the scatter of
0.34~dex found by \citet{shen08} between Mg\,{\sc ii} and \civ\ FWHM
based BH mass estimates, and significantly larger than the scatter of
0.22~dex they found between Mg\,{\sc ii} and H$\beta$ FWHM based BH
masses.

We find that the residuals between the \civ\ and H$\beta$ and
H$\alpha$ based mass estimates are not strongly correlated with the UV
or optical continuum luminosities, redshift or Eddington ratio, but we
find a strong dependence on the ratio of the UV to optical continuum
luminosities. If we correct for this color dependence, the agreement
between the \civ\ and Balmer line estimates is remarkably good, with
an average scatter of 0.18~dex, almost a factor of 2 smaller. We find
the scatter is smallest --- approximately 0.1~dex --- when using the
H$\beta$ line and the $\sigma_l$ characterization of \civ\ rather than
its FWHM. This observed correlation could be caused by i) reddening,
ii) host contamination, or iii) an object-dependent SED shape. The
slope we observe is somewhat steeper than that expected in any of
these cases, and may suggest a luminosity component to the line-width
characterization of the broad emission lines. A larger sample is
needed to accurately determine the slope of this correlation and
determine its nature with certainty. More generally, the comparison
shows that many of the problems in comparing \civ\ and Balmer line BH
mass estimates are associated with the continuum luminosities rather
than any potential physical complexities with the \civ\ lines. When we
compare the line-widths directly instead of the BH masses, we find
that the width of \civ\ is well correlated with those of the Balmer
lines once the correction based on the ratio of the continuum
luminosities is applied.

Our conclusions are unchanged if we add 45 additional, but
heterogeneously analyzed, \civ\ and H$\beta$ estimates from
\citet{vestergaard06}, \citet{netzer07} and \citet{dietrich09}. We
used the published FWHM of both emission lines and rest-frame UV and
optical continuum luminosities of these sources, but the mass
calibrations used for our sample. There is a clear linear correlation
between the BH mass estimates, and the residuals are again correlated
with the ratio of the continuum luminosities. The residuals are not
correlated with either continuum luminosity alone, redshift, BH mass
or Eddington ratio. We also find for this heterogeneous sample that
the width of \civ\ is well correlated with that of H$\beta$,
particularly after we eliminate the objects with low $S/N$
\civ\ spectra. Relatively high S/N spectra are essential to obtaining
accurate line widths.

In summary, our results show that \civ\ is a good BH mass estimator but
with small prescription-dependent offsets.  The correlation of the
mass residuals with the continuum slope could be a bias in either or
both of the estimators.  Determining the ``blame'' would require an
independent mass estimate, but its existence should not be a surprise
given that quasar SEDs are not universal \citep[e.g.,][]{yip04}.  More
generally, unless we are to believe that all properties of AGN are
determined by a single quantity, the black hole mass, both
single-epoch mass estimates and reverberation-mapping radius estimates
must depend on additional parameters.  That the black hole mass seems
to dominate is convenient, but the excess scatter in mass and radius
estimates beyond the measurement uncertainties requires either that
the error estimates are incorrect or is evidence for additional
parameters. One possibility is that radiation pressure plays a
significant role \citep{marconi08} and it could easily affect
different lines in different ways. While there has been considerable
recent effort to expand the range of black hole masses included in
these studies \citep[e.g.,][]{kaspi07,bentz09,botti10}, it is equally
important to expand the range in other physical parameters such as
spectral shape and Eddington ratio in order to better search for these
additional correlations.

\acknowledgments

We would like to thank Jenny E. Greene, Christopher Onken, Chien
Y. Peng, Kristen Sellgren, Marianne Vestergaard and Linda Watson for
their help and suggestions that improved our work. We thank
F.~Courbin, E.~Mediavilla, V.~Motta, L.J.~Goicoechea, S.~Sluse,
J.L.~Tonry, L.~Wisotzki and J.~Mu\~noz for sending us their optical
spectra of Q2237+030, SDSS1138+0314, Q0957+561, HE1104--1805,
B1422+231, and SBS0909+532. We thank F.~Harrison for helping us obtain
an optical spectrum of SDSS1151+0340. We would also like to thank all
the people in the LUCIFER science demonstration time team that did not
participate directly in this work. We thank the anonymous referee for
suggestions that help improve our work. R.J.A. was supported in part
by an appointment to the NASA Postdoctoral Program at the Jet
Propulsion Laboratory, administered by Oak Ridge Associated
Universities through a contract with NASA. C.S.K. is supported by NSF
grants AST-0708082 and AST-1009756. B.M.P., M.D. and R.W.P. are
supported by NSF grant AST-1008882. P.M. is supported by NSF grant
AST-0705170. This research has made use of the NASA/IPAC Extragalactic
Database (NED) which is operated by the Jet Propulsion Laboratory,
California Institute of Technology, under contract with the National
Aeronautics and Space Administration.

\appendix

\section{Notes on Individual Objects}\label{sssec:notes}

\begin{figure}
  \begin{center}
    \epsscale{0.9}
    \plotone{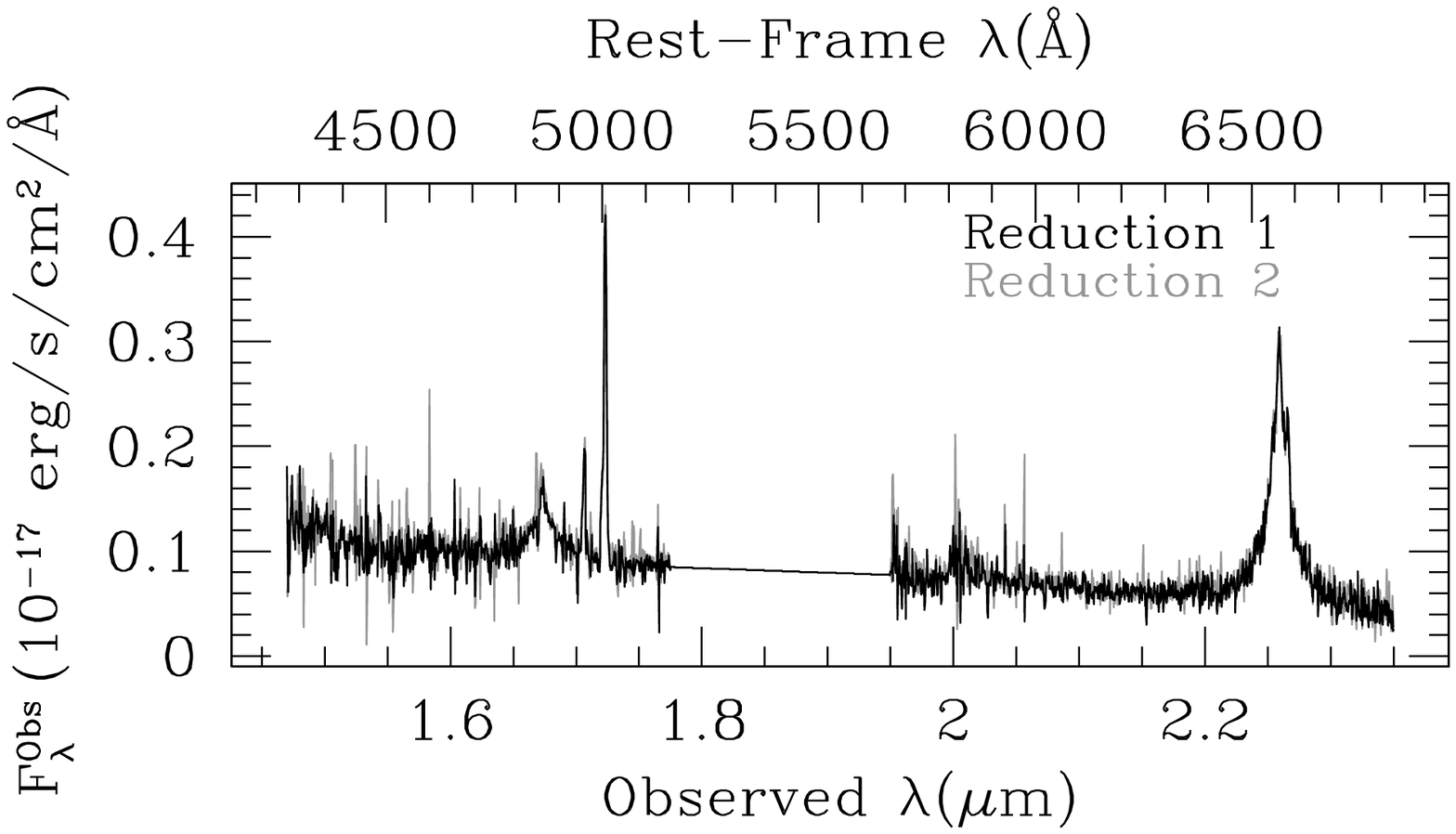}
    \plotone{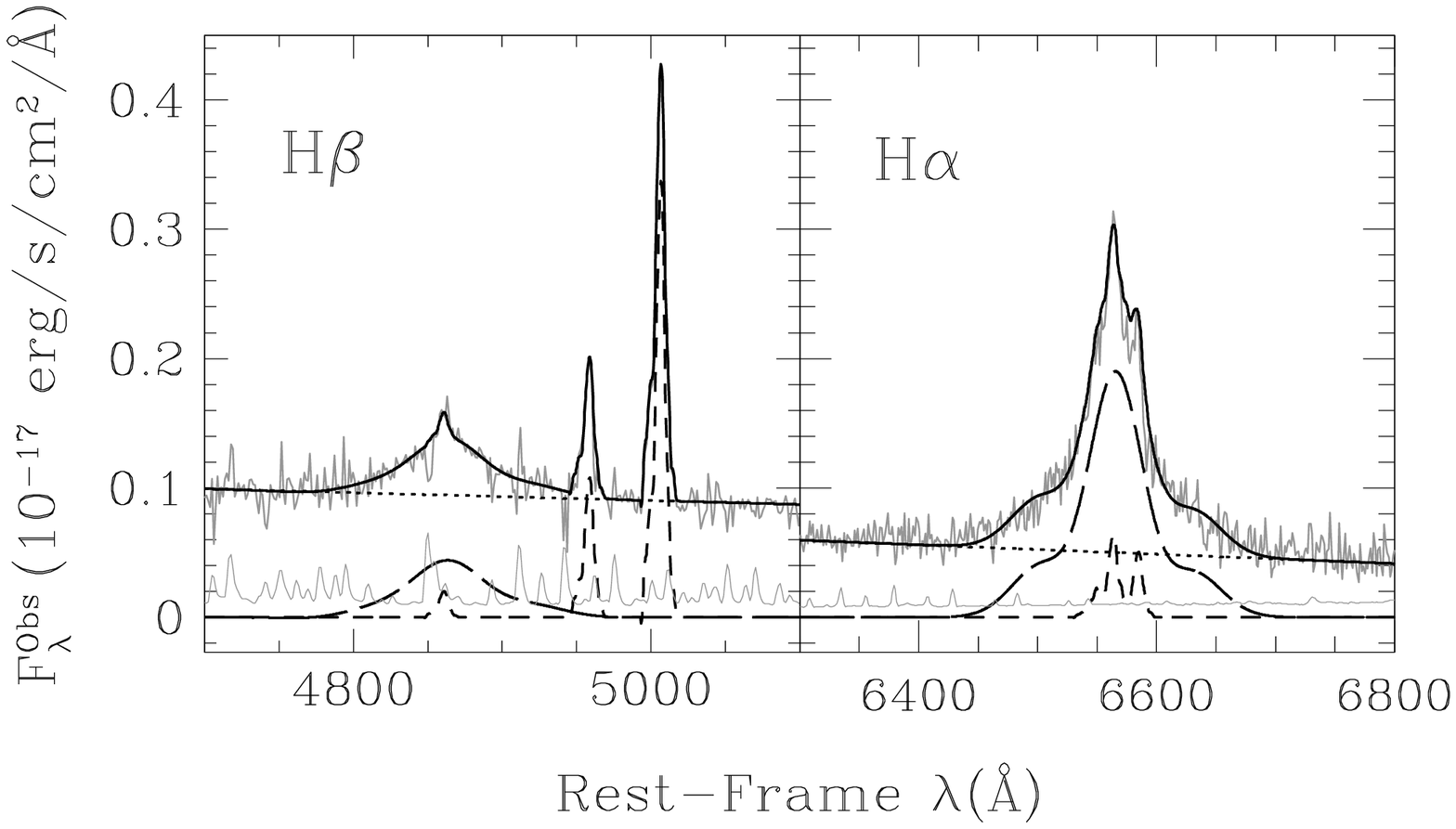}
    \caption{({\it{Top}}) LUCIFER {\it{H-}} and {\it{K-}}band spectra
    of SDSS1138+0314. The black solid line shows the spectrum obtained
    by performing the sky subtraction with the median combination of
    the sky frames while the gray line shows that obtained by using
    the modified version of the COSMOS software described in the
    text. ({\it{Bottom Left}}) Spectral region around
    H$\beta$. Overlaid on top are the best fit continuum ({\it{black
    dotted line}}) and narrow ({\it{black short-dashed line}}) and
    broad line components ({\it{black long-dashed line}}), as well as
    their sum ({\it{black solid line}}) and the error spectrum
    ({\it{thin gray solid line}}). ({\it{Bottom Right}}) Same as
    bottom left but for H$\alpha$.}
    \label{fg:spec}
    \epsscale{1}
  \end{center}
\end{figure}

\begin{figure}
  \begin{center}
    \plotone{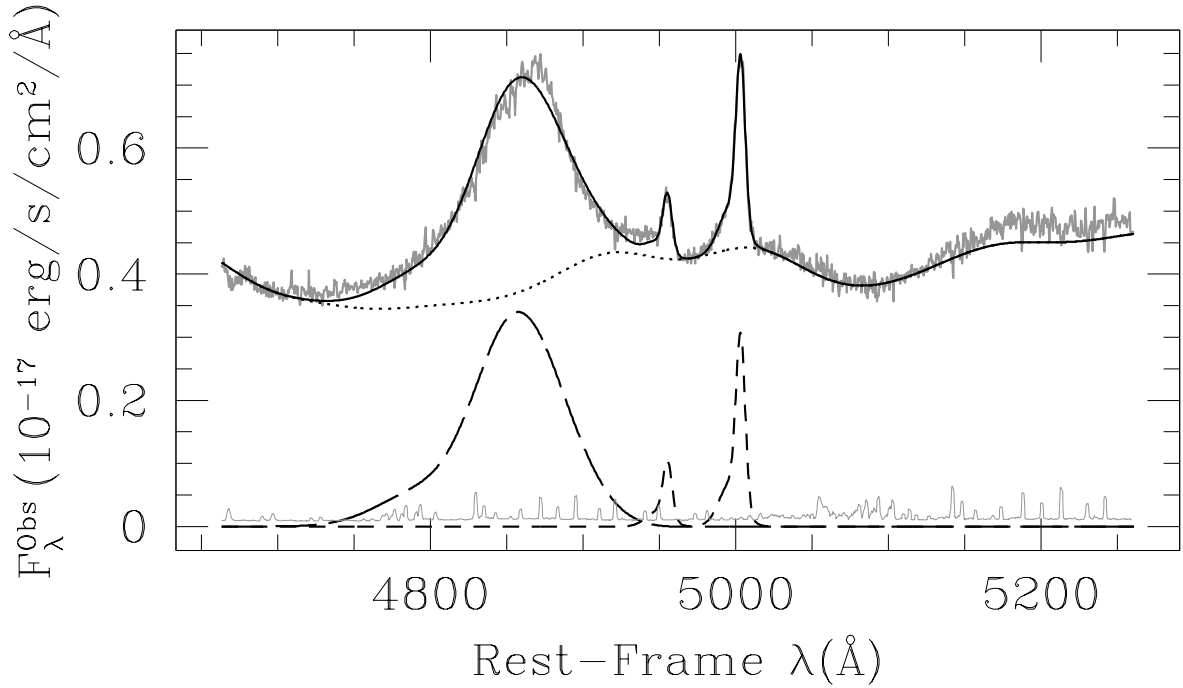}
    \caption{LUCIFER {\it{J-}}band spectrum of HS0810+2554 ({\it{gray
    solid line}}). Overlaid are the best fit continuum and FeII
    emission ({\it{black dotted line}}), narrow line emission
    ({\it{black short-dashed line}}), broad line component ({\it{black
    long-dashed line}}) and their sum ({\it{black solid line}}), as
    well as the error spectrum ({\it{thin gray solid line}}).}
    \label{fg:spec_hs0810}
  \end{center}
\end{figure}

\begin{figure}
  \begin{center}
    \epsscale{0.9}
    \plotone{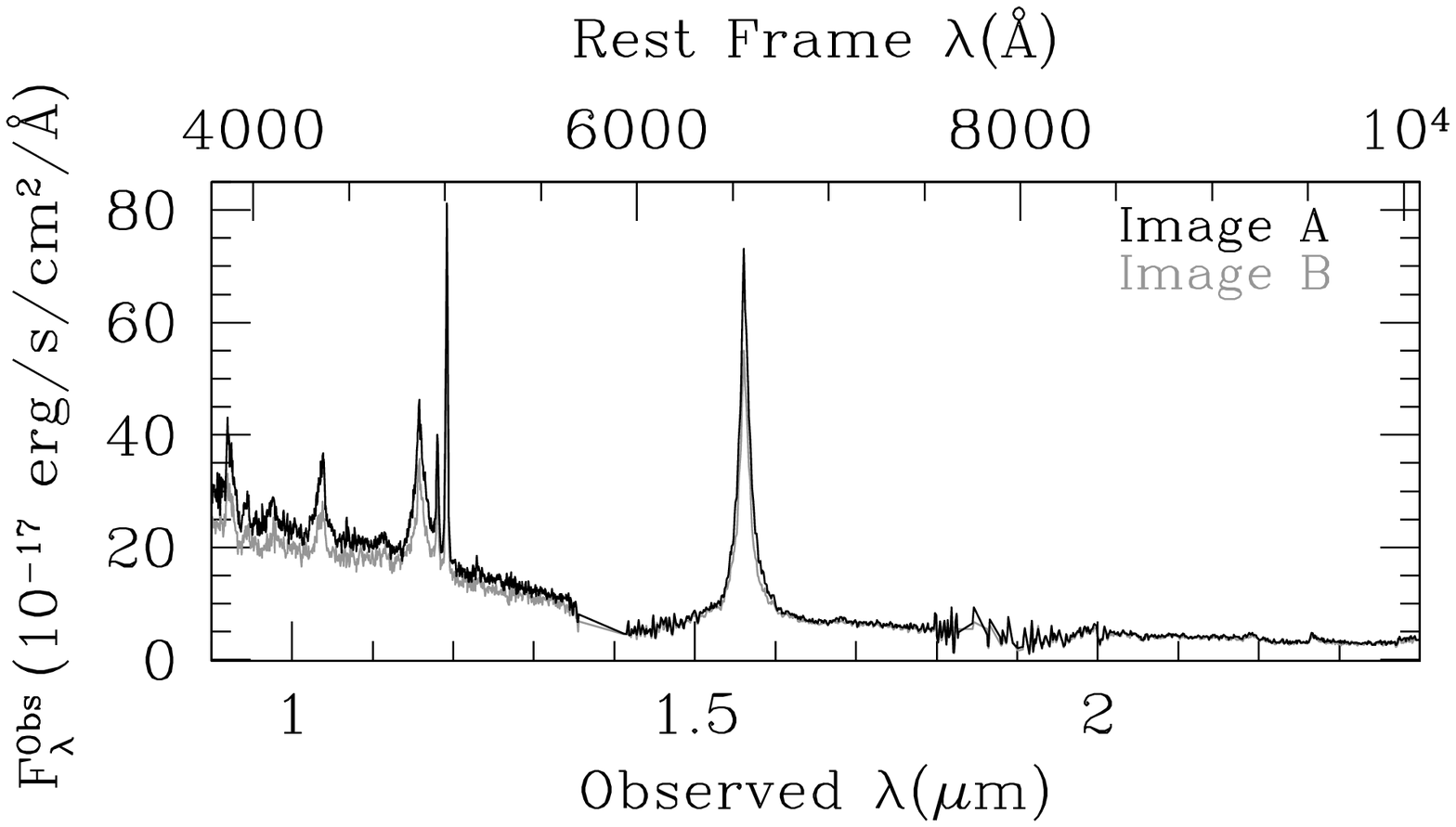}
    \epsscale{1.05}
    \plotone{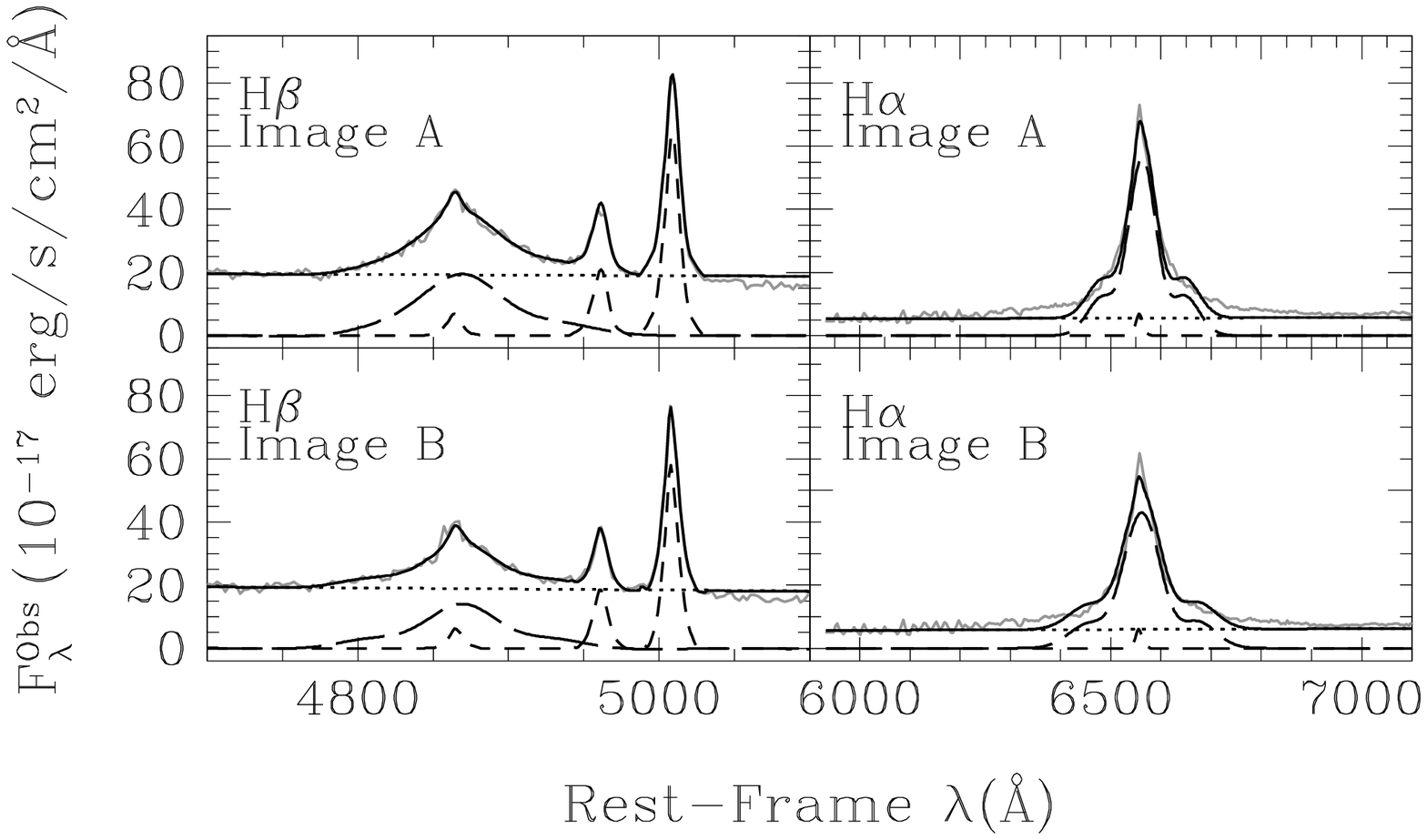}
    \epsscale{1}
    \caption{LIRIS near-IR spectra of images A and B of SBS0909+532
    obtained by \citet{mediavilla10}. The top panel shows the complete
    spectrum while the bottom four panels show the spectral regions
    around H$\alpha$ and H$\beta$ of each quasar image.  Overlaid on
    top are the best fit continuum and narrow- and broad-line
    components, as well as their overall sum, using the same
    line-styles as in Figure \ref{fg:spec}.}
    \label{fg:spec_sbs0909}
  \end{center}
\end{figure}

In this section we discuss some details of our line-width and
continuum measurements of individual objects. All UV/optical spectra,
as well as the continuum and line-profile fits, are shown in Figure
\ref{fg:all_uv_opt_spec}. LUCIFER spectra of SDSS1138+0314 and
HS0810+2554 are shown in Figures \ref{fg:spec} and
\ref{fg:spec_hs0810}, while the LIRIS spectra of SBS0909+532 are shown
in Figure \ref{fg:spec_sbs0909}.

\medskip
\noindent{\it{HS0810+2554 ---}} The \Civ\ profile of HS0810 shows a
small amount of absorption near the peak of the line.  We interpolate
over this region before making the line-width measurements and fitting
the GH polynomial. Our results are consistent with or without the
interpolation, as the absorption is weak and only seen near the very
peak of the line. To fit the continuum and emission-line features that
blended with the \Hbeta\ emission of HS0810+2554, a power-law
continuum and \feii\ broad-emission line template were fit to the
spectrum based on the continuum regions listed in Table
\ref{tab:nir_lineboundaries} and the rest-frame optical \feii\
template of \citet{Boroson92} from observations of I\,Zw1
\citep[see][for more details]{Wills85, Dietrich02, Dietrich05}. Narrow
\ob\ emission was then removed by creating a template from a
two-component Gaussian fit to the [\oiii]\,$\lambda 5007$ narrow line
and then scaling it to [\oiii]\,$\lambda 4959$ based on standard
emission line ratios.  We could not remove narrow \Hbeta\ emission
because such a component was not obvious in the
spectrum\footnote{\citet{greene10} are similarly unable to isolate a
narrow component in their observations of HS0810+2554}. After
subtracting these components, the remaining broad \Hbeta\ emission was
fit with a Gauss-Hermite polynomial, and the FWHM and line dispersion
were measured from this fit as described in
\S\ref{ssec:lw_method}. The deblended spectrum of HS0810+2554, showing
each component including the GH fit, is shown in Figure
\ref{fg:spec_hs0810}. Our \Hbeta\ FWHM measured from the LUCIFER
spectrum of \citet{mogren10} is consistent with that of
\citet{greene10}.

\medskip
\noindent{\it{SBS0909+532 --- }} We use the combined UV-optical-NIR
spectrum of \citet{mediavilla10} of images A and B of this object,
based on a combination of {\it{HST}} STIS and WHT INTEGRAL and LIRIS
observations. The UV section of the spectrum is shown in Figure
\ref{fg:all_uv_opt_spec} while the NIR section is shown in Figure
\ref{fg:spec_sbs0909}. The \civ\ profile of SBS0909+532 showed a small
absorption trough near observed frame 3600\AA\ and we interpolated
over this region before measuring widths.  The SBS0909+532 \civ\
profile shape is `peaky' with broader wings at the base, and our GH
fitting procedure was unable to satisfactorily fit this line profile,
so we estimate errors based on the original spectra instead of a GH
polynomial fit. For this object we only measure the UV continuum
luminosity on image A, as image B shows clear differential reddening
with respect to A. We note that the \citet{peng06} mass quoted by
\citet{greene10} is based on Mg{\sc II}, so using the \civ\ line-width
measurements given in Table \ref{tab:velocities} provides the first
estimate of a \civ-based black hole mass for this object. For the IR
spectra of \citet{mediavilla10}, shown in Figure
\ref{fg:spec_sbs0909}, we removed narrow-line components from the IR
spectra using the [\oiii]\,\lam5007 line as a template and scaling it
to the other narrow lines using standard emission-line ratios between
lines of the same atomic species and basing the strength of the Balmer
narrow lines on the ratio of [\oiii]\,\lam5007/H$\beta$ determined by
inspection. We are not as confident in our narrow-line subtraction for
this object as for the others because (1) we see residuals near the
peak of H$\beta$, and (2) the exact strength of H$\alpha$ is uncertain
because narrow-line emission remains present after subtraction. The
exact level of the residuals for H$\alpha$ is unclear, since
increasing the fraction of emission by as much as a factor of 2 does
not result in an obviously improved subtraction. In the case of
H$\beta$, the residuals are not larger than expected based on the
$S/N$ of the images, but for H$\alpha$ we report uncertainties
determined from difference between the widths determined with or
without the narrow-line subtraction. This results in an H$\alpha$ FWHM
uncertainty several times larger than would be estimated by our Monte
Carlo simulations. Comparable $\sigma_l$ uncertainties are measured
using both methods, because the line dispersion is far less dependent
on the presence of a narrow-line component \citep[see][]{denney09}. We
measure the H$\beta$ line-widths from the GH fits to the profile and
the H$\alpha$ line-widths directly from the data because the GH
polynomials did not accurately fit the line profile. Image B may have
a residual sky line peak just blueward of the H$\beta$ narrow-line
component.  The presence of this emission has little effect on our
fits, however, since we measure consistent line widths if we
interpolate under this emission to remove it. Our H$\alpha$ and
H$\beta$ widths are consistent with those reported by
\citet{greene10}.

\medskip
\noindent{\it{Q0957+561 --- }} We use the {\it{HST}} STIS UV spectrum
of both images obtained by \citet{goicoechea05}. The rather strange
\Civ\ line profiles in this object may indicate that there is
absorption and/or that the profile shapes in individual images are
affected by microlensing from the lens galaxy.  However, since there
was no definite source of uncertainty to correct for, we simply
measured the observed line widths from each spectrum.

\medskip
\noindent{\it{HE1104--1805 --- }} We use the EFOSC1 ESO 3.6m telescope
UV/optical spectrum of \citet{wisotzki95}. The \Civ\ profile shows a
small amount of absorption near the peak of the line, similar to that
of the HS0810+2554 \Civ\ profile.  We therefore apply the same
treatment to this line as to the HS0810+2554 profile, and find
similarly consistent results with or without interpolation.

\medskip
\noindent{\it{PG1115+080 --- }} Due to the severe absorption, both
narrow and broad, in the \Civ\ line profile in this object, we could
not measure the \Civ\ line-width directly from the data.  However, by
masking out the absorption regions, we made a reasonable GH fit to
this line profile, from which we measured the line widths given in
Table \ref{tab:velocities}.

\medskip
\noindent{\it{SDSS1138+0314 ---}} To measure the width of \civ\ and
the UV continuum luminosity we use the FORS1 VLT spectra of images B
and C obtained by \citet{eigenbrod06}. The \Civ\ line profile not only
shows absorption in the blue side of the line, but is also
particularly narrow and `peaky' with a broad base, similar to
SBS0909+523. We were unable to reasonably approximate the profile
shape with a sixth-order Gauss-Hermite polynomial. However, since the
$S/N$ of this spectrum was very high (see Table
\ref{tab:uv_opt_lineboundaries}), we interpolated over the absorption
with a 2nd order polynomial, measured the line width directly from the
interpolated data, and used this interpolated spectrum and the error
spectrum formed with equation (\ref{eqn:widtherror}) to derive
uncertainties in the \Civ\ width measurement. Because of the combined
effects of absorption and the narrow line profile (i.e., where the
absorption could be masking the true width), we treat our \civ\ widths
as lower limits.  At rest-frame optical wavelengths, the difficulty in
removing the blended narrow-line components of \Halpha\ and \niidb\
combined with our attempt to accurately fit the emission-line peak
(often underestimated with line profile fits) led to an overestimate
of the flux between the \Halpha\ and \nii\,\lam6583 narrow lines. This
overestimate does not significantly affect our width
measurements. This object was not part of the \citet{greene10} sample.

\medskip
\noindent{\it{H1413+117 --- }} This object is a BAL QSO and therefore
a large portion of the \Civ\ line profile is completely absorbed on
the blue side.  Hence, we adopt the \Civ\ width measured from only the
red side of the line, and we consider this to be a lower limit on the
width.

\medskip
\noindent{\it{B1422+231 --- }} We use the LRIS Keck II UV/optical
spectrum of \citet{tonry98}. We interpolated over the two small
absorption troughs near $\sim$6875\AA\ and $\sim$7020\AA\ before
measuring the \Civ\ widths directly from the data. Our treatment of
these regions did not affect the resulting GH fit to the data.  From
the \citet{greene10} data, we cannot assess the reliability of their
fit to the H$\beta$ profile, because they plotted the \Hbeta\ spectrum
of HE1104--1805 in place of the spectrum of B1422+231. In order to be
conservative, we therefore flag B1422+231 as one of the objects with
possible problems in the sample of \citet{greene10}.

\medskip
\noindent{\it{FBQ1633+3134 --- }} There is evidence for absorption in
the blue side of \civ, however, it is not clear that a reliable
interpolation could be made across this possible absorption.  We
measure the line width as is, and treat this measurement as a lower
limit.

\medskip
\noindent{\it{Q2237+030 --- }} We use the FORS1 VLT UV/optical spectra
of images C and D obtained by \citet{eigenbrod08}. We follow the same
prescription as for SDSS1138+0314 and Q0957+561 and use an average of
the \civ\ line-widths of each image to estimate $M_{\rm BH}$. This
object shows \Civ\ absorption in the red side of the line.  We
interpolate over this absorption with a third order polynomial before
measuring the line width and fitting the GH polynomial to the data.
The interpolation creates a peak slightly higher than that observed in
the original spectrum, but makes for a much more symmetric line
profile, which is more typical of the core of \Civ\ line profiles,
than a linear or quadratic interpolation. This increase in the assumed
line peak decreases our line-width measurements, but not
significantly.

\begin{deluxetable}{l c c c c c c c c c c c c c}

\tablecaption{Lens Magnifications and Continuum Luminosities\label{tab:magnifications}}

\tablehead{
  \multicolumn{1}{l}{Object}& 
  \colhead{Ref.}& 
  \colhead{$z$}& 
  \colhead{}& 
  \multicolumn{4}{c}{Image Magnification} & 
  \colhead{}& 
  \colhead{$m_H$\tablenotemark{\dagger}}& 
  \colhead{}& 
  \multicolumn{3}{c}{$\log \lambda L_{\lambda}\ /\ \rm erg\ \rm s^{-1}$}\\
  \cline{5-8}
  \cline{12-14}
  \colhead{}& 
  \colhead{}& 
  \colhead{}& 
  \colhead{}& 
  \colhead{A}& 
  \colhead{B}& 
  \colhead{C}& 
  \colhead{D}&
  \colhead{}& 
  \colhead{(mag)}& 
  \colhead{}& 
  \colhead{1350\AA}& 
  \colhead{1450\AA}& 
  \colhead{5100\AA} 
}

\tablewidth{0pt}
\tablecolumns{14}

\startdata
Q0142--100      &    $a$  & 2.72  & &\phn3.3  &\phn0.4  & \nodata & \nodata & & 16.56 & & 46.83   & 46.76 & 46.27 \\
SDSS0246--0825  &    $a$  & 1.69  & &   26.9  &\phn8.9  & \nodata & \nodata & & 20.39 & & \nodata & 44.53 & 44.59 \\
HS0810+2554     & \nodata & 1.51  & &   47.2  &   51.1  &   13.5  &    7.7  & & 18.72 & & \nodata & 44.44 & 44.84 \\
SBS0909+532     &    $b$  & 1.38  & &\phn1.7  &\phn1.5  & \nodata & \nodata & & 15.18 & & 46.08   & 46.05 & 46.31 \\
Q0957+561       &    $c$  & 1.41  & &\phn3.1  &\phn1.7  & \nodata & \nodata & & 16.51 & & 46.31   & 46.25 & 45.79 \\
HE1104--1805    &    $d$  & 2.32  & &   16.2  &\phn2.3  & \nodata & \nodata & & 18.52 & & 46.15   & 46.09 & 45.38 \\
PG1115+080      &    $a$  & 1.72  & &   19.6  &   18.7  &\phn3.2  &    4.9  & & 19.13 & & \nodata & 45.47 & 44.93 \\
SDSS1138+0314   &    $e$  & 2.44  & &\phn7.3  &\phn3.7  &\phn5.2  &    6.9  & & 20.65 & & 44.83   & 44.77 & 44.81 \\
H1413+117       &    $a$  & 2.55  & &\phn8.2  &\phn6.8  &\phn6.8  &    3.4  & & 18.05 & & 45.73   & 45.78 & 45.63 \\
B1422+231       &    $f$  & 3.62  & &\phn6.6  &\phn8.2  &\phn4.3  &    0.3  & & 16.55 & & 46.83   & 46.74 & 46.42 \\
FBQ1633+3134    & \nodata & 1.52  & &\phn2.7  &\phn0.7  & \nodata & \nodata & & 16.85 & & 45.65   & 45.64 & 45.72 \\
Q2237+030       &    $g$  & 1.69  & &\phn4.9  &\phn4.3  &\phn2.2  &    4.1  & & 16.83 & & \nodata & 45.53 & 45.98 \\
\enddata

\tablecomments{\,Literature UV/optical spectra obtained from the
following references: $a)$ SDSS DR7 \citep{abazajian09}, $b)$
\citet{mediavilla10}, $c)$ \citet{goicoechea05}, $d)$
\citet{wisotzki95}, $e)$ \citet{eigenbrod06}, $f)$ \citet{tonry98},
$g)$ \citet{eigenbrod08}.}

\tablenotetext{\dagger}{Unmagnified quasar $H$-band magnitudes.}

\end{deluxetable}

\begin{deluxetable}{lccccccccc}
\rotate
\tablecolumns{10}
\tablewidth{0pt}
\tablecaption{\civ\ Emission Line and Continuum Region Boundaries\label{tab:uv_opt_lineboundaries}}
\tablehead{
\colhead{}&
\colhead{}&
\multicolumn{3}{c}{Prescription A}&
\colhead{}&
\multicolumn{3}{c}{Prescription B}&
\colhead{}
\\
\cline{3-5}
\cline{7-9}
\colhead{}&
\colhead{}&
\colhead{Blue Cont.}&
\colhead{Red Cont.}&
\colhead{Broad Line}&
\colhead{}&
\colhead{Blue Cont.}&
\colhead{Red Cont.}&
\colhead{Broad Line}&
\colhead{Res}
\\
\colhead{Object}&
\colhead{$S/N$\tablenotemark{a}}&
\colhead{(\AA)}&
\colhead{(\AA)}&
\colhead{(\AA)}&
\colhead{}&
\colhead{(\AA)}&
\colhead{(\AA)}&
\colhead{(\AA)}&
\colhead{(\AA)}
}

\startdata
Q0142--100        &  45 &  5401--5468 &  6250--6343 &  5570--5954 & &  5440--5500   &5900--5960   & 5500--5900 & 2.6\\
SDSS0246--0825    &  17 &  3980--4030 &  4519--4586 &  4028--4306 & &  3980--4030   &4250--4300   & 4050--4250 & 2.6\\
HS0810+2554       &  19 &  3610--3690 &  4217--4280 &  3758--4018 & &  3620--3685   &3985--4005   & 3715--3975 & 15.2\\
SBS0909+532--A    &  71 &  3420--3470 &  4070--4120 &  3564--3810 & &  3420--3470   &3820--3850   & 3560--3820 & 3.5\\
SBS0909+532--B    &  16 &  3420--3470 &  4070--4120 &  3564--3810 & &  3420--3470   &3820--3850   & 3560--3820 & 3.5\\
Q0957+561--A      &  66 &  3434--3543 &  4049--4109 &  3609--3858 & &  3550--3600   &3810--3830   & 3655--3805 & 3.5\\
Q0957+561--B      &  28 &  3434--3543 &  4049--4109 &  3609--3858 & &  3550--3600   &3810--3830   & 3655--3805 & 3.5\\
HE1104--1805      & 221 &  4731--4880 &  5578--5661 &  4971--5314 & &  4870--4910   &5300--5340   & 4970--5290 & \nodata\\
PG1115+080        &  86 &  3897--4020 &  4595--4663 &  4095--4378 & &  4000--4031   &4415--4455   & 4085--4415 & 2.2\\
SDSS1138+0314--B  &  78 &  4904--5059 &  5782--5868 &  5153--5509 & &  4974--5020   &5520--5540   & 5020--5520 & \nodata\\
SDSS1138+0314--C  &  47 &  4904--5059 &  5782--5868 &  5153--5509 & &  4974--5020   &5520--5540   & 5020--5520 & \nodata\\
H1413+117         &  43 &  5073--5233 &  5981--6070 &  5331--5698 & &  5140--5190   &5630--5680   & 5210--5615 & 2.3\\
B1422+231         & 270 &  6598--6806 &  7778--7894 &  6933--7411 & &  6690--6740   &7400--7450   & 6830--7400 & \nodata\\
FBQ1633+3134      &  52 &  3350--3412 &  4234--4297 &  3773--4034 & &  3350--3412   &4020--4060   & 3795--3980 & 13.0\\
Q2237+030--C      &  36 &  3840--3962 &  4528--4595 &  4035--4314 & &  3870--3910   &4275--4315   & 4040--4265 & \nodata\\
Q2237+030--D      &  54 &  3840--3962 &  4528--4595 &  4035--4314 & &  3870--3910   &4275--4315   & 4040--4265 & \nodata\\
\enddata

\tablenotetext{a}{The $S/N$ quoted is the $S/N$ per pixel averaged over
all continuum regions listed for each object.}  

\end{deluxetable}

\begin{deluxetable}{l c c c c c c c c}

\rotate

\tablecaption{Velocity Widths\label{tab:velocities}}

\tablehead{
  \colhead{}&
  \multicolumn{8}{c}{Broad Emission Line Velocity Widths / $10^3~\rm km~\rm s^{-1}$}\\
  \colhead{Object}&
  $\rm FWHM_{\civm}$--p.A & 
  $\rm FWHM_{\civm}$--p.B & 
  $\sigma_{l,\civm}$--p.A &
  $\sigma_{l,\civm}$--p.B &
  $\rm FWHM_{\rm H\beta}$\tablenotemark{\dagger} & $\sigma_{l,\rm H\beta}$ &
  $\rm FWHM_{\rm H\alpha}$\tablenotemark{\dagger} & $\sigma_{l,\rm H\alpha}$}

\tabletypesize{\scriptsize}
\tablewidth{0pt}
\tablecolumns{9}

\startdata
Q0142--100      & \phs5.20 $\pm$ 0.18 & \phs4.75 $\pm$ 0.22 & \phs3.67 $\pm$ 0.04 & \phs3.67  $\pm$ 0.07  &  2.70 $\pm$ 0.60\tablenotemark{*}  &      \nodata      &  3.80 $\pm$ 0.30\tablenotemark{*}  &     \nodata    \\
SDSS0246--0825  & \phs4.43 $\pm$ 0.24 & \phs4.40 $\pm$ 0.22 & \phs3.09 $\pm$ 0.06 & \phs2.51  $\pm$ 0.04  &  2.50 $\pm$ 0.60                   &      \nodata      &  2.50 $\pm$ 0.20                   &     \nodata    \\
HS0810+2554     & \phs3.68 $\pm$ 0.14 & \phs3.53 $\pm$ 0.16 & \phs3.40 $\pm$ 0.05 & \phs2.85  $\pm$ 0.08  &  4.40 $\pm$ 0.06                   &  2.21 $\pm$ 0.04  &  3.80 $\pm$ 0.00                   &     \nodata    \\
SBS0909+532     & \phs2.38 $\pm$ 0.08 & \phs2.36 $\pm$ 0.08 & \phs2.90 $\pm$ 0.04 & \phs2.82  $\pm$ 0.05  &  3.95 $\pm$ 0.17                   &  2.20 $\pm$ 0.05  &  3.06 $\pm$ 0.34                   & 5.24 $\pm$ 0.04\\
Q0957+561       & \phs3.68 $\pm$ 0.25 & \phs3.47 $\pm$ 0.08 & \phs3.27 $\pm$ 0.09 & \phs2.272 $\pm$ 0.007 &  3.30 $\pm$ 0.90                   &      \nodata      &  3.00 $\pm$ 0.20                   &     \nodata    \\
HE1104--1805    & \phs6.08 $\pm$ 0.35 & \phs5.75 $\pm$ 0.05 & \phs3.84 $\pm$ 0.10 & \phs2.897 $\pm$ 0.004 &  3.80 $\pm$ 0.90                   &      \nodata      &  4.70 $\pm$ 0.20\tablenotemark{*}  &     \nodata    \\
PG1115+080      & \phs4.98 $\pm$ 0.18 & \phs4.67 $\pm$ 0.13 & \phs3.68 $\pm$ 0.04 & \phs3.40  $\pm$ 0.04  &  4.40 $\pm$ 0.20                   &      \nodata      &  4.00 $\pm$ 0.10                   &     \nodata    \\
SDSS1138+0314   &  $>$2.02 $\pm$ 0.15 &  $>$1.99 $\pm$ 0.18 &  $>$3.12 $\pm$ 0.04 &  $>$2.40  $\pm$ 0.11\ &  3.93 $\pm$ 0.30                   &  2.08 $\pm$ 0.18  &  2.57 $\pm$ 0.04                   & 1.90 $\pm$ 0.05\\
H1413+117       &  $>$2.62 $\pm$ 0.95 &  $>$2.54 $\pm$ 0.37 &  $>$3.78 $\pm$ 0.15 &  $>$1.82  $\pm$ 0.07  &  6.70 $\pm$ 1.90                   &      \nodata      &  5.30 $\pm$ 0.80                   &     \nodata    \\
B1422+231       & \phs5.81 $\pm$ 0.16 & \phs5.56 $\pm$ 0.02 & \phs3.70 $\pm$ 0.03 & \phs3.321 $\pm$ 0.006 &  6.10 $\pm$ 2.20\tablenotemark{*}  &      \nodata      &      \nodata                       &     \nodata    \\
FBQ1633+3134    &  $>$4.71 $\pm$ 0.18 &  $>$4.40 $\pm$ 0.16 &  $>$3.83 $\pm$ 0.06 &  $>$2.20  $\pm$ 0.06  &  4.60 $\pm$ 0.90\tablenotemark{*}  &      \nodata      &  4.10 $\pm$ 0.70                   &     \nodata    \\
Q2237+030       & \phs3.96 $\pm$ 0.18 & \phs3.78 $\pm$ 0.12 & \phs3.49 $\pm$ 0.07 & \phs2.51  $\pm$ 0.05  &  3.80 $\pm$ 1.40                   &      \nodata      &  4.80 $\pm$ 0.60\tablenotemark{*}  &     \nodata    \\
\enddata

\tablenotetext{\dagger}{All H$\alpha$ and H$\beta$ line width
measurements correspond to those in Table 1 of \citet{greene10},
except for SDSS1138+0314, SBS0909+523 and the H$\beta$ widths of
HS0810+2554.}

\tablenotetext{*}{Group II line-widths. See \S\ref{ssec:lw_method} for
details.}

\end{deluxetable}

\clearpage
\begin{deluxetable}{lcccccc}

\tablecolumns{7}
\tablewidth{0pt}
\tablecaption{NIR Emission Line and Continuum Region Boundaries\label{tab:nir_lineboundaries}}
\tablehead{
\colhead{}&
\colhead{Emission}&
\colhead{}&
\colhead{Blue Cont.}&
\colhead{Red Cont.}&
\colhead{Broad Line}&
\colhead{Res}\\
\colhead{Object}&
\colhead{Line}&
\colhead{$S/N$\tablenotemark{a}}&
\colhead{(\AA)}&
\colhead{(\AA)}&
\colhead{(\AA)}&
\colhead{(\AA)}
}

\startdata
HS0810+2554\tablenotemark{a}& \Hbeta\  &  36 & 4680--4710   &5080--5120   & 4710--4960 & 8.0\\
SBS0909+532--A              & \Hbeta\  &  68 & 11205--11310 &11840--11855 &11380--11840& \nodata\\
SBS0909+532--B              & \Hbeta\  &  22 & 11205--11310 &11840--11855 &11380--11840& \nodata\\
SBS0909+532--A              & \Halpha\ &  25 & 14210--14440 &17500--17700 &14865--16650& \nodata\\
SBS0909+532--B              & \Halpha\ &  18 & 14210--14440 &17500--17700 &14865--16650& \nodata\\
SDSS1138+0314               & \Hbeta\  &  12 & 16150--16300 &17485--17623 &16300--17100& 8.0\\
SDSS1138+0314               & \Halpha\ &   8 & 21780--21850 &23270--23310 &22135--23040& 8.0\\
\enddata

\tablenotetext{a}{Rest frame wavelengths are used here because line
boundaries were chosen after the deblending procedure that transfers the
spectrum into the rest frame.}

\end{deluxetable}

\begin{deluxetable}{l  c  c  c  c  c  c  c}

\rotate

\tablecaption{H$\beta$, H$\alpha$ and CIV BH Mass Estimates\label{tab:all_objs}}

\tablehead{ 
  \colhead{}& 
  \multicolumn{7}{c}{$\log M_{\rm BH}/M_{\odot}$}\\ 
  \cline{2-8}
  \colhead{Object}& 
  $\rm FWHM_{\civm}$--p.A & 
  $\rm FWHM_{\civm}$--p.B & 
  $\sigma_{l,\civm}$--p.A &
  $\sigma_{l,\civm}$--p.B &
  $\rm FWHM_{\rm H\beta}$ & 
  $\sigma_{l,\rm H\beta}$ &
  $\rm FWHM_{\rm H\alpha}$
}

\tabletypesize{\small}
\tablewidth{0pt}
\tablecolumns{8}

\startdata
Q0142--100     &  9.59 $\pm$  0.32                      &  9.51 $\pm$  0.33                      & 9.36 $\pm$  0.29                      &  9.36 $\pm$  0.29                      &  8.94 $\pm$  0.30\tablenotemark{*} &      \nodata      &  9.33 $\pm$  0.23\tablenotemark{*} \\
SDSS0246--0825 &  8.24 $\pm$  0.33                      &  8.23 $\pm$  0.33                      & 7.99 $\pm$  0.29                      &  7.81 $\pm$  0.29                      &  8.00 $\pm$  0.31                  &      \nodata      &  8.08 $\pm$  0.23                  \\
HS0810+2554    &  8.03 $\pm$  0.32                      &  7.99 $\pm$  0.33                      & 8.02 $\pm$  0.29                      &  7.87 $\pm$  0.29                      &  8.62 $\pm$  0.22                  &  8.54 $\pm$ 0.17  &  8.65 $\pm$  0.22                  \\
SBS0909+532    &  8.51 $\pm$  0.32                      &  8.51 $\pm$  0.32                      & 8.76 $\pm$  0.29                      &  8.73 $\pm$  0.29                      &  9.29 $\pm$  0.23                  &  9.29 $\pm$ 0.17  &  9.15 $\pm$  0.24                  \\
Q0957+561      &  9.02 $\pm$  0.33                      &  8.97 $\pm$  0.32                      & 8.98 $\pm$  0.29                      &  8.67 $\pm$  0.29                      &  8.86 $\pm$  0.33                  &      \nodata      &  8.87 $\pm$  0.23                  \\
HE1104--1805   &  9.37 $\pm$  0.33                      &  9.32 $\pm$  0.32                      & 9.04 $\pm$  0.29                      &  8.79 $\pm$  0.29                      &  8.77 $\pm$  0.30                  &      \nodata      &  9.05 $\pm$  0.23\tablenotemark{*} \\
PG1115+080     &  8.83 $\pm$  0.32                      &  8.78 $\pm$  0.32                      & 8.64 $\pm$  0.29                      &  8.57 $\pm$  0.29                      &  8.66 $\pm$  0.23                  &      \nodata      &  8.68 $\pm$  0.22                  \\
SDSS1138+0314  &  7.71 $\pm$  0.33\tablenotemark{\ddag} &  7.69 $\pm$  0.33\tablenotemark{\ddag} & 8.15 $\pm$  0.29\tablenotemark{\ddag} &  7.93 $\pm$  0.29\tablenotemark{\ddag} &  8.50 $\pm$  0.23                  &  8.47 $\pm$ 0.19  &  8.22 $\pm$  0.22                  \\
H1413+117      &  8.41 $\pm$  0.45\tablenotemark{\ddag} &  8.39 $\pm$  0.35\tablenotemark{\ddag} & 8.80 $\pm$  0.29\tablenotemark{\ddag} &  8.17 $\pm$  0.29\tablenotemark{\ddag} &  9.39 $\pm$  0.33                  &      \nodata      &  9.29 $\pm$  0.26                  \\
B1422+231      &  9.69 $\pm$  0.32                      &  9.65 $\pm$  0.32                      & 9.37 $\pm$  0.29                      &  9.27 $\pm$  0.29                      &  9.72 $\pm$  0.38\tablenotemark{*} &      \nodata      &      \nodata                       \\
FBQ1633+3134   &  8.88 $\pm$  0.32\tablenotemark{\ddag} &  8.82 $\pm$  0.32\tablenotemark{\ddag} & 8.77 $\pm$  0.29\tablenotemark{\ddag} &  8.29 $\pm$  0.29\tablenotemark{\ddag} &  9.11 $\pm$  0.28\tablenotemark{*} &      \nodata      &  9.11 $\pm$  0.27                  \\
Q2237+030      &  8.67 $\pm$  0.33                      &  8.63 $\pm$  0.32                      & 8.63 $\pm$  0.29                      &  8.34 $\pm$  0.29                      &  9.08 $\pm$  0.39                  &      \nodata      &  9.38 $\pm$  0.25\tablenotemark{*} \\
\enddata

\tablecomments{All BH masses correspond to those obtained from
eqns. (\ref{eq:m_hbeta}), (\ref{eq:m_halpha}) and
(\ref{eq:mass_civ}). None of the corrections discussed in
\S\S\ref{ssec:civ_sigma_fwhm} and \ref{ssec:civ_balmer_comp} have been
applied.}

\tablenotetext{*}{Based on group II line-width. See
\S\ref{ssec:lw_method} for details.}

\tablenotetext{\ddag}{Should be considered as lower bound. See
Appendix \ref{sssec:notes} for details.}

\end{deluxetable}

\begin{deluxetable}{l l l  c c c  c  c c c}

\tablecaption{Correlations of BH Mass Residuals\label{tab:residuals}}

\tablehead{Variable & \civ\  & Balmer & \multicolumn{3}{c}{Group I Estimates} & & \multicolumn{3}{c}{Group I \& II Estimates}\\
                    & Width  & Line   & $r_s$ & $N$ & $P_{\rm ran}$           & &  $r_s$ & $N$ & $P_{\rm ran}$}

\tabletypesize{\scriptsize}
\tablewidth{0pt}
\tablecolumns{10}

\startdata
$M_{\rm BH}$ Balmer lines          & $\sigma_l$     & H$\beta$        &  --0.214 &    7 &    0.644 & &  --0.183 &    9 &    0.637\\
                                   &                & H$\alpha$       &\phn0.100 &    5 &    0.873 & &  --0.048 &    8 &    0.911\\
                                   & FWHM           & H$\beta$        &  --0.500 &    7 &    0.253 & &  --0.333 &    9 &    0.381\\
                                   &                & H$\alpha$       &  --0.400 &    5 &    0.505 & &  --0.143 &    8 &    0.736\\
\\
$\lambda L_{\lambda}(1350\rm\AA)$  & $\sigma_l$     & H$\beta$        &\phn0.357 &    7 &    0.432 & &\phn0.467 &    9 &    0.205\\
                                   &                & H$\alpha$       &\phn0.500 &    5 &    0.391 & &\phn0.571 &    8 &    0.139\\
                                   & FWHM           & H$\beta$        &\phn0.143 &    7 &    0.760 & &\phn0.367 &    9 &    0.332\\
                                   &                & H$\alpha$       &\phn0.100 &    5 &    0.873 & &\phn0.429 &    8 &    0.289\\
\\
$\lambda L_{\lambda}(5100\rm\AA)$  & $\sigma_l$     & H$\beta$        &  --0.214 &    7 &    0.644 & &  --0.067 &    9 &    0.865\\
                                   &                & H$\alpha$       &  --0.300 &    5 &    0.624 & &  --0.048 &    8 &    0.911\\
                                   & FWHM           & H$\beta$        &  --0.500 &    7 &    0.253 & &  --0.233 &    9 &    0.546\\
                                   &                & H$\alpha$       &  --0.700 &    5 &    0.188 & &  --0.214 &    8 &    0.610\\
\\
Redshift                           & $\sigma_l$     & H$\beta$        &\phn0.607 &    7 &    0.148 & &\phn0.517 &    9 &    0.154\\
                                   &                & H$\alpha$       &\phn0.500 &    5 &    0.391 & &\phn0.500 &    8 &    0.207\\
                                   & FWHM           & H$\beta$        &\phn0.750 &    7 &    0.052 & &\phn0.583 &    9 &    0.099\\
                                   &                & H$\alpha$       &\phn0.600 &    5 &    0.285 & &\phn0.619 &    8 &    0.102\\
\\
$L/L_{\rm Edd}$                    & $\sigma_l$     & H$\beta$        &  --0.036 &    7 &    0.939 & &\phn0.200 &    9 &    0.606\\
                                   &                & H$\alpha$       &  --0.300 &    5 &    0.624 & &\phn0.024 &    8 &    0.955\\
                                   & FWHM           & H$\beta$        &  --0.286 &    7 &    0.534 & &\phn0.067 &    9 &    0.865\\
                                   &                & H$\alpha$       &  --0.200 &    5 &    0.747 & &  --0.071 &    8 &    0.867\\
\\
\civ\ Blueshift                    & $\sigma_l$     & H$\beta$        &\phn0.536 &    7 &    0.215 & &\phn0.033 &    9 &    0.932\\
                                   &                & H$\alpha$       &\phn0.100 &    5 &    0.873 & &  --0.214 &    8 &    0.610\\
                                   & FWHM           & H$\beta$        &\phn0.679 &    7 &    0.094 & &\phn0.133 &    9 &    0.732\\
                                   &                & H$\alpha$       &\phn0.600 &    5 &    0.285 & &\phn0.143 &    8 &    0.736\\
\\
\civ\ assymetry                    & $\sigma_l$     & H$\beta$        &\phn0.429 &    7 &    0.337 & &\phn0.117 &    9 &    0.765\\
                                   &                & H$\alpha$       &\phn0.300 &    5 &    0.624 & &  --0.048 &    8 &    0.911\\
                                   & FWHM           & H$\beta$        &  --0.393 &    7 &    0.383 & &  --0.333 &    9 &    0.381\\
                                   &                & H$\alpha$       &\phn0.300 &    5 &    0.624 & &  --0.476 &    8 &    0.233\\
\\
$\lambda L_{\lambda}(1350\rm\AA)/\lambda L_{\lambda}(5100\rm\AA)$
                                   & $\sigma_l$     & H$\beta$        &\phn0.929 &    7 &    0.003 & &\phn0.883 &    9 &    0.002\\
                                   &                & H$\alpha$       &\phn1.000 &    5 &    0.000 & &\phn0.809 &    8 &    0.015\\
                                   & FWHM           & H$\beta$        &\phn0.750 &    7 &    0.052 & &\phn0.767 &    9 &    0.016\\
                                   &                & H$\alpha$       &\phn0.700 &    5 &    0.188 & &\phn0.857 &    8 &    0.007\\
\enddata

\tablecomments{The table shows the correlation strength of the BH mass
residuals as a function of each different variable, quantified by the
Spearman rank order coefficient, $r_s$. Results are shown for group I
and the combination of groups I and II measurements. In each case, $N$
indicates the number of QSOs used to estimate the correlation strength
and $P_{\rm ran}$ indicates the probability of observing such a
correlation by chance if the variables are uncorrelated.}

\end{deluxetable}

\begin{deluxetable}{l l  c c  c  c c}

\tablecaption{Linear Fits to Correlations of BH Mass Residuals with AGN color\label{tab:color_residuals}}

\tablehead{\civ\  & Balmer & \multicolumn{2}{c}{Group I Estimates}   & & \multicolumn{2}{c}{Group I \& II Estimates}\\
           Width  & Line   & $a$ & $b$                               & & $a$ & $b$ }

\tabletypesize{\small}

\tablewidth{0pt}
\tablecolumns{7}

\startdata
Prescription A\\
$\sigma_l$  &  H$\beta  $  &  0.64 $\pm$ 0.13 & --0.13 $\pm$ 0.06 & & 0.68 $\pm$ 0.16 & --0.13 $\pm$ 0.08\\
            &  H$\alpha $  &  0.58 $\pm$ 0.15 & --0.11 $\pm$ 0.06 & & 0.57 $\pm$ 0.11 & --0.17 $\pm$ 0.06\\
 FWHM       &  H$\beta  $  &  0.89 $\pm$ 0.25 & --0.20 $\pm$ 0.12 & & 0.95 $\pm$ 0.22 & --0.19 $\pm$ 0.11\\
            &  H$\alpha $  &  0.75 $\pm$ 0.30 & --0.20 $\pm$ 0.13 & & 0.79 $\pm$ 0.15 & --0.23 $\pm$ 0.08\\
\\
Prescription B\\
$\sigma_l$  &  H$\beta  $  &  0.60 $\pm$ 0.11 & --0.18 $\pm$ 0.05 & & 0.68 $\pm$ 0.17 & --0.16 $\pm$ 0.08\\
            &  H$\alpha $  &  0.51 $\pm$ 0.14 & --0.14 $\pm$ 0.06 & & 0.58 $\pm$ 0.16 & --0.22 $\pm$ 0.08\\
 FWHM       &  H$\beta  $  &  0.86 $\pm$ 0.25 & --0.23 $\pm$ 0.12 & & 0.91 $\pm$ 0.22 & --0.22 $\pm$ 0.10\\
            &  H$\alpha $  &  0.72 $\pm$ 0.30 & --0.23 $\pm$ 0.13 & & 0.76 $\pm$ 0.16 & --0.27 $\pm$ 0.08\\
\\
Combined Sample\\
 FWHM       &  H$\beta  $  &  0.82 $\pm$ 0.18 & --0.40 $\pm$ 0.07 & & 0.85$\pm$ 0.18 & --0.40$\pm$ 0.07\\
\enddata

\tablecomments{The fits discussed in \S\ref{ssec:civ_balmer_comp}
correspond those performed using the prescription B \civ\
line-widths. Fits obtained using prescription A measurements are shown
for completeness. The fits to the combined sample discussed in
\S\ref{sec:comp_others} are also reported here.}

\end{deluxetable}

\begin{deluxetable}{l l l  c c c  c  c c c}

\tablecaption{Correlations of Line-Width Estimates\label{tab:lw_residuals}}

\tablehead{\civ\ Prescription & \civ\  & Balmer & \multicolumn{3}{c}{Group I Estimates} & & \multicolumn{3}{c}{Group I \& II Estimates}\\
                    & Width  & Line   & $r_s$ & $N$ & $P_{\rm ran}$           & &  $r_s$ & $N$ & $P_{\rm ran}$}

\tabletypesize{\small}
\tablewidth{0pt}
\tablecolumns{10}

\startdata
\multicolumn{10}{l}{Without color correction}\\
Prescription A                      & $\sigma_l$     & H$\beta$        &\phn0.214 &    7 &    0.644 & &\phn0.300 &    9 &    0.433\\
                                    &                & H$\alpha$       &\phn0.700 &    5 &    0.188 & &\phn0.786 &    8 &    0.021\\
                                    & FWHM           & H$\beta$        &  --0.143 &    7 &    0.760 & &\phn0.000 &    9 &    1.000\\
                                    &                & H$\alpha$       &\phn0.300 &    5 &    0.624 & &\phn0.500 &    8 &    0.207\\
Prescription B                      & $\sigma_l$     & H$\beta$        &\phn0.679 &    7 &    0.094 & &\phn0.333 &    9 &    0.381\\
                                    &                & H$\alpha$       &\phn0.900 &    5 &    0.037 & &\phn0.524 &    8 &    0.183\\
                                    & FWHM           & H$\beta$        &  --0.143 &    7 &    0.760 & &\phn0.000 &    9 &    1.000\\
                                    &                & H$\alpha$       &\phn0.300 &    5 &    0.624 & &\phn0.500 &    8 &    0.207\\
Combined Sample                     & FWHM           & H$\beta$        &\phn0.346 &   52 &    0.012 & &\phn0.342 &   54 &    0.011\\
Combined Sample ($S/N>10$)          & FWHM           & H$\beta$        &\phn0.551 &   31 & $1.3\times10^{-3}$& &\phn0.539 &   33 & $1.2\times10^{-3}$\\
\\
With color correction & & & & & & & & &\\
Prescription A                      & $\sigma_l$     & H$\beta$        &\phn0.571 &    7 &    0.180 & &\phn0.550 &    9 &    0.125\\
                                    &                & H$\alpha$       &\phn0.900 &    5 &    0.037 & &\phn0.905 &    8 &    0.002\\
                                    & FWHM           & H$\beta$        &\phn0.357 &    7 &    0.432 & &\phn0.383 &    9 &    0.308\\
                                    &                & H$\alpha$       &\phn0.400 &    5 &    0.505 & &\phn0.738 &    8 &    0.037\\
Prescription B                      & $\sigma_l$     & H$\beta$        &\phn0.929 &    7 &    0.003 & &\phn0.467 &    9 &    0.205\\
                                    &                & H$\alpha$       &\phn0.900 &    5 &    0.037 & &\phn0.452 &    8 &    0.260\\
                                    & FWHM           & H$\beta$        &\phn0.357 &    7 &    0.432 & &\phn0.383 &    9 &    0.308\\
                                    &                & H$\alpha$       &\phn0.400 &    5 &    0.505 & &\phn0.667 &    8 &    0.071\\
Combined Sample                     & FWHM           & H$\beta$        &\phn0.326 &   52 &    0.018 & &\phn0.323 &   54 &    0.017\\
Combined Sample ($S/N>10$)          & FWHM           & H$\beta$        &\phn0.602 &   31 & $3.4\times10^{-4}$& &\phn0.587 &   33 & $3.3\times10^{-4}$\\
\enddata

\tablecomments{The table shows the correlation strength of the
  \civ\ and Balmer line-width estimates for both \civ\ line-width
  measurement prescription, quantified by the Spearman rank order
  coefficient, $r_s$. Results are shown for group I and the
  combination of groups I and II measurements. In each case, $N$
  indicates the number of lenses used to estimate the correlation
  strength and $P_{\rm ran}$ indicates the probability of observing
  such a correlation by chance if the variables are uncorrelated.}

\end{deluxetable}

\newpage 

\begin{figure}
  \begin{center}
    \includegraphics[width=0.49\textwidth]{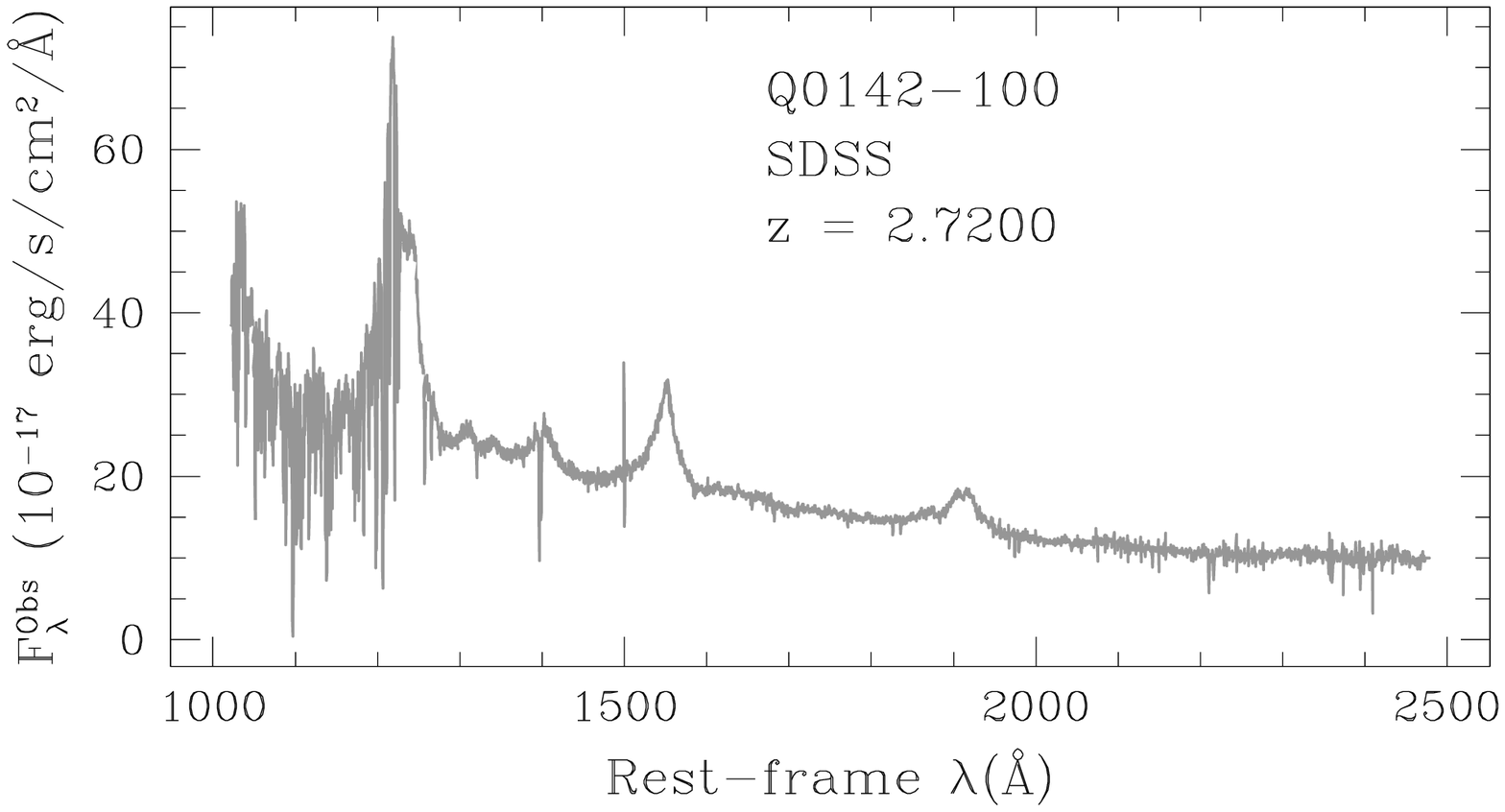}
    \includegraphics[width=0.49\textwidth]{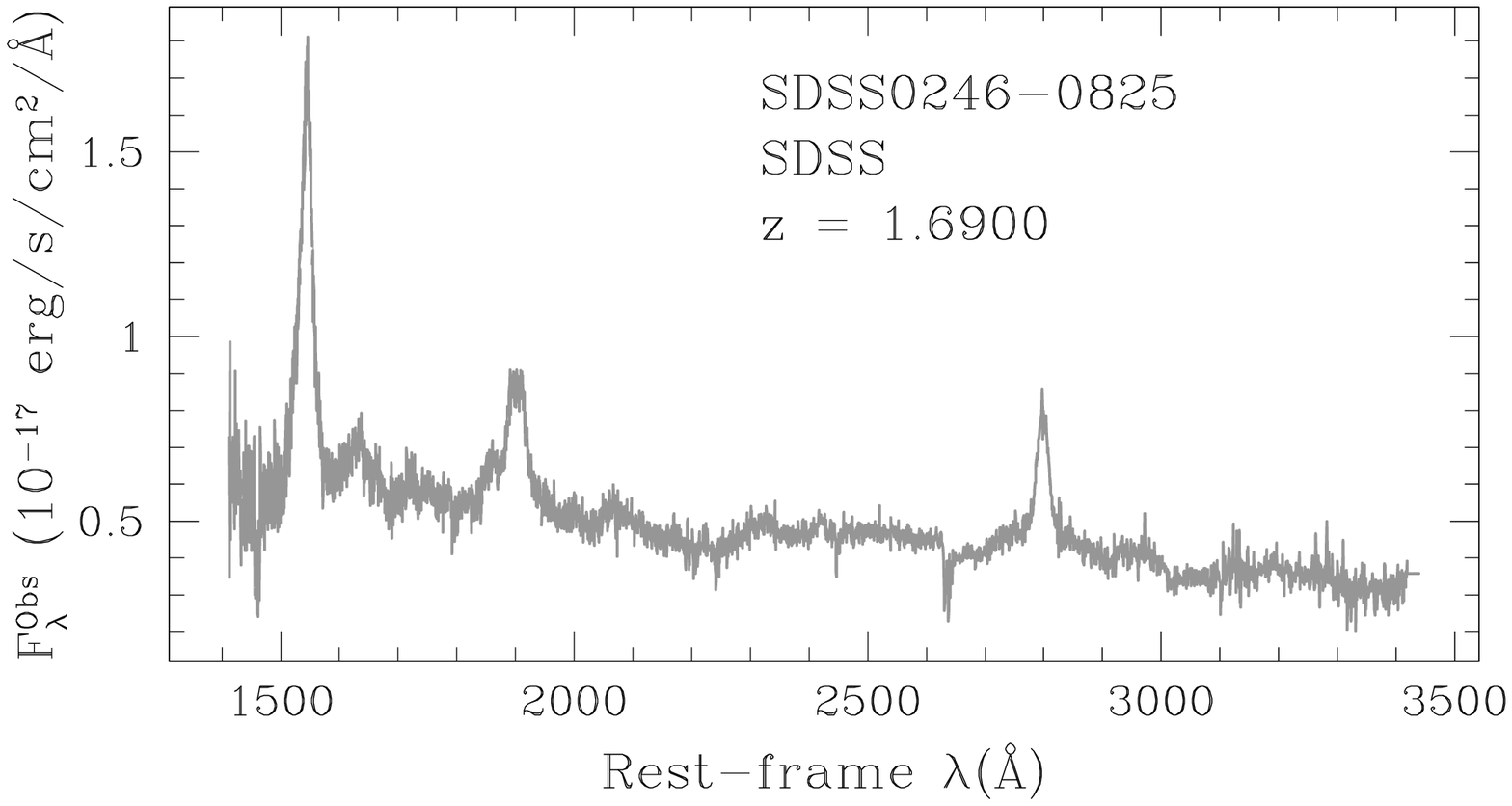}
    \includegraphics[width=0.49\textwidth]{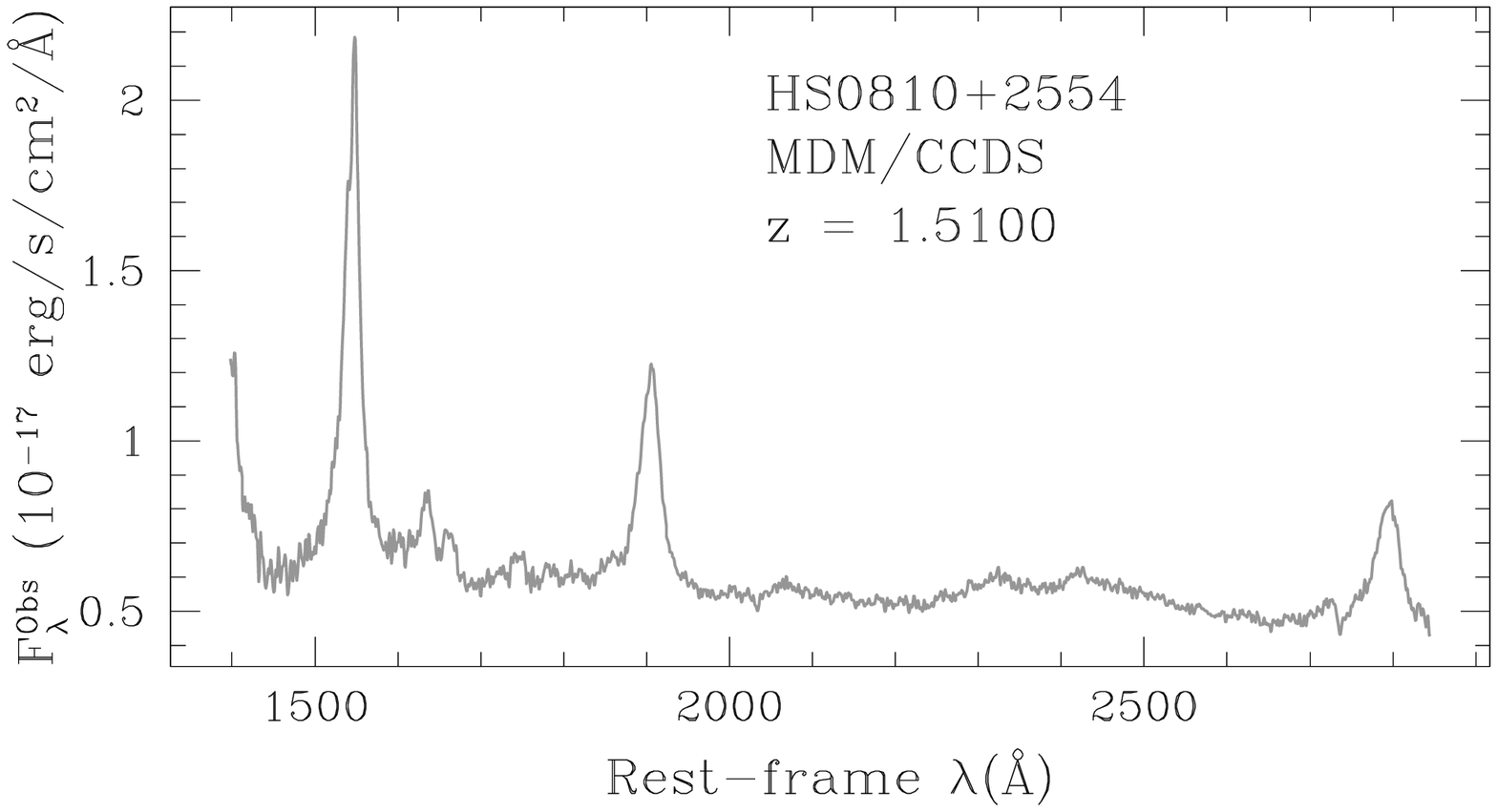}
    \includegraphics[width=0.49\textwidth]{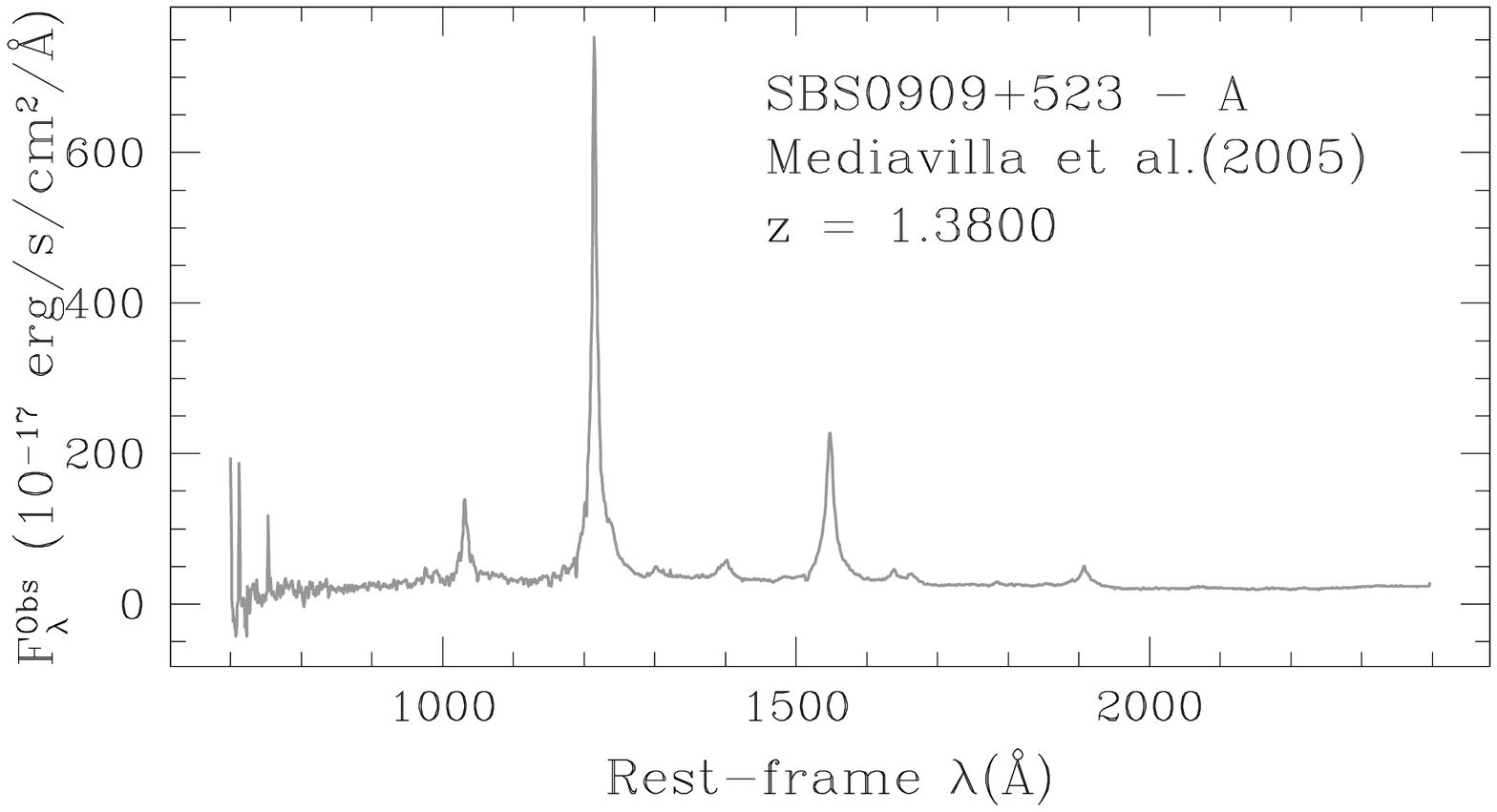}
    \includegraphics[width=0.49\textwidth]{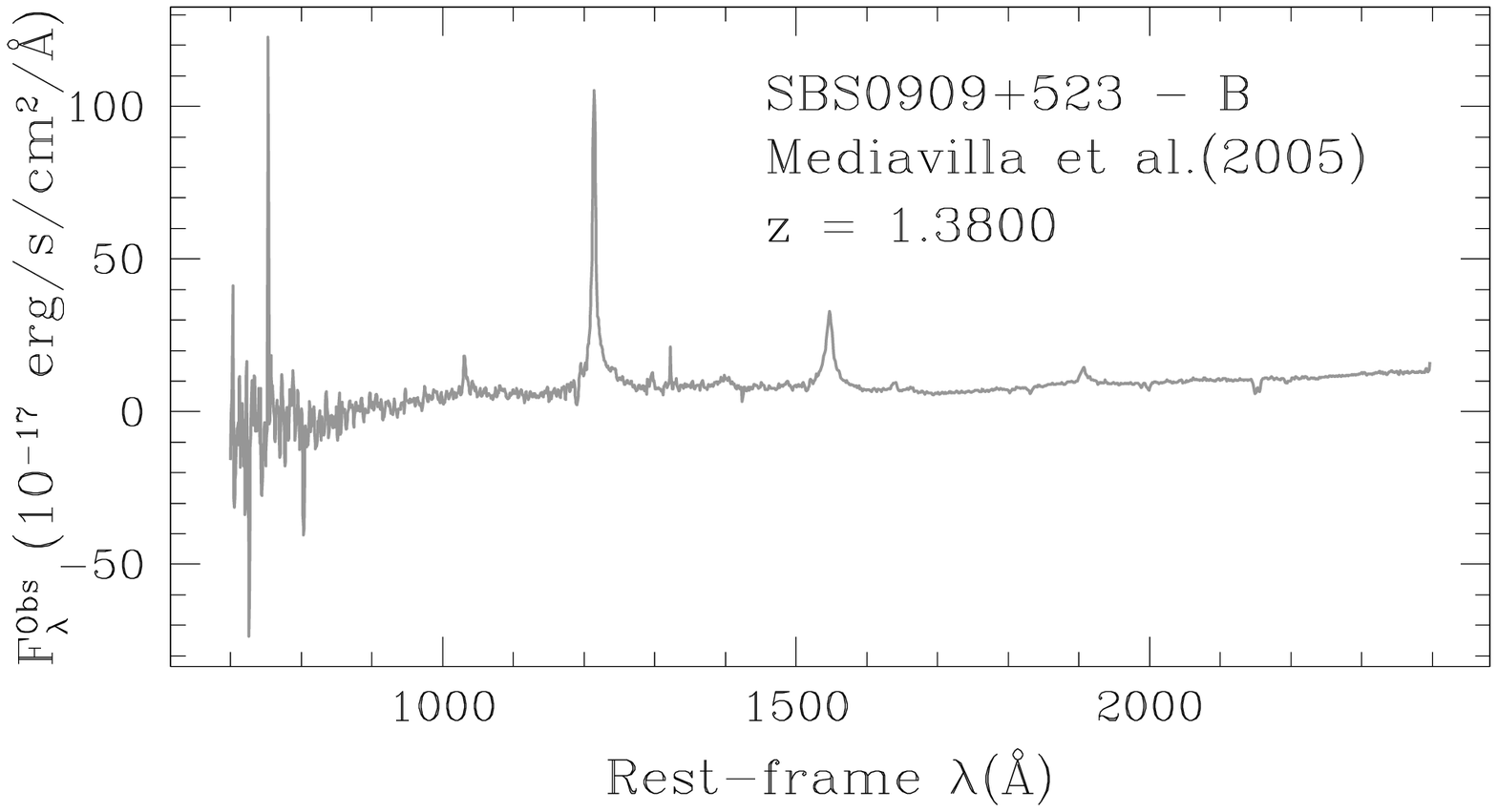}
    \includegraphics[width=0.49\textwidth]{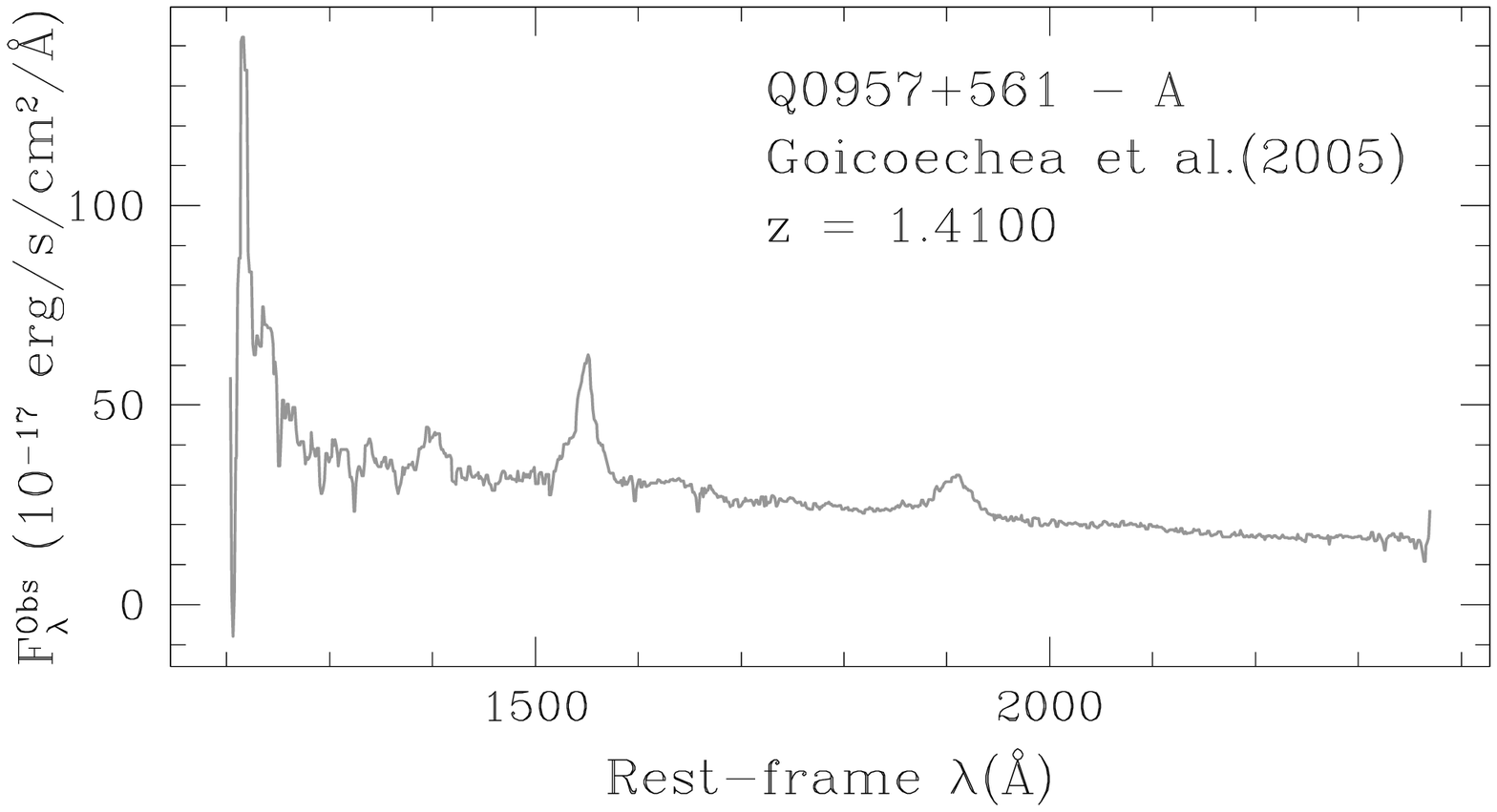}
    \includegraphics[width=0.49\textwidth]{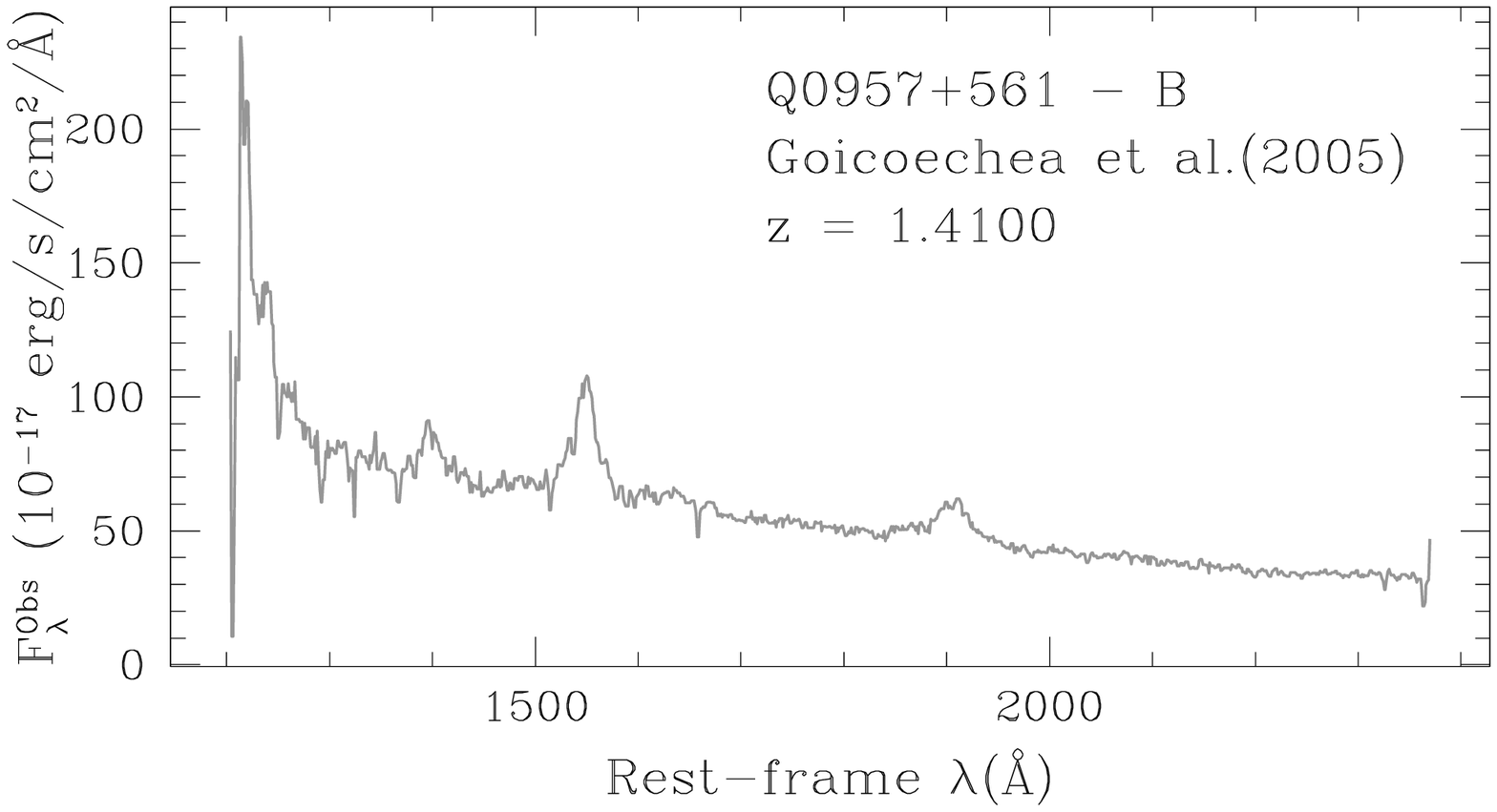}
    \includegraphics[width=0.49\textwidth]{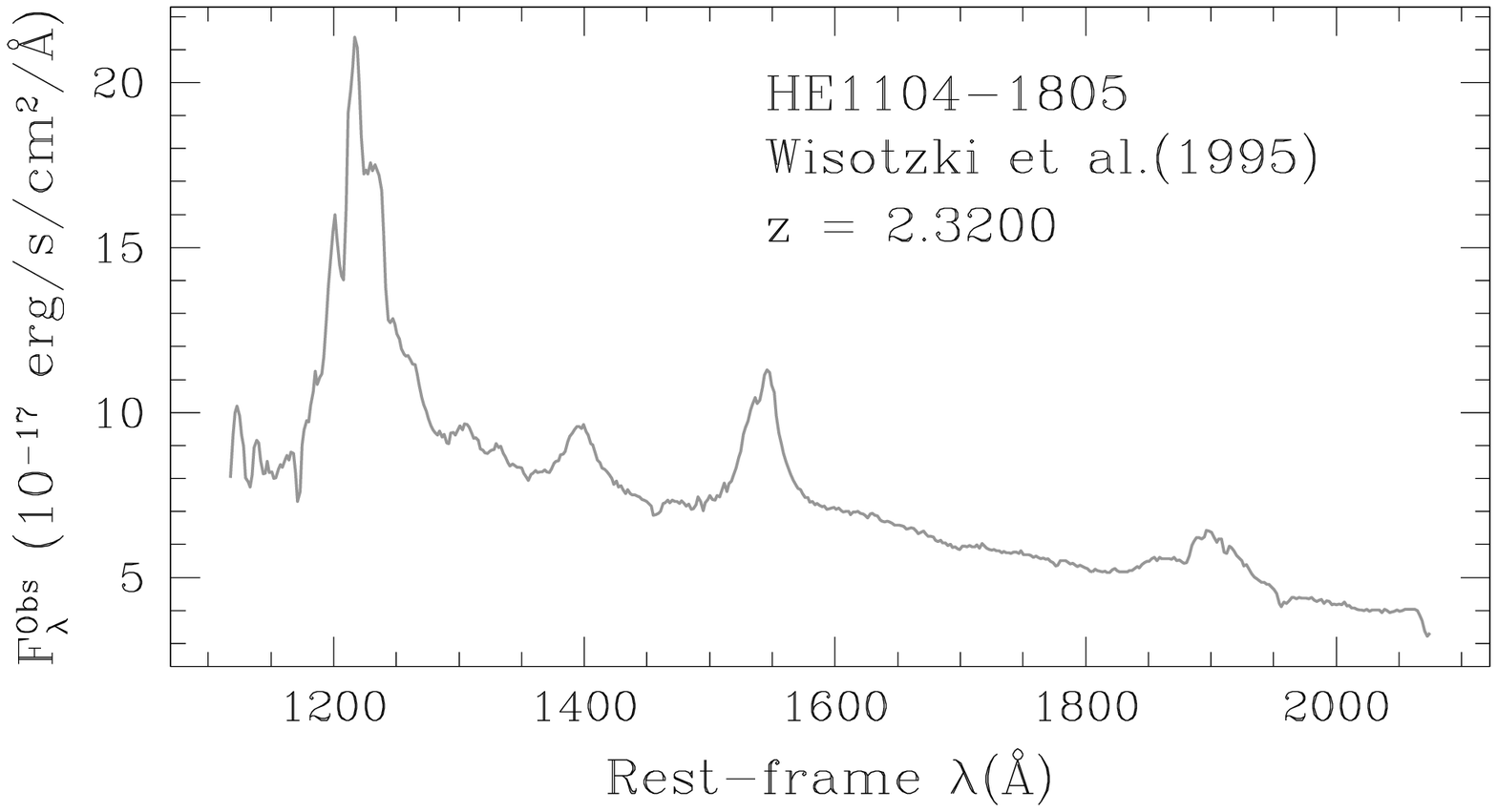}

    \caption{UV/optical spectra of all objects used in this study. For
    each object the panel shows the full spectrum, the source from
    which it was obtained and its redshift.}
    \label{fg:all_uv_opt_spec}
  \end{center}
\end{figure}

\begin{figure}
  \begin{center}
    \includegraphics[width=0.49\textwidth]{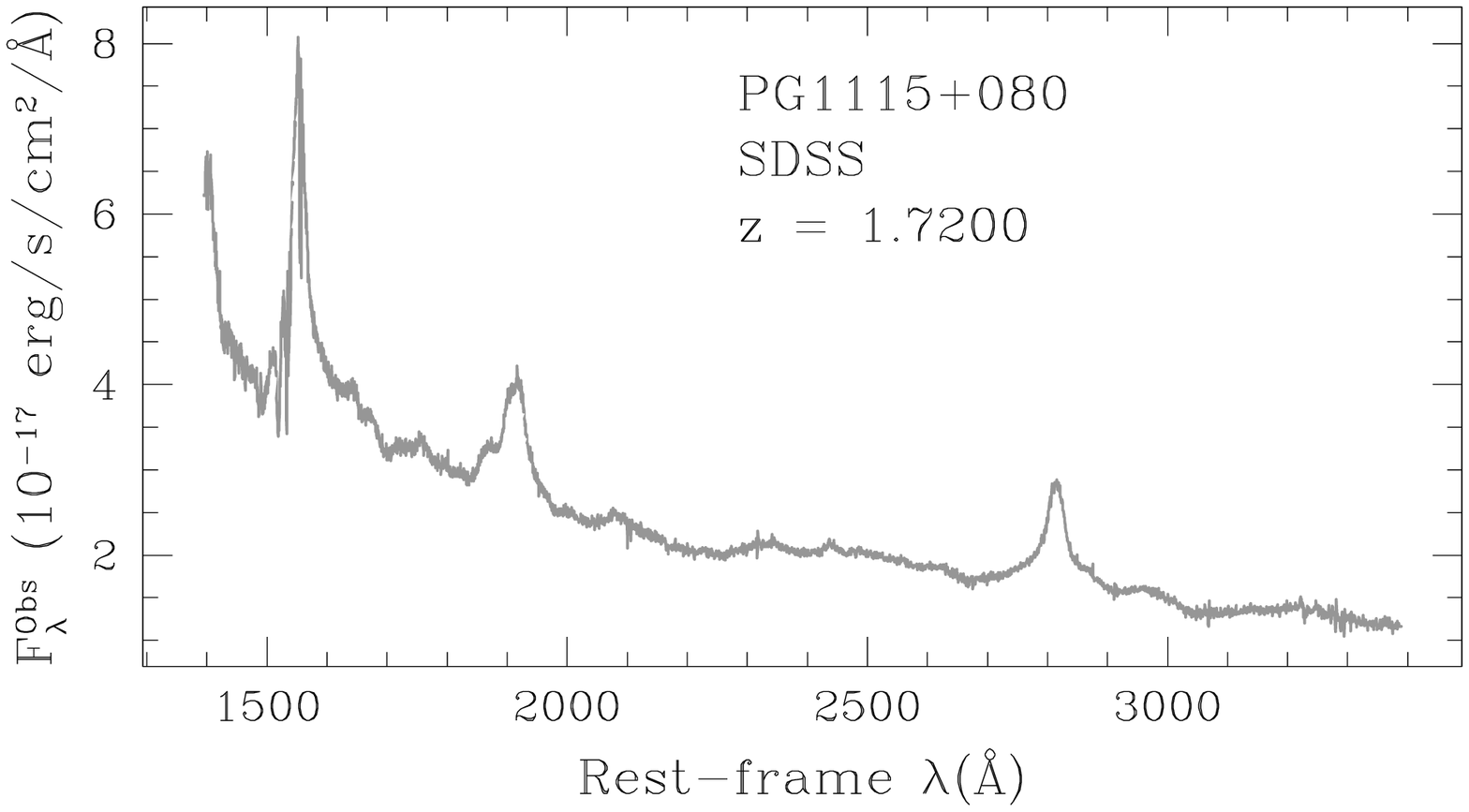}
    \includegraphics[width=0.49\textwidth]{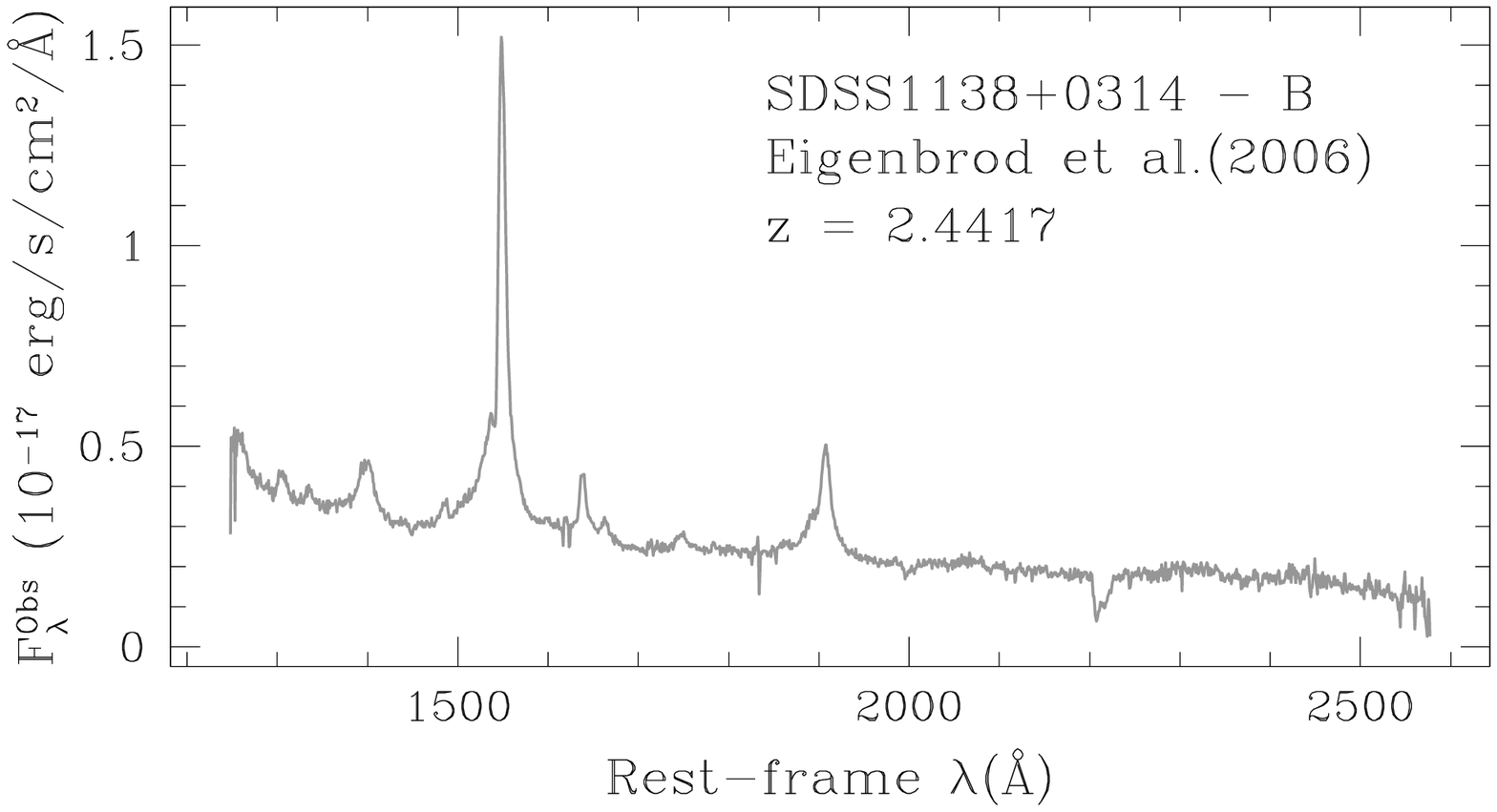}
    \includegraphics[width=0.49\textwidth]{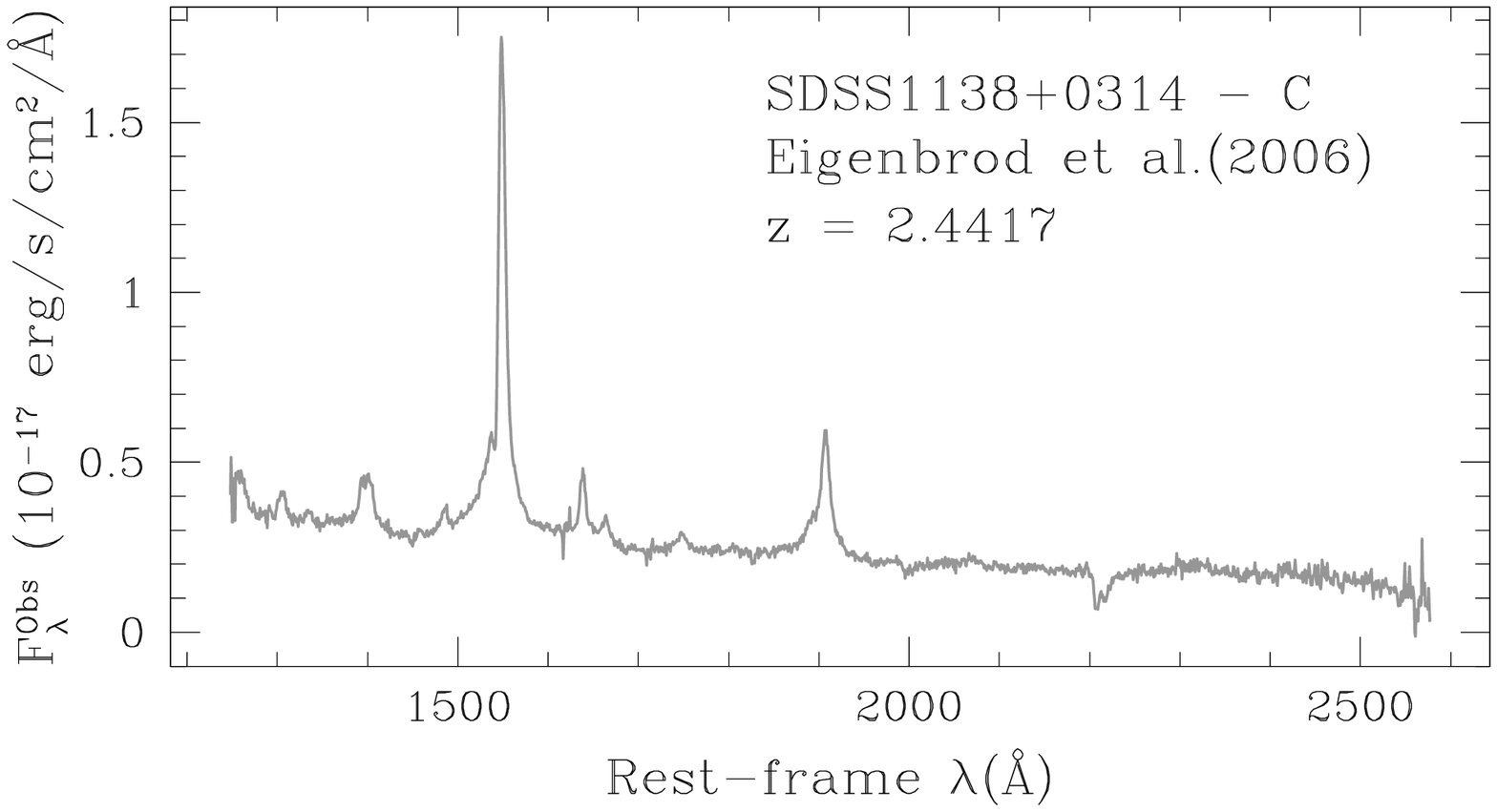}
    \includegraphics[width=0.49\textwidth]{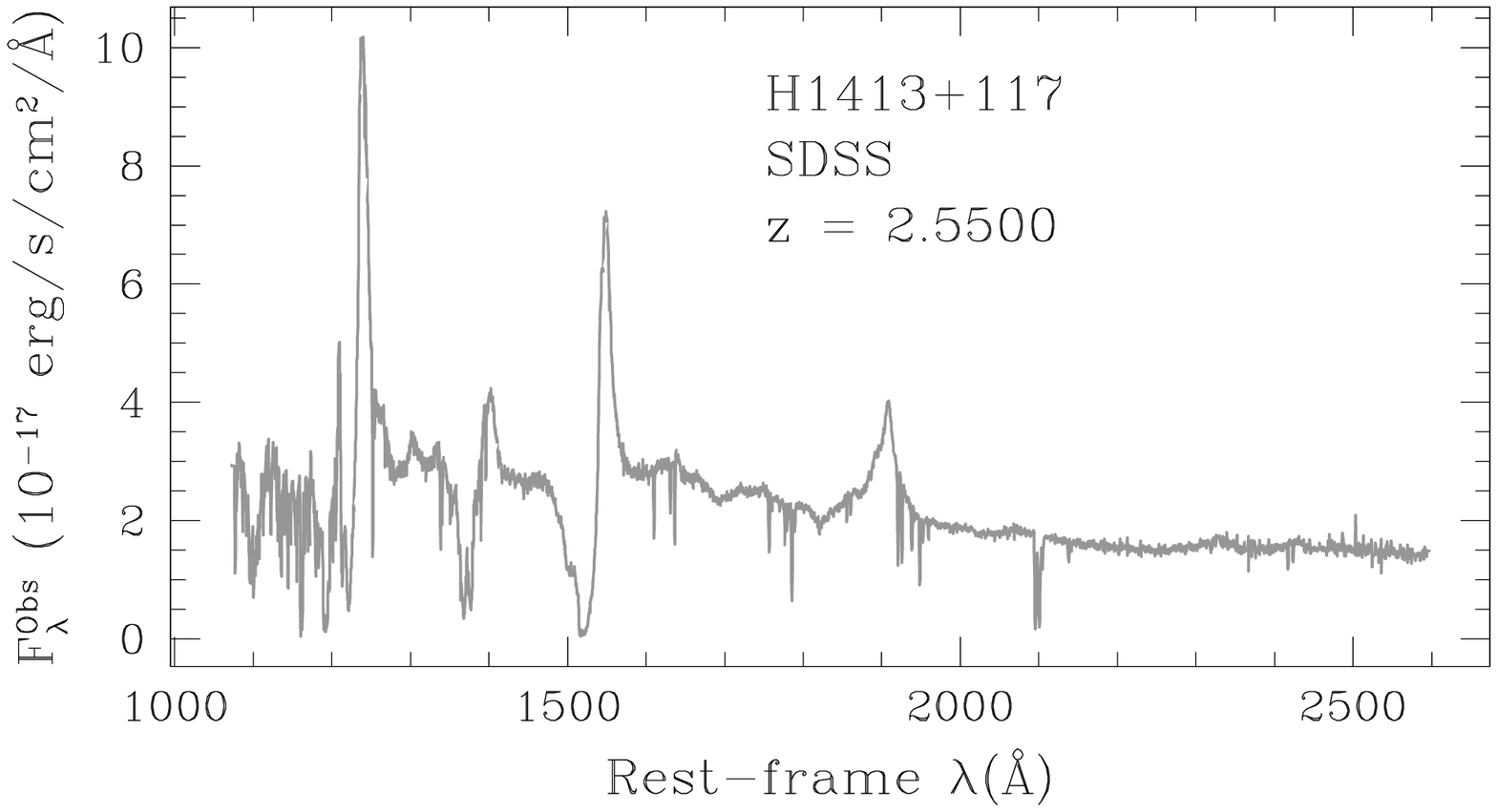}
    \includegraphics[width=0.49\textwidth]{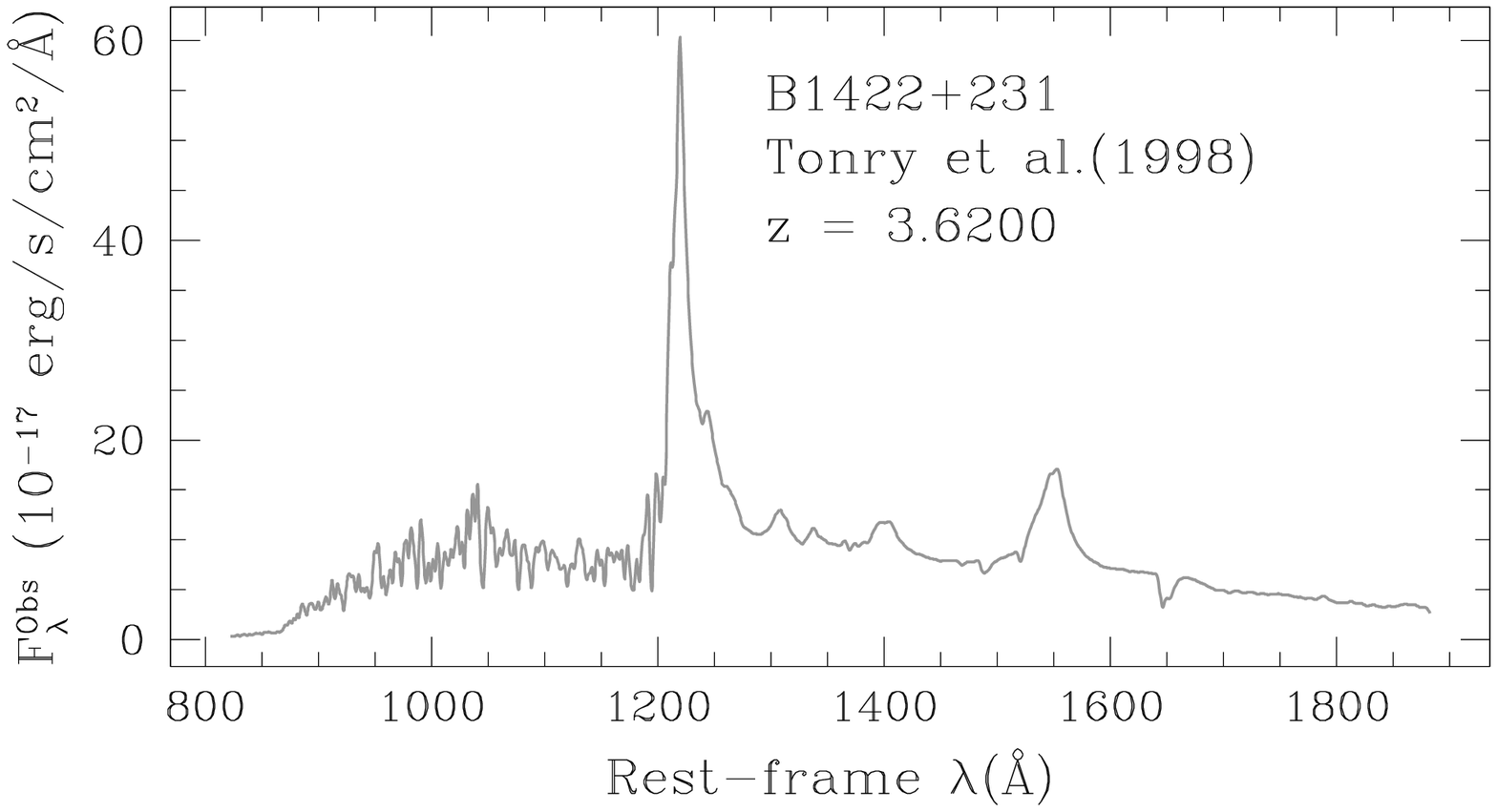}
    \includegraphics[width=0.49\textwidth]{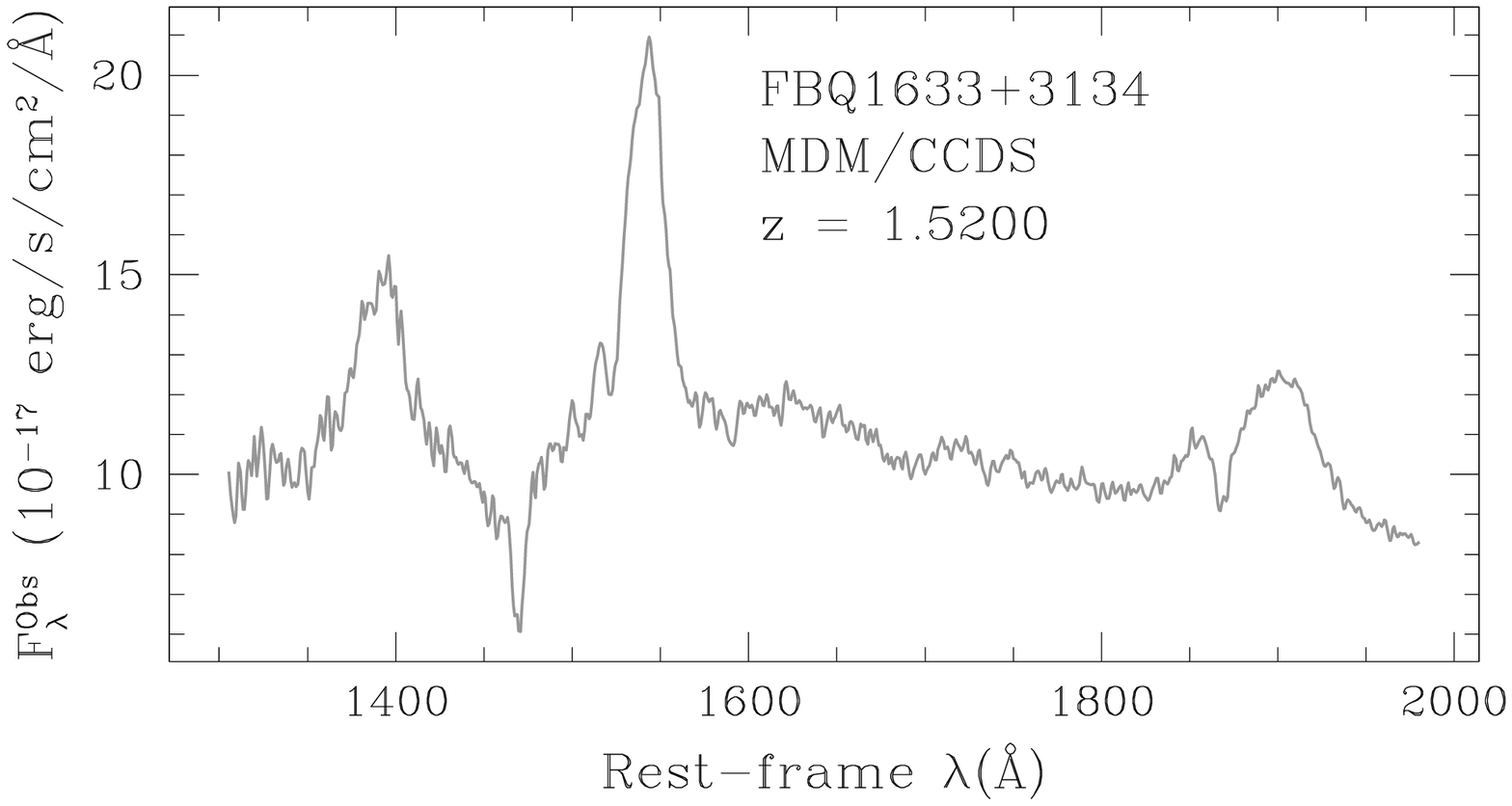}
    \includegraphics[width=0.49\textwidth]{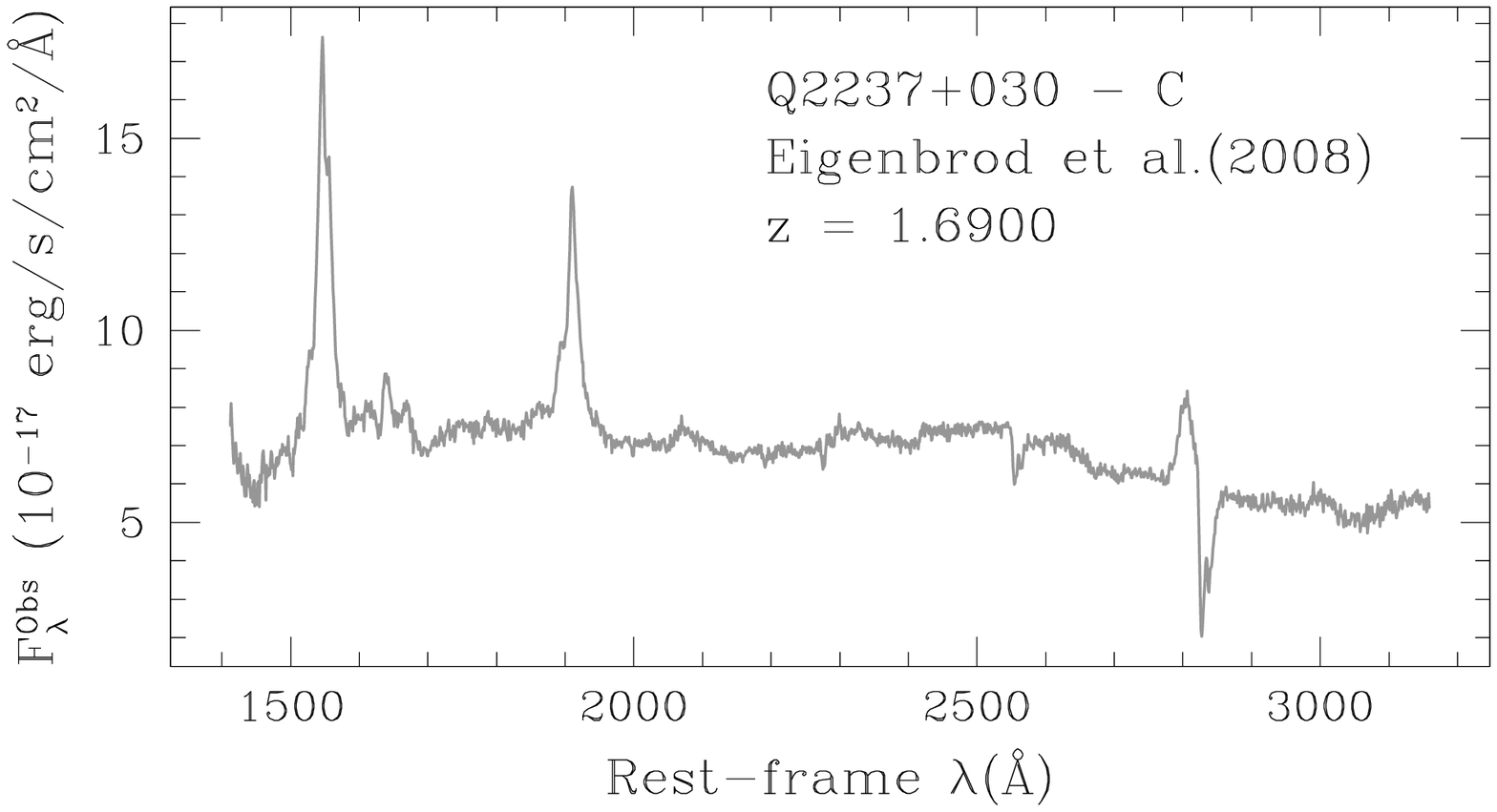}
    \includegraphics[width=0.49\textwidth]{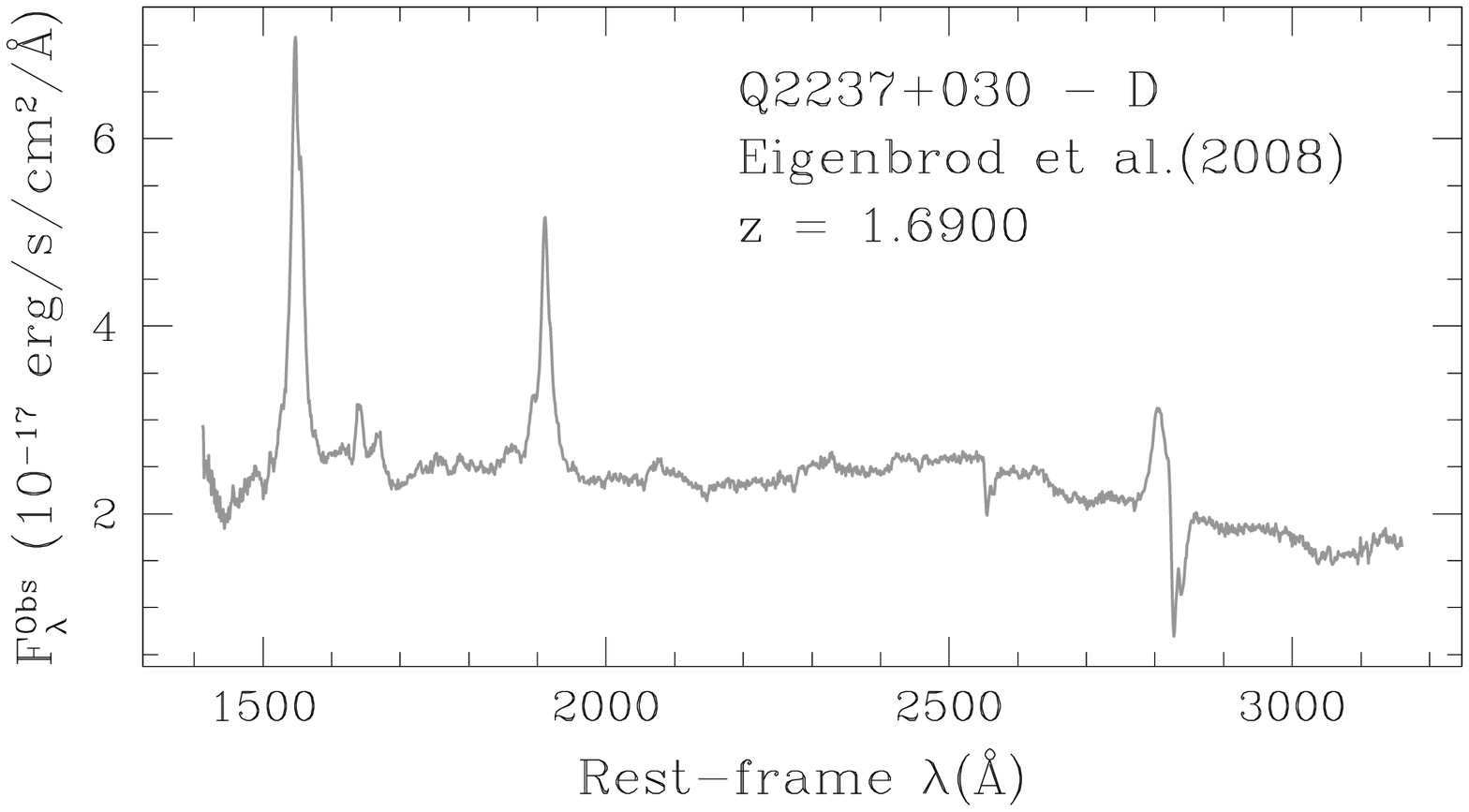}

    \figurenum{4}
    \caption{{\it{Continued}}}
  \end{center}
\end{figure}

\begin{figure}
  \begin{center}
    \plotone{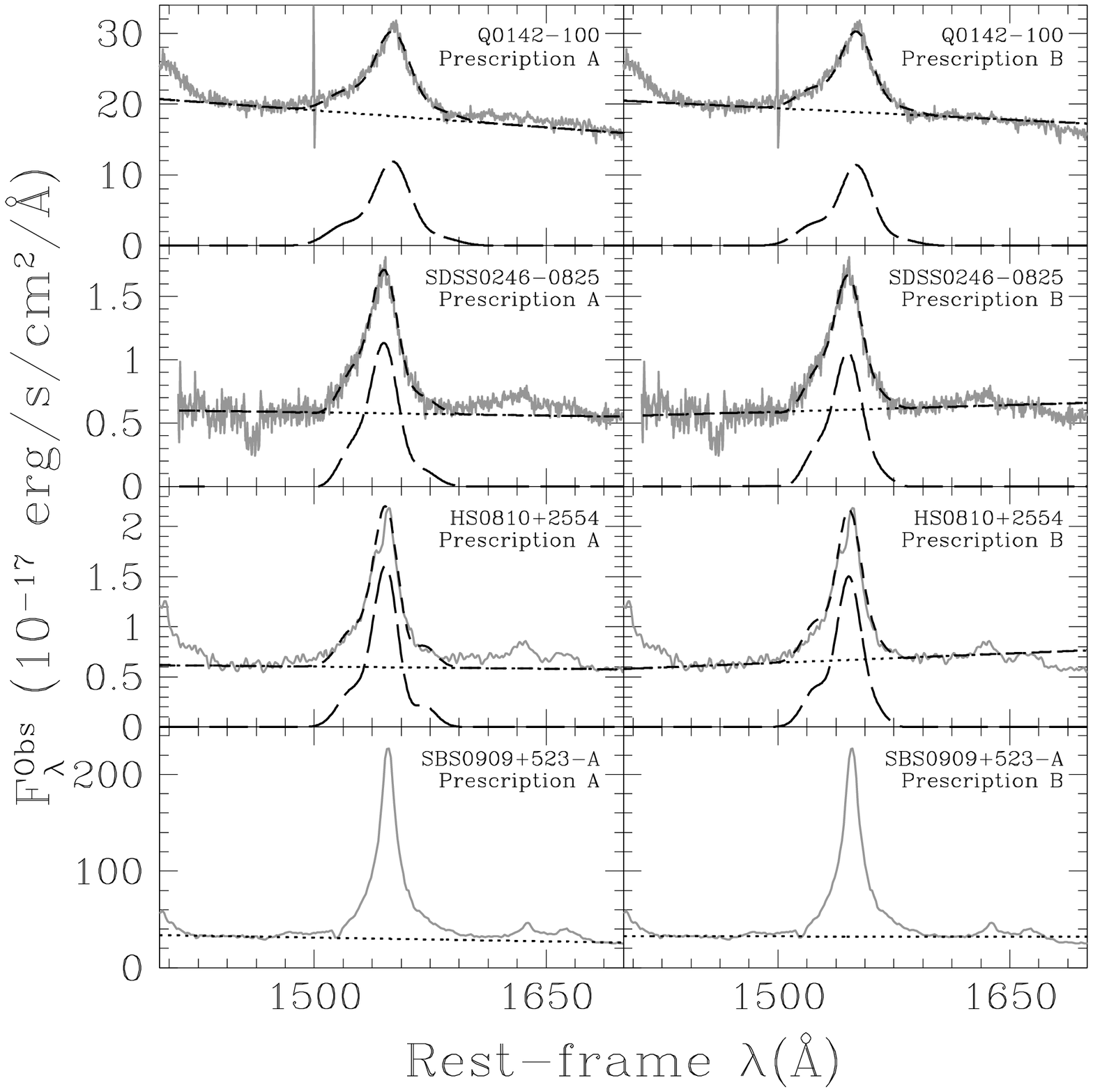}
    \caption{The Figure shows the region of each UV/optical spectrum
    around \civ\ along with the best fit continuum ({\it{dotted
    line}}) around the line, the best fit line profile ({\it{long
    dashed}}) and the addition of both ({\it{short dashed}}). For each
    spectrum, the fits obtained using prescription A are shown in the
    left panel while those obtained following prescription B are shown
    in the right panel. We could not obtain good fits for
    SDSS1138+0314 and SBS0909+532 \civ\ lines (see Appendix
    \ref{sssec:notes} for details).}
    \label{fg:all_uv_opt_fits}
  \end{center}
\end{figure}

\begin{figure}
  \begin{center}
    \plotone{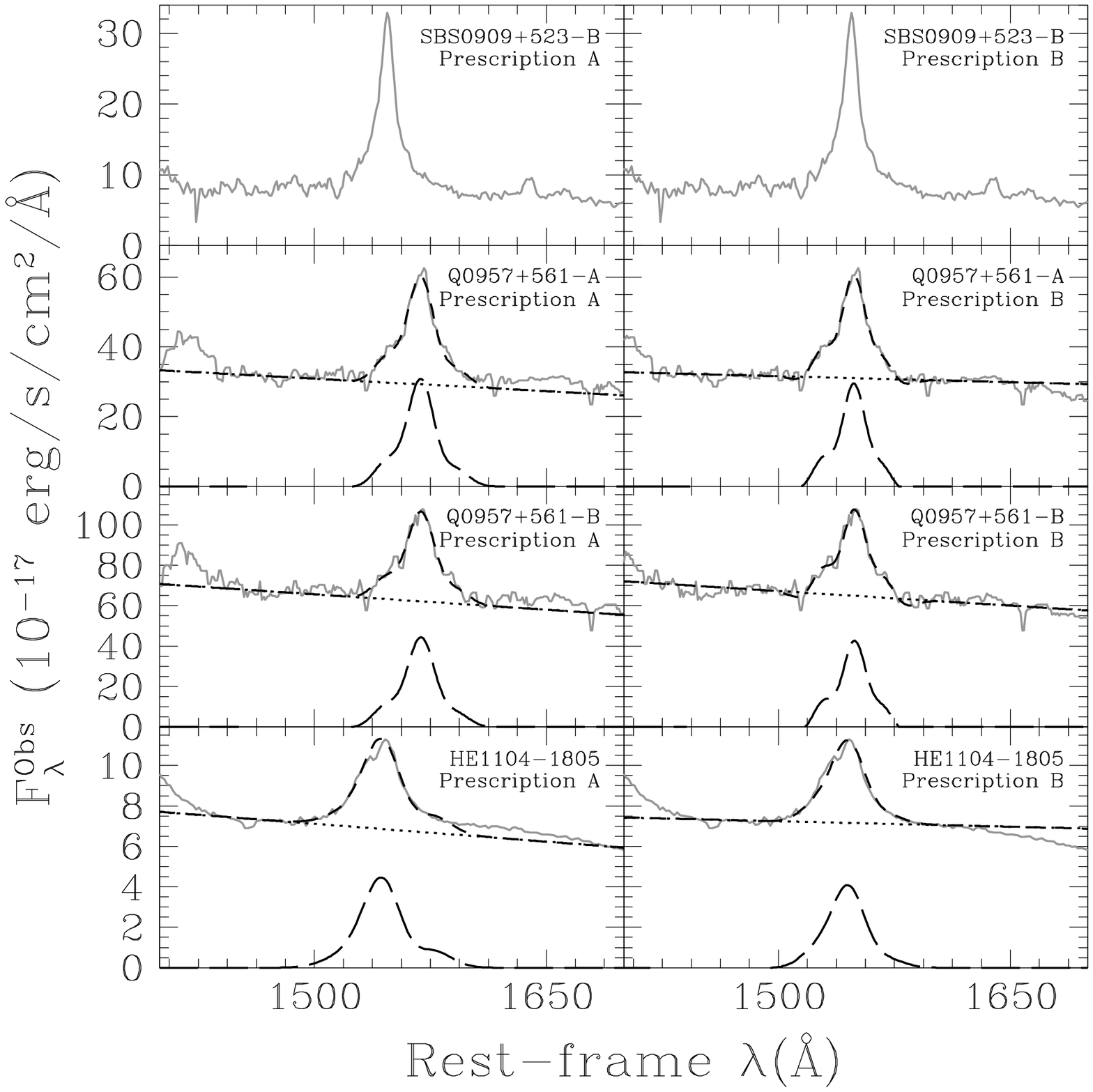}
    \figurenum{5}
    \caption{{\it{Continued}}}
  \end{center}
\end{figure}

\begin{figure}
  \begin{center}
    \plotone{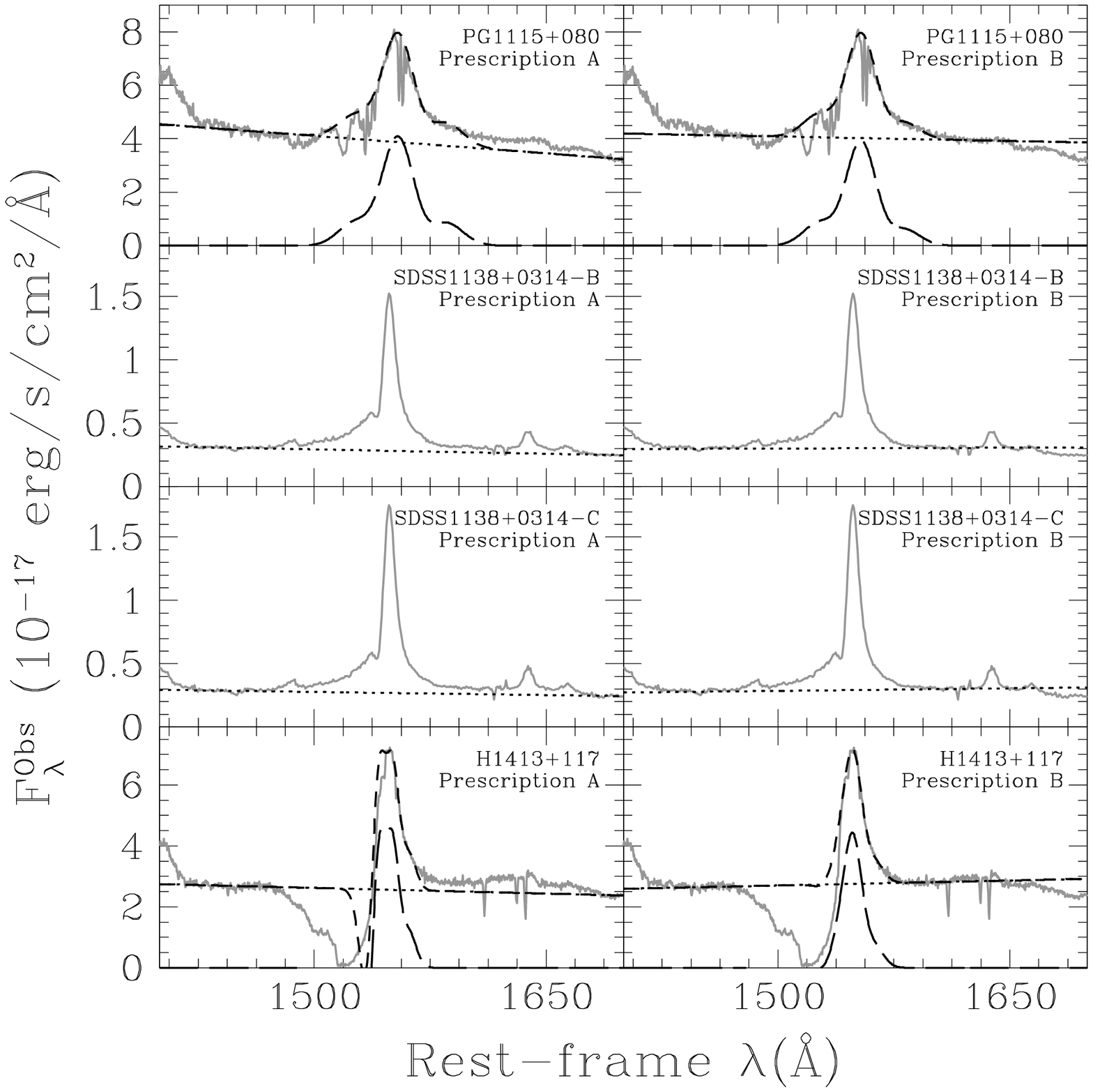}
    \figurenum{5}
    \caption{{\it{Continued}}}
  \end{center}
\end{figure}

\begin{figure}
  \begin{center}
    \plotone{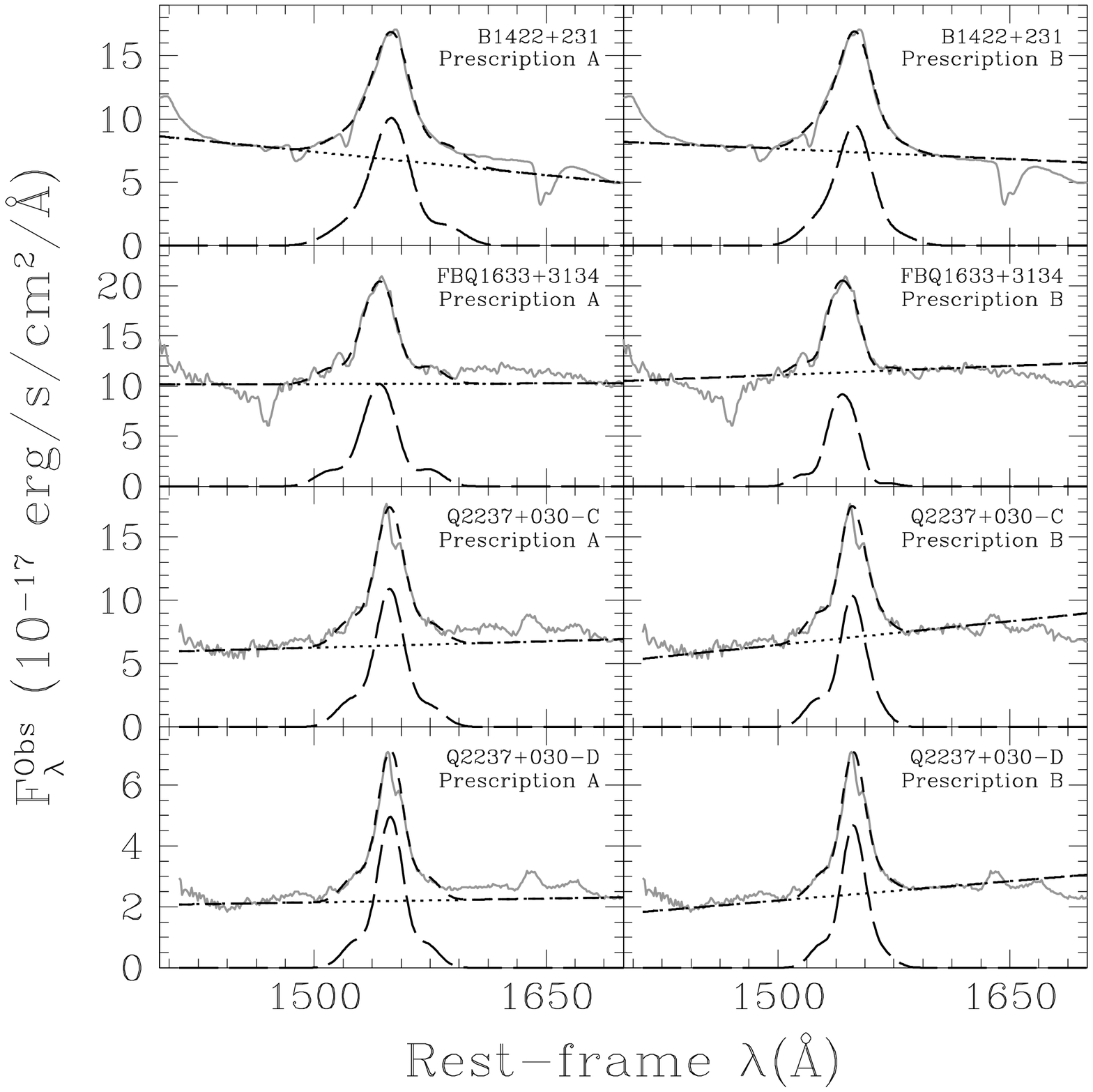}
    \figurenum{5}
    \caption{{\it{Continued}}}
  \end{center}
\end{figure}

\clearpage 

\clearpage

\end{document}